\documentclass[lettersize,journal]{IEEEtran}
\usepackage{amsmath,amsfonts}
\usepackage{array}
\usepackage{hyperref}
\usepackage{balance}
\usepackage{textcomp}
\usepackage{stfloats}
\usepackage{url}
\usepackage{verbatim}
\usepackage{graphicx}
\usepackage{cite}
\usepackage{subfigure}
\usepackage{threeparttable}
\usepackage{colortbl}
\usepackage{multirow}
\usepackage{color}
\usepackage{xcolor}
\usepackage{tabularx}
\usepackage{graphicx}
\usepackage[ruled]{algorithm2e}
\newcommand{\xu}[1]{{\leavevmode\color{black}#1}}
\begin{document}

\newcommand{\name}{AutoRec}
\newtheorem{theorem}{Theorem}
\newtheorem{lemma}{Lemma}
\newtheorem{proof}{Proof}[section]

\title{AutoRec: Accelerating Loss Recovery for Live Streaming in a Multi-Supplier Market}

\author
{
    Tong~Li,~\IEEEmembership{Member,~IEEE,}
    Xu~Yan,~
    Bo~Wu,~
    Cheng~Luo,~
    Fuyu~Wang,~\\
    Jiuxiang~Zhu,~
    Haoyi~Fang,~
    Xinle~Du,~
    Ke~Xu,~\IEEEmembership{Fellow,~IEEE}
        
    \thanks
    {
        %This work is supported in part by the National Key Research and Development Program of China (No.2022YFB3102301), the China National Funds for Distinguished Young Scientists (No.62425201), and the NSFC Projects (No.62202473, No.61932016, No.62132011, and No.62221003). 
        This work is supported in part by the National Natural Science Foundation of China under No. 62572473, No. 62202473, No.62441230 and No.62272466,  the Science Fund for Creative Research Groups of the National Natural Science Foundation of China under No. 62221003, the Key Program of the National Natural Science Foundation of China under No. 61932016 and No. 62132011, the National Science Foundation for Distinguished Young Scholars of China under No. 62425201, the Tencent Basic Platform Technology Rhino-Bird Focused Research Program. \textit{(Corresponding author: Bo Wu)}

        Tong Li, Xu Yan, Jiuxiang Zhu, Haoyi Fang are with the Key Laboratory of Data Engineering and Knowledge Engineering, and also with the School of Information, Renmin University of China, Beijing 100872, China (e-mail: \{tong.li, yanxu2, jiuxiangzhu, 2021201680\}@ruc.edu.cn).
        
        Bo Wu, Cheng Luo, and Fuyu Wang are with the Department of Cloud Architecture and Platform, Tencent Technologies. Email: wub14@tsinghua.org.cn, \{lancelotluo, ivanfywang\}@tencent.com
        
        % Xinle Du is with 2012 Labs, Huawei, Shenzhen 518129, China (e-mail: xinledu18.thucsnet@gmail.com).
        Xinle Du is with the Department of Computer Science and Technology, Tsinghua University, Beijing 100084, China (e-mail: xinledu18.thucsnet@gmail.com).
        
        % Xiangyu Gao is with the Institute for Network Sciences and Cyberspace, Tsinghua University, Beijing 100084, China (e-mail: gaoxy21@mails.tsinghua.edu.cn).
        
        % Hanlin Huang is with the Department of Computer Science and Technology, Tsinghua University, Beijing 100084, China (e-mail: hhl21@mails.tsinghua.edu.cn).
        
        % Zhuotao Liu is with the Institute for Network Sciences and Cyberspace, Tsinghua University, Beijing 100084, China, and also with Zhongguancun Laboratory, Beijing 100094, China (e-mail: zhuotaoliu@tsinghua.edu.cn).
        
        % Mowei Wang is with Datacom Production Line, Huawei, Shenzhen 518129, China (e-mail: wangmowei@huawei.com).
        
        Ke Xu is with the Department of Computer Science and Technology, Tsinghua University, Beijing 100084, China,
        and also with Zhongguancun Laboratory, Beijing 100094, China (e-mail: xuke@tsinghua.edu.cn).
    }
    
}

%\title{Toward Timeliness-Enhanced Loss Recovery \\for Large-Scale Live Streaming}




\maketitle

% \begin{abstract}
% This document describes the most common article elements and how to use the IEEEtran class with \LaTeX \ to produce files that are suitable for submission to the IEEE.  IEEEtran can produce conference, journal, and technical note (correspondence) papers with a suitable choice of class options. 
% \end{abstract}

\begin{abstract}
Due to the limited permissions for upgrading dual-side (i.e., server-side and client-side) loss tolerance schemes from the perspective of CDN vendors in a multi-supplier market, modern large-scale live streaming services are still using the automatic-repeat-request (ARQ) based paradigm for loss recovery, which only requires server-side modifications. In this paper, we first conduct a large-scale measurement study with up to 50 million live streams. We find that loss shows \emph{dynamics} and live streaming contains frequent \emph{on-off mode switching} in the wild. We further find that the recovery latency, enlarged by the ubiquitous retransmission loss, is a critical factor affecting live streaming's client-side QoE (e.g., video freezing). 
We then propose an enhanced recovery mechanism called AutoRec, which can transform the disadvantages of on-off mode switching into an advantage for reducing loss recovery latency without any modifications on the client side. 
AutoRec allows users to customize overhead tolerance and recovery latency tolerance and adaptively adjusts strategies as the network environment changes to ensure that recovery latency meets user demands whenever possible while keeping overhead under control.
% AutoRec also adopts an online learning-based policy to fit the dynamics of loss, balancing the tradeoff between recovery latency and overhead incurred. 
% AutoRec based on the user-customizable overhead accelerates loss recovery to meet user demands whenever pos-
% sible while keeping the overhead controllable. 
% Specifically, it adopts an \emph{online learning-based policy} to dynamically set the number of replicas. 
% It sets the number of replicas based on the user-customizable overhead tolerance, recovery latency tolerance, and the current network environment.
% This assures loss recovery is accelerated to meet user demands whenever
% possible while keeping the overhead controllable
We implement AutoRec upon QUIC and evaluate it via testbed and real-world commercial services deployments. 
The experimental results demonstrate the practicability and profitability of AutoRec.

%, in which the average 
% 95th-percentile 
%times and duration of client-side video freezing can be lowered by 11.4\% and 5.2\%, respectively. 

\end{abstract}


\section{Introduction}

\IEEEPARstart{I}{nternet} live services such as Youtube Live, TikTok Live, and Twitch have gradually become a fundamental element for enriching daily life and work \cite{livestreamingmarket,videogame}. 
This underscores the critical need to enhance live streaming transmission performance 
The ubiquitous packet loss is an essential factor affecting client-side quality-of-experience (QoE) \cite{li2022livenet,kim2020neural}, which will introduce head-of-line (HOL) blocking and even incur long-time video freezing if the available frames in the player buffer are all consumed. 
Therefore, loss tolerance control matters in live-streaming services. 

\begin{figure}[t]
  \centering
  \includegraphics[width=6.37cm]{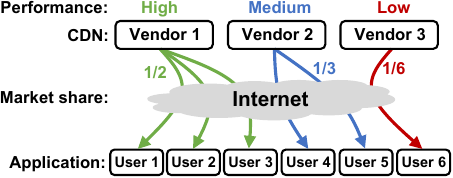}\\
  \vspace{-2mm}
  \caption{An example of a multi-supplier market for CDN vendors in large-scale live streaming services.}\label{market_share_fig}
\vspace{-0.4cm}
\end{figure}

The existing loss tolerance schemes mainly focus on designing dual-side (i.e., service-side CDN and client-side application) control policies, including Forward Error Correction (FEC) \cite{chan2006video,michel2019quic,michael2023tambur,meng2024hairpin}, multi-path retransmissions \cite{zheng2021xlink,chen2018fuso}, semi-reliable transmissions \cite{palmer2021voxel,pauly2018unreliable}, and application-level controls \cite{vamanan2012deadline,zhou2021deadline,zhang2015more}.
However, as shown in Fig.~\ref{market_share_fig}, live-streaming application operators (e.g., TikTok Live) usually apply the Multi-Supplier Strategy~\cite{arda2006inventory} in the CDN market (see \S \ref{sec:strategy_sec}). As a result, it is the CDN vendor's duty that optimize the transmission performance (e.g., loss tolerance), according to which the application operators will choose the better-performed CDN vendors to carry more traffic (i.e., larger market share). In this case, only server-side sending policies can be adjusted by the selected CDN vendors, which lack the proper authority to synchronize client-side control rules. 
% Thus, the above-mentioned arts for loss tolerance control suffer from deployment issues in the multi-supplier CDN market.
Thus, the aforementioned loss tolerance schemes face significant deployment challenges in a multi-supplier CDN market.

In this case, most modern CDN vendors only apply the automatic-repeat-request (ARQ) paradigm \cite{bertsekasandr1992gallager,li2021acknowledgment} to control loss tolerance as the commercial solution, which retransmits only one replica of the lost packet when a loss is detected. 
However, we find that the legacy ARQ-based loss recovery is far from satisfactory according to our performed large-scale measurements. For example, the proportion of connections with maximum retransmission times of two or more exceeds 43\%. Among them, a considerable portion of the connections has certain packets that are retransmitted even more than 10 times (\S \ref{sec:background_motivation}).
The retransmission loss enlarges the loss recovery latency by 123.2 ms on average and 279.3 ms in the worst case. This enlarged recovery latency further increases the probability of empty buffer space on the client side, thereby increasing the risk of video freezing (\S\ref{loss_recovery_quality_sec}). Thus from a philosophical standpoint, it is worth asking:
Can we accelerate loss recovery solely through CDN servers without modifying clients? How can it be addressed with controllable overhead?

\IEEEpubidadjcol

Our key insight is that the on-off mode switching ubiquitously occurs in current live streams (\S \ref{onoff_sec}), where the bandwidth during the "off" periods is not fully utilized.
Our measurements show that each live stream spends 464 ms in off-mode per second and enters off-mode 20 times per second on average. 
It is well-studied that the on-off traffic pattern is not conducive to transmission control \cite{benson2010understanding,benson2010network,cheng2022estimation,sivaraman2014experimental,rajasekaran2022congestion,floyd2000equation}. 
However, we argue that it can act as an advantage for the loss tolerance control of live streaming. In this paper, we present \texttt{\name{}}. This enhanced loss recovery mechanism can transform the disadvantages of frequent on-off mode switching into an advantage of loss tolerance control in live streaming. 

% A straightforward way of AutoRec is to directly reinject a fixed number of replicas of loss packets once stepping into off-streaming mode. 
A naive implementation of \name{} is to directly reinject a fixed number of replicas of loss packets upon entering the off-mode.
However, this approach faces two challenges. \textbf{First}, the fixed settings of redundant replicas cannot adapt well to the dynamics of packet loss, while excessive reinjection of packets results in non-trivial recovery overhead. 
\textbf{Second}, according to our testbed experiments, the off-mode can be unevenly distributed throughout an entire live session, and the sending of replicas might be delayed without timely off-mode entry. To tackle these issues, we propose a two-step solution: Redundancy Adaption and Reinjection Control.

\emph{Redundancy Adaption} intelligently determines how many replicas of the lost packets should be reinjected. 
% Specifically, it adopts an \emph{online learning-based policy} to dynamically set the number of replicas. 
It sets the number of replicas based on the user-customizable overhead tolerance, recovery latency tolerance, and the network environment.
This assures loss recovery is accelerated to meet user demands whenever
possible while keeping the overhead controllable.
% send the least number of replicas that adapt to the dynamics of packet loss.
% This assures a minimized redundancy overhead while accelerating loss recovery. 

\emph{Reinjection Control} determines when to retransmit the replicas. Generally, it enables the replica reinjection during the off-modes. \name{} further adopts the \emph{opportunistic reinjection} to trigger loss reinjection even lacking the desired opportunity of off-mode. This assures that each replica can be reinjected in time even under uneven distribution of off-modes.

We implement the \name{} prototype in the user-space QUIC protocol and deploy it on both testbed and real-network CDN proxy for 6 months. The experimental results demonstrate the practicability and profitability of \name{}, in which the average times and duration of client-side video freezing can be lowered by 10.10\% and 4.74\%, respectively. 

The rest of the paper is organized as follows: \S \ref{sec:background} introduces the multi-supplier CDN market and its impact on CDN vendors' accelerating loss recovery. \S \ref{motivation_sec} motivates our work with a large-scale measurement study. Then, the high-level architecture and design details of \name{} are depicted in \S \ref{framework_sec} and \S \ref{details_sec}, respectively. \S \ref{implement_sec} describes the detailed implementation of \name{}. \S \ref{experimental_sec} gives the experimental evaluations of \name{}. \S \ref{discussion_sec} presents a discussion of \name{}’s overhead and its interaction with QUIC’s native mechanisms. \S \ref{relatedwork_sec} overviews the related work and \S \ref{conclusion_sec} concludes the paper. 

%A straightforward way of AutoRec is to directly reinject loss duplicates once stepping into off-mode for accelerating loss recovery of live streams. 

%However, to better apply AutoRec without significant side effects, several challenges that cannot be ignored should be first overcome, as follows. 
%(i) The pre-configured or unchanged reinjection policies cannot well adapt to the dynamics of packet loss (\S \ref{sec:background_motivation}), where some live streams might be plagued by worse recovery timeliness or recovery overhead. 
%(ii) The off-mode is unevenly distributed throughout an entire live session, in which more recovery tasks might encounter fewer off-mode-based reinjection opportunities (\S \ref{onoff_sec}). 
%(iii) Excessive reinjection packets for optimizing unsatisfied \textsf{recovery latency} (\S \ref{loss_recovery_quality_sec}) will occupy (or even exhaust) transmission resources, resulting in non-trivial recovery overhead (e.g., lower goodput and extra traffic cost). 

\vspace{-1mm}

\section{Background}
\label{background_sec}
\label{sec:background}
In this section, we first introduce the Multi-Supplier Strategy in the modern CDN market (\S \ref{sec:strategy_sec}). We then discuss the appeal of accelerating loss recovery (\S \ref{sec:freezing_sec}), along with the restriction on the loss recovery mechanism (\S \ref{sec:fec_sec}) from the perspective of CDN vendors in the multi-supplier CDN market.

\vspace{-2mm}
\subsection{Multi-Supplier CDN Market}\label{sec:strategy_sec}
% Live streaming service providers typically subscribe to multiple CDN vendors simultaneously, periodically assessing the performance of these services through various weighted QoE metrics. Based on this evaluation, they reallocate their subscription shares according to service performance.
Live-streaming application operators typically subscribe to multiple CDN vendors simultaneously and periodically evaluate the performance of these vendors' services based on weighted QoE metrics, then reallocate subscription shares accordingly. 
For example, as shown in Fig. \ref{market_share_fig}, a live-streaming application operator subscribes to the services of three CDN vendors: Vendor 1, Vendor 2, and Vendor 3. In the previous performance evaluation period conducted by the live-streaming application operator, Vendor 1 had the highest performance, Vendor 2 had medium performance, and Vendor 3 had the lowest performance. Therefore, the live-streaming application operator will allocate 1/2 of all service requests to Vendor 1, 1/3 to Vendor 2, and 1/6 to Vendor 3. For live-streaming application operators, the Multi-Supplier Strategy not only allows for a comparative analysis of different CDN vendors' service performance, facilitating the selection of higher-performing CDN vendors but also continually drives CDN vendors to enhance their service performance, leading to better CDN services for the operators themselves. Thus, live-streaming application operators are motivated to adopt the Multi-Supplier Strategy, leading to the formation of a multi-supplier CDN market where multiple CDN vendors serve the same video live-streaming application operator.

\vspace{-4mm}

\subsection{Appeal of Accelerating Loss Recovery}
\label{sec:freezing_sec}
% \section{Video Freezing Measurements}

\begin{figure}[tbp]
\centering
\subfigure[Area-variant video freezes]{
\includegraphics[width=3.8cm,height=2.22cm,angle=0]{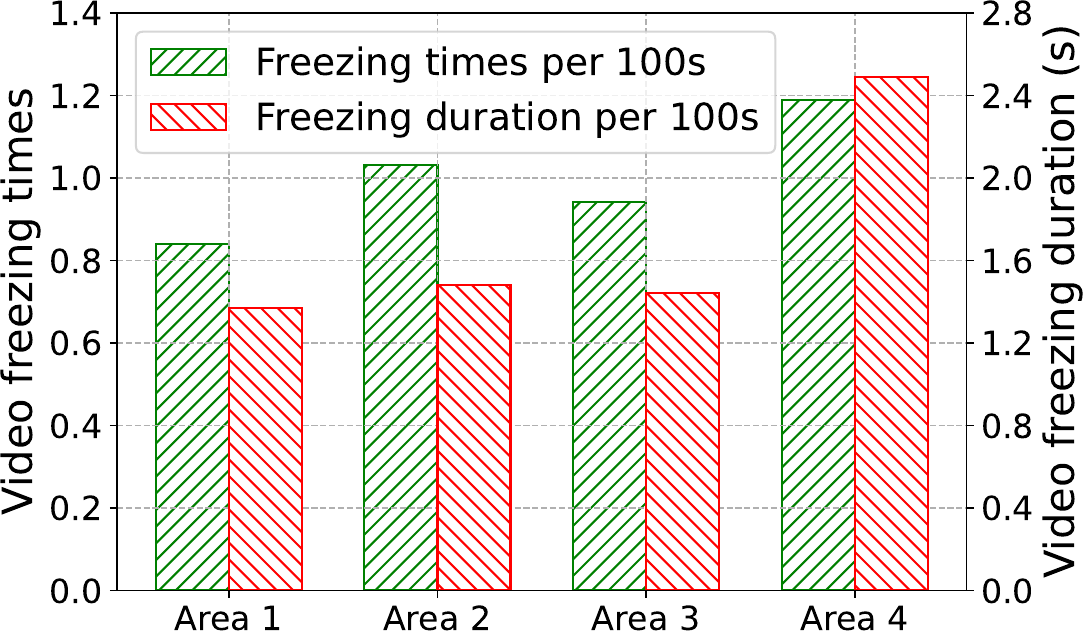}
\label{l7_freezing_area_fig}
}%\hspace{1mm}
\subfigure[Time-variant video freezes]{
\includegraphics[width=3.8cm,height=2.22cm,angle=0]{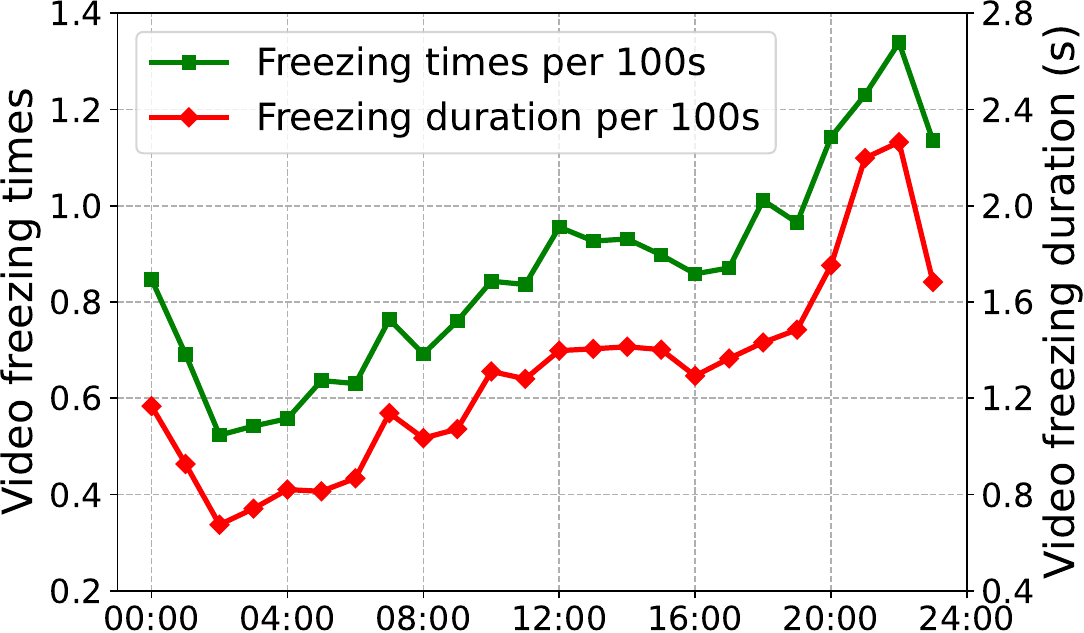}
\label{l7_freezing_time_fig}
}
\vspace{-2mm}
\caption{The times and duration of L7-player freezes.}\label{l7_player_freezing_fig}
\vspace{-0.4cm}
\end{figure}
\vspace{1ex} \noindent \textbf{Loss recovery influences video freezing.} 
% Live streaming has a high requirement for data timeliness that can affect client-side QoE, in which the traffic data that fails to reach receivers in time will introduce video freezing. 
Live streaming is highly sensitive to data timeliness, as traffic that fails to reach receivers on time results in video freezing, thereby affecting client-side QoE.
% To better depict data timeliness, both freezing frequency and duration on the client side are usually employed by live platforms to evaluate their subscribed CDN performance. 
% We (as a CDN vendor) can learn video freezing occurs once the player buffer becomes empty, i.e., no video frame can be uploaded to the player buffer before all cached frames in the player have been consumed. 
Video freezing occurs when the player buffer becomes empty, that is, when no video frame can be uploaded before all cached frames have been consumed.
% This is because (i) the player rendering speed exceeds network goodput from senders to receivers; 
One important reason for the player buffer becoming empty is that the lost data cannot be recovered in time, causing longer intra-stream HOL blocking time, especially in the common designs of TCP and QUIC protocol. 
% In this paper, we mainly focus on enhancing the timeliness of loss recovery to optimize video freezing. 
This paper focuses on enhancing the timeliness of loss recovery to reduce video freezing.

\vspace{1ex} \noindent \textbf{Video freezing influences profit.} In the multi-supplier CDN market, for CDN vendors, improving their QoE metrics can lead to a larger share of subscriptions, resulting in higher profits. The video freezing metrics (i.e., video freezing times per 100 seconds and video freezing duration per 100 seconds) are important QoE indicators for evaluating the service performance of CDN providers. Thus, the CDN vendors are motivated to optimize video freezing metrics.

\vspace{1ex} \noindent \textbf{The current video freezing is suboptimal.}
We conduct large-scale measurements and collect one-year values of video freezing metrics from a famous live platform in four areas (i.e., Southeast Asia, Latin America, the Middle East, and a selected country). 
% For more convenient presentations, we aggregate the measurement results and show every-100s freezing times and duration for each live stream, as Fig.~\ref{l7_player_freezing_fig} shows.  
% we can observe that the current video freezing is less than perfect. 
% Fig.~\ref{l7_freezing_area_fig} shows the variations in the measured values of video freezing metrics across different areas, while Fig.~\ref{l7_freezing_time_fig} illustrates the changes in the measured values of video freezing metrics over time in area 1.
Fig.~\ref{l7_freezing_area_fig} presents the spatial variations of video freezing metrics across different regions, while Fig.~\ref{l7_freezing_time_fig} depicts their temporal variations in Area 1.
% As Fig.~\ref{l7_freezing_area_fig} shows, for every 100s\footnote{Our measurements also show the average QUIC connection lifetime of each live stream is 99.6s.}, clients suffer from video freezing of over 1.3s (worse still, $\sim$2.5s in Area 4), on average (as Fig.~\ref{l7_freezing_area_fig} shows), in which nearly 1-time freeze usually makes customers intolerable. 
As shown in Fig.~\ref{l7_freezing_area_fig}, over each 100 s interval\footnote{The average QUIC connection lifetime of each live stream is 99.6 s.}, clients experience on average more than 1.3 s of video freezing (up to $\sim$2.5 s in Area 4), where even a single freeze event can significantly degrade user experience.
% Furthermore, as Fig.~\ref{l7_freezing_time_fig} shows, video freezes of more than 2s or 2 times often appear after 20:00.
Furthermore, Fig.~\ref{l7_freezing_time_fig} indicates that video freezing events exceeding 2 s or occurring more than twice are frequently observed after 20:00.

\vspace{-4mm}
\subsection{Restriction on the Loss Recovery Mechanism}
\label{sec:fec_sec}
In the multi-supplier CDN market, CDN vendors face difficulties in deploying loss recovery mechanisms that require modifications on both ends, which poses challenges for the universal deployment of mechanisms like FEC in production networks.
The live-streaming client software is designed and developed by the live-streaming application operators themselves, which includes proprietary information, so its source code is prohibited from being accessed or modified by CDN vendors. Therefore, loss recovery mechanisms that require modifications on both ends need the cooperation of live-streaming application operators. 
% In a multi-supplier market, if live-streaming application operators develop adaptations for the loss recovery mechanisms upgrades of a specific CDN vendor, it would not only affect the generality of their software protocols and introduce potential risks for interaction with other CDN vendors but also increase labor costs. 
In a multi-supplier market, developing adaptations for a specific CDN vendor may compromise the generality of their software protocols, pose interaction risks with other vendors, and increase operational costs.
Consequently, live-streaming application operators have no incentive to develop adaptations for the mechanism upgrades of a specific CDN vendor.

\section{Measurement Study}
\label{motivation_sec}

\begin{figure}[t!]
\subfigure[PDF]
{
    % \label{violin}
    \includegraphics[width=4.0cm,height=2.4cm]{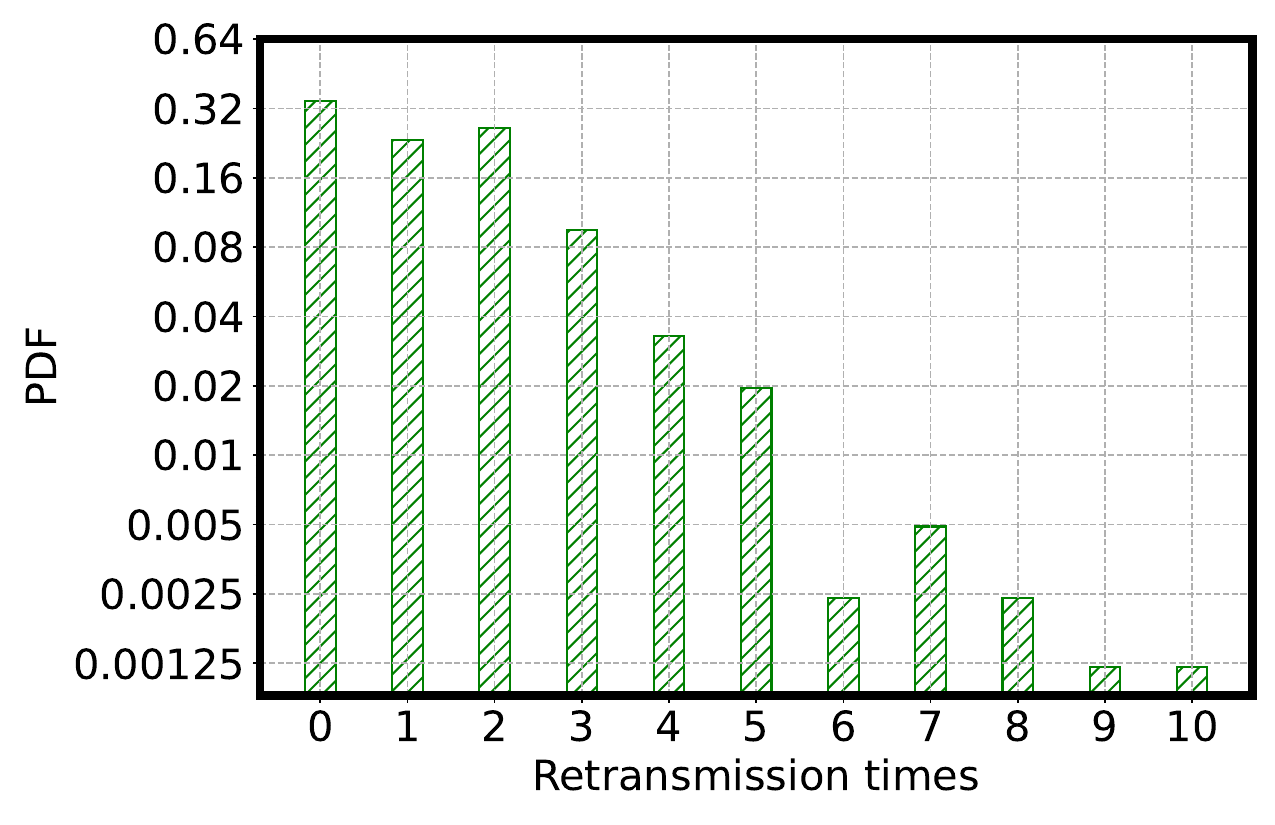}
}
% \hspace{-0.4cm}
\subfigure[CDF]
{
    % \label{loss_rate}
    \includegraphics[width=4.0cm,height=2.35cm]{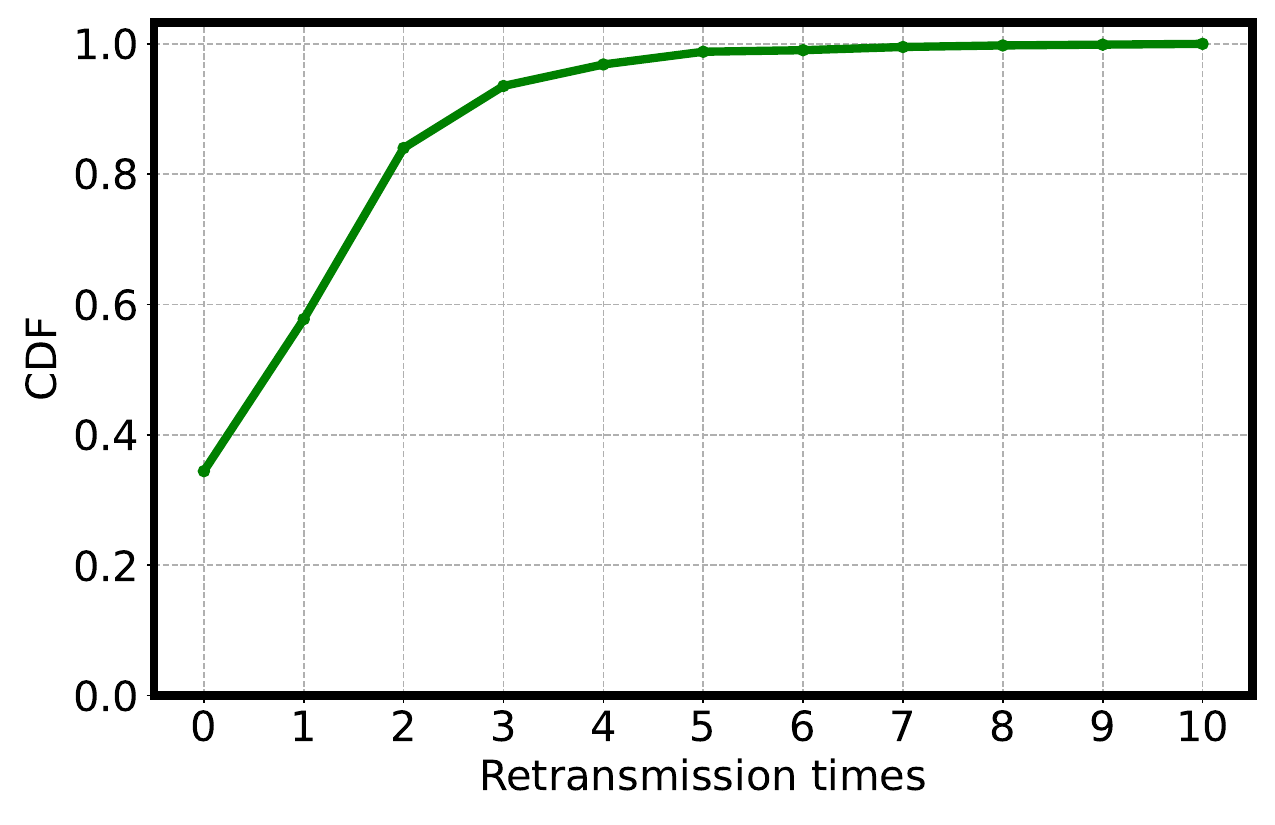}
}
\vspace{-0.3cm}
\caption{The maximum retransmission times in the wild.}
\label{retransmission_again}
\vspace{-0.4cm}
\end{figure} 
\begin{figure}[t!]
\subfigure[Loss distribution in different regions.]
{
    \label{violin}
    \includegraphics[width=4.0cm,height=2.5cm]{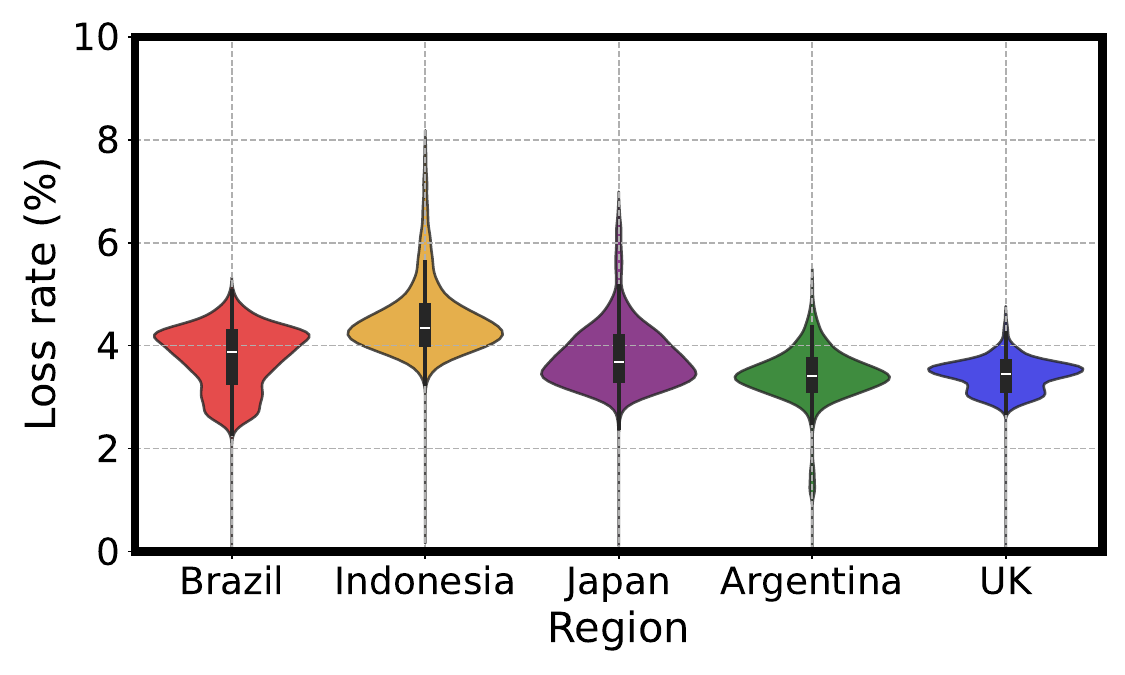}
}
% \hspace{-0.4cm}
\subfigure[Loss dynamics over time.]
{
    \label{loss_rate}
    \includegraphics[width=4.0cm,height=2.5cm]{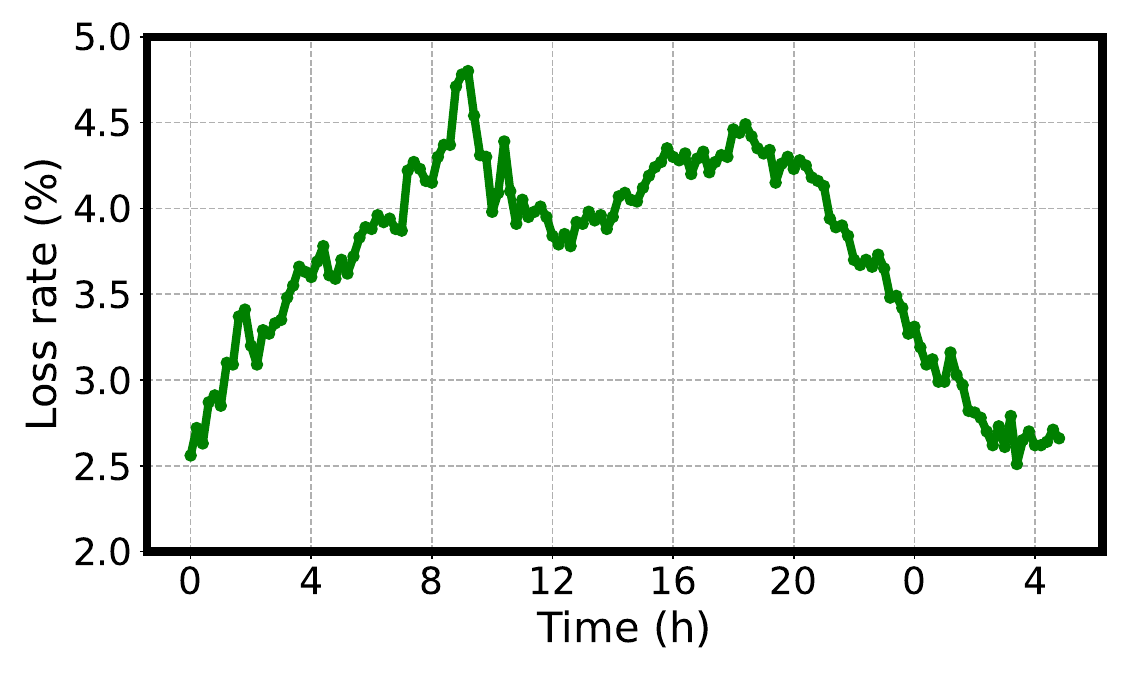}
}
\vspace{-0.2cm}
\caption{Examples of loss dynamics in the wild.} \label{loss_feature_wild}
\vspace{-0.4cm}
\end{figure}

In this section, we conduct large-scale measurements\footnote{The large-scale measurements are performed on the server side that runs NGINX architecture with QUIC BBR.} to motivate our work, in which the characteristics of loss (\S \ref{sec:background_motivation}) and live streaming (\S \ref{onoff_sec}) in the wild are further explored. 
We then analyze the performance of live streaming in the presence of packet loss, particularly retransmission loss (\S \ref{loss_recovery_quality_sec}).

%In this section, we conducted large-scale measurements, revealing the characteristics of loss (\S \ref{sec:background_motivation}) and live streaming (\S \ref{onoff_sec}). Furthermore, the results demonstrate that the timeliness of data due to packet loss recovery in current live streams is not as satisfactory as expected (\S \ref{loss_recovery_quality_sec}).

% \vspace{-4mm}

\subsection{Characteristics of Loss}\label{sec:background_motivation}

% remove by yanxu
% Each log corresponds to a QUIC connection, which contains the size of a connection, sequence number, and packet sending and loss information of multiple QUIC streams. 

% Logs from the production network are gathered from a random sampling conducted over a two-week period.
% We conduct an analysis of these logs, documenting specific details pertaining to packet loss, including the frequency of loss events and the magnitude of each loss. Our measurements encompass over 200,000 connections, spanning a variety of application scenarios and regions worldwide.

We analyzed the characteristics of loss using real-world logs spanning up to two weeks. These logs documented specific details related to packet loss, including the frequency of loss events and the magnitude of each loss. 
% Our measurements encompass over 200,000 connections, spanning a variety of application scenarios and regions worldwide. The findings are as follows.
Our measurements cover over 200,000 connections across diverse application scenarios and geographic regions.
% \begin{figure}
%     \centering
%     \includegraphics[width=0.45\textwidth]{figures/qvis.pdf}
%     \caption{An example of loss burstiness in the wild.}
%     \label{qvis}
%     \vspace{-0.2cm}
% \end{figure}

% \begin{figure}[t!]
% \hspace*{-0.3cm}
% \subfigure[Burst size of loss]
% {
%     \label{continuous_loss_data}
% %\begin{minipage}[b]{1.7in}
%     \includegraphics[width=3.8cm,angle=0]{figures/burstloss.pdf}
% %\end{minipage}}
% }
% \hspace*{-0.3cm}
% \subfigure[Maximum retransmission times]
% {
%     \label{retransmission_again}
% %\begin{minipage}[b]{1.7in}
%     \includegraphics[width=3.8cm,angle=0]{figures/retransmission_again.pdf}
% %\end{minipage}}
% }
% \caption{ (a) Burst loss distribution in the wild. The $x$-axis is the number of continuously lost packets. (b) The distribution of maximum retransmission times in the wild.} \label{loss_situation}
% \vspace{-0.4cm}
% \end{figure}

% \begin{figure}[t!]
% \hspace*{-0.3cm}   
% \includegraphics[width=5.0cm,angle=0]{figures/retransmission_again.pdf}
% \caption{The distribution of maximum retransmission times in the wild.}
% \label{retransmission_again}
% \vspace{-0.4cm}
% \end{figure}

\vspace{1ex} \noindent \textbf{Observation \#1: Retransmission loss is ubiquitous.}
A packet may undergo multiple retransmissions before the receiver correctly receives it. 
% We can delve deeper into the number of times each packet was retransmitted before successful reception. 
The number of retransmissions per packet before successful reception can be further analyzed.
For any given connection, the maximum retransmission times can be calculated as the highest count of retransmissions across all packets within that connection. Fig.~\ref{retransmission_again} illustrates the distribution, i.e., PDF (Probability Density Function) and CDF (Cumulative Distribution Function) of the maximum retransmission times in the production network. The results show that the proportion of connections with maximum retransmission times of two or more exceeds 43\%. Among them, a considerable portion of the connections have certain packets that are retransmitted even more than 10 times. Traditional ARQ mechanism focuses solely on promptly retransmitting lost packets after each loss event, but it overlooks the total number of retransmissions and the total time required for the retransmitted data to be successfully received by the receiver. Therefore, the traditional ARQ mechanism hardly meets the timeliness requirements for data in certain scenarios (e.g., live streaming).
%Such retransmission loss is detrimental to both delay-sensitive and throughput-insensitive applications, as we will discuss next. 

\vspace{1ex} \noindent \textbf{Observation \#2: Loss shows dynamics.}
We then investigate the distribution of the loss rate (every 5 minutes) for each connection across various global regions. 
The results are depicted in Fig.~\ref{violin}. 
% While some regions exhibit similar average loss rates, their packet loss rate deviations differ (i.e., the violin shapes are different). 
While some regions exhibit similar average loss rates, their packet loss rate deviations differ, as indicated by the shapes of the violin plots.
For instance, both Brazil and Japan have an average loss rate of 3.78\%, but Japan has the highest packet loss rate of 7.1\% while Argentina has 5.7\%. Fig.~\ref{loss_rate} further illustrates how the packet loss rate evolves. Specifically, the loss rate is dynamic, varying between 2\% and 5\% over 24 hours. This confirms that packet loss exhibits dynamic behavior in real-world scenarios. This also reveals that an adaptive loss tolerance scheme should adapt to differentiated conditions and the dynamic nature of networks. 

% \xu{Whether or not to remove the discussion of packet loss bursts here.}
% \subsubsection{Loss Shows Burstiness}\label{subsection:burstiness}

% Fig.~\ref{qvis} gives an example of loss burstiness (denoted by red blocks) in a randomly selected connection. We further explore the burst loss distribution in the production network. Fig.~\ref{continuous_loss_data} shows the results. Surprisingly, the probability of only one packet being lost is not as high as imagined (i.e., 37.2\% ), instead, the probability of losing multiple packages ($\ge 2$) together accounts for a larger proportion (i.e., 62.8\%). Specifically, as illustrated in the sub-figure of Fig.~\ref{continuous_loss_data}, the $90^{th}$, $95^{th}$, and $99^{th}$ percentile burst size of loss (i.e., the number of continuously lost packets) is 13, 27, and 125, respectively, showing extremely bursty packet loss. Numerous factors contribute to the emergence of the phenomenon of loss burstiness. Here we provide readers with a duo of potential explanations worthy of contemplation. First, it might be due to the queue management strategy~\cite{michael2023tambur} or traffic handling policy~\cite{cardwell2017bbr} employed by the Internet Service Provider (ISP). Moreover, burst packet loss can also be caused by wireless interference in unstable access networks of end users~\cite{baltrunas2016investigating}.

\begin{figure}[t]
  \centering
  \includegraphics[width=6.37cm]{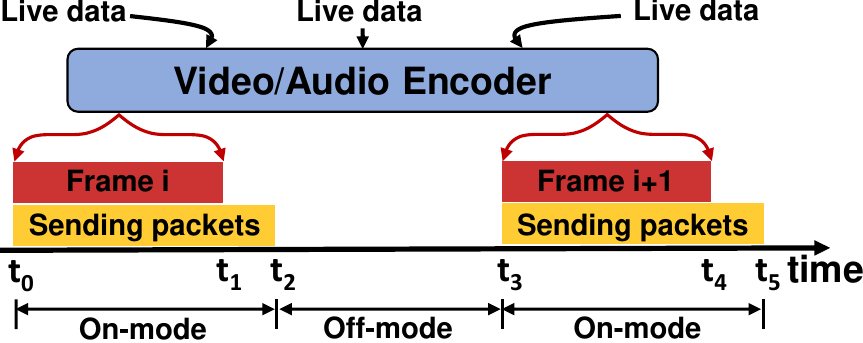}\\
  \vspace{-2mm}
  \caption{The sketch of on-off mode in live streams.}\label{on_off_fig}
\vspace{-0.4cm}
\end{figure}

\begin{table*}[tbp]
  \small
  \centering
  \caption{Live-streaming \textsf{recovery latency} measurement results.}
  \label{lrq_table}
  \resizebox{\textwidth}{!}{
  \begin{threeparttable}
  \begin{tabular}{|c|c|c|c|c|c|c|c|c|c|c|}
   \hline
   \hline
   \multirow{2}*{metrics
   }&\multirow{2}*{value
   }&\multicolumn{4}{c|}{\textsf{loss rate}}&\multicolumn{4}{c|}{\textsf{SRTT(ms)}}\\
   \cline{3-10}
   &&0\% $\sim$ 3\%&3\% $\sim$ 10\%&10\% $\sim$ 30\%&30\% $\sim$ 50\%&0 $\sim$ 50&50 $\sim$ 200&200 $\sim$ 500&500 $\sim$ 2000\\
   \hline
   \hline
   % \rowcolor{red!10}\textsf{RDR}&6.8\%&1.4\%&4.9\%&13.1\%&23.5\%&6.0\%&9.0\%&9.6\%&8.7\%\\
   % \hline
   \rowcolor{red!10}\textsf{ \textsf{recovery latency}} (ms)\tnote{*}&123.2&26.8&94.5&186.4&530.2&68.2&170.2&273.6&418.0\\
   \hline
   \rowcolor{green!10}\textsf{ \textsf{maximum recovery latency}} (ms)&279.3&51.8&204.9&451.3&1288.4&149.2&391.3&698.8&960.9\\
   \hline
   % Goodput&&&&&&&\\
   % \hline
   \hline
  \end{tabular}
  \begin{tablenotes}
  \footnotesize
    \item[*] The displayed \textsf{recovery latency} (\textsf{maximum recovery latency}) only records the each-stream's average (maximum) \textsf{recovery latency} of the lost data, whose recovery requires two or more retransmissions. 
  \end{tablenotes}
  \end{threeparttable}
  }
 \vspace{-0.2cm}
 \end{table*}

\begin{figure}[tbp]
% \vspace{-3mm}
\centering
\subfigure[Off-mode times and duration.]{
\includegraphics[width=3.8cm,height=2.22cm,angle=0]{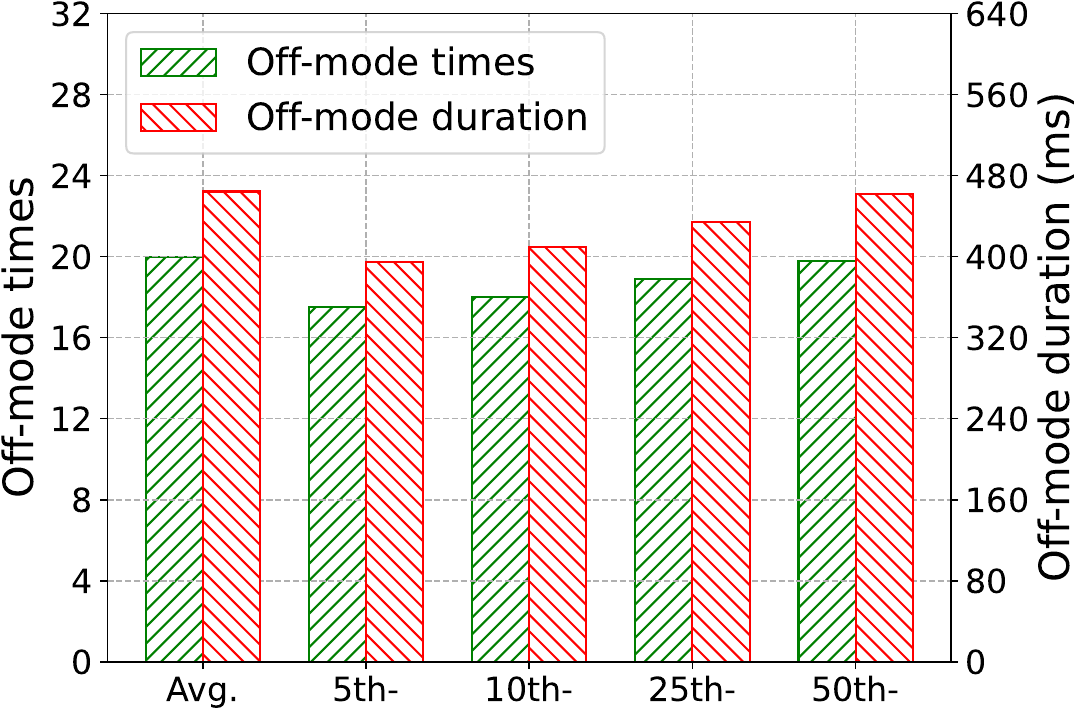}
\label{off_times_duration_fig}
}
\subfigure[Off-mode distribution.]{
\includegraphics[width=3.8cm,height=2.22cm,angle=0]{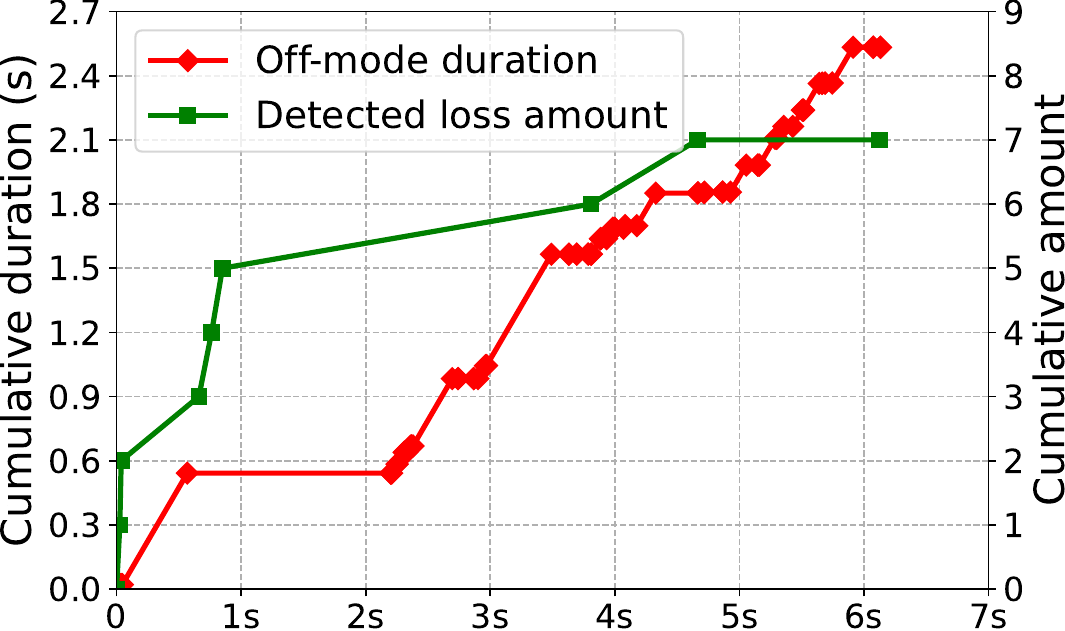}
\label{off_distribution_fig}
}
\vspace{-0.2cm}
\caption{On-off mode measurements and experiments.}\label{retran_ratio_evaluation_fig}
\label{off_state_fig}
\vspace{-0.4cm}
\end{figure}  

% \vspace{-3mm}
\subsection{Characteristics of Live Streaming}
\label{onoff_sec}
Unlike video-on-demand or file traffic, the current live streams frequently and extensively reveal off-mode, in which a sender\footnote{Since we collect and analyze data as well as conduct experimental evaluations all from the CDN edge server, without special instructions, the term "sender" refers to the CDN edge server instead of the original streaming source.} temporarily has no data (i.e., becomes application-limited \cite{cheng2022estimation}\cite{iyengar2020quic}) for continuous transmissions after sending one or more frames. 
% This is actually caused by the following factors. 
On the one hand, the new generation rate of live data might be slower (e.g., than the sending rate), requiring senders to wait until their sending queues are replenished.
For CDN vendors, the data generation rate reflects the traffic transmission rate from anchors to servers, which can be easily affected by real-time network status. 
On the other hand, the live data will be encoded into video or audio frames based on selected frame rate(s) 
% (\textsf{R}$_\textsf{fram}$) 
and bitrate 
% (\textsf{R}$_\textsf{bit}$) 
before its transmission, which can also introduce time intervals between adjacent frames. 
If one frame has been sent out while the follow-up frame has not yet been encoded, the CDN sender enters off-mode. 
% keep application limited. 

As Fig. \ref{on_off_fig} shows, the data of live streaming is encoded (by video/audio encoder) into frame i which is delivered to the sending queue (from $t_{0}$ to $t_{1}$) and is sent out before $t_{2}$. 
In this case, the sender must wait (from $t_{2}$ to $t_{3}$) for frame i+1 which will be encoded based on the follow-up live data. 
In this paper, the on-off mode that appears on the sender side can be recognized with the following conditions.
\begin{itemize}
\item \textbf{On-mode:} A mode that occurs when enough data exists in the sending queue (i.e., $t_{0} \sim$ $t_{2}$ and $t_{3} \sim$ $t_{5}$ in Fig. \ref{on_off_fig}), which can be sent by a sender at the subsequent time. 
\item \textbf{Off-mode:} A mode that occurs when no data can be obtained for traffic transmissions (i.e., $t_{2} \sim $ $t_{3}$ in Fig. \ref{on_off_fig}), making senders have to wait for the follow-up frame. 
% \footnote{Actually, if $t_{2}$ $\geq$ $t_{3}$ in Fig. \ref{on_off_fig}, i.e., the follow-up frame i+1 is delivered to sending queue before the data of frame i is all send out, off-streaming mode will not occur from $t_{0}$ to $t_{5}$.}. 
\end{itemize}

% In \S \ref{on_off_measurement_sec}, we demonstrate the on-off mode are commonly existing in current live streams through our large-scale network measurements, and further discover this mode is actually unfriendly for existing transport controls, which can be learned through our testbed experiments. 

We make large-scale measurements and gain the following observations to further explore the on-off mode of live streaming. 

\vspace{1ex} \noindent \textbf{Observation \#1: The on-off mode switching commonly exists in live streaming.}  
% Fig. \ref{off_times_duration_fig} shows the measured times and the duration of off-streaming mode, where we can learn the off-mode times and duration of 5\%, 10\% and 50\% of live streams can experience over 6800, 3100 and 290 off-streaming mode with the duration of 173s, 80s and 12s, respectively. 
% On average, 1800-times off-streaming mode with a time length of 48s will appear in each live stream.  
Fig. \ref{off_times_duration_fig} presents our measurements of the duration each stream spends in off-mode every second, as well as the frequency of each stream entering off-mode on a per-second basis, indicating that more than 95\% of the live streams can be in off-mode for 394 ms per second and enter off-mode 17 times per second.
On average, each live stream spends 464 ms in off-mode per second and enters off-mode 20 times per second.

\vspace{1ex} \noindent \textbf{Observation \#2: The off-modes can be unevenly distributed throughout the live-streaming lifetime.} 
Fig. \ref{off_distribution_fig} depicts the cumulative values of both off-mode duration and detected loss amount in a connection of our testbed experiments. 
We can find (i) off-mode mainly occurs after 2 s while only 2 out of 7 packets are detected lost during this period; 
(ii) most packet losses (5 of 7) occur with minimal off-mode exposure within the first second of the measured stream, during which the stream remains in on-mode when the 4th to 6th losses are detected.
\vspace{-3mm}
\subsection{Live Streaming Performance under Loss}
\label{loss_recovery_quality_sec}
% \xu{lack the essential of enhancing loss recovery}
% \xu{add Tunit to figure 6}
\begin{figure}[tbp]
\centering
\subfigure[Loss recovery (\textsf{K = 1}).]{
\includegraphics[width=3.4cm, angle=0]{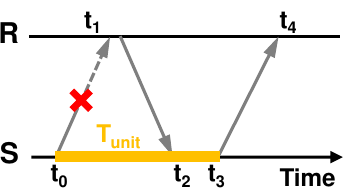}
\label{retrans_fig_1}
}\hspace{2mm}
\subfigure[Loss recovery (\textsf{K = 2}).]{
\includegraphics[width=3.7cm, angle=0]{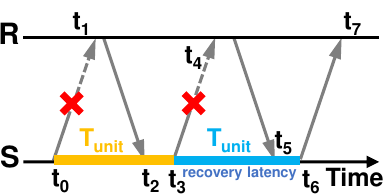}
\label{retrans_fig_2}
}
\vspace{-0.2cm}
\caption{Loss recovery between sender (S) and receiver (R).}\label{retrans_fig}
\vspace{-0.4cm}
\end{figure}
Live streaming requires high data timeliness which can affect client-side QoE. Due to the characteristics of packet loss and the limitations of traditional ARQ (\S \ref{sec:background_motivation}), the timeliness issues caused by packet loss in live streaming are critical. We introduce \emph{recovery latency} to measure the timeliness of packet loss recovery and conduct large-scale measurements on live streams in real production networks. 

\emph{Recovery latency} is defined as the duration from when any data is detected lost to when resending a recovery packet \emph{that will be successfully received}.  
\emph{Loss detection time}, denoted as 
$T_{unit}$, is defined as the time elapsed for a data packet from being sent to being retransmitted.
Recovery latency consists of zero or more $T_{unit}$, reflecting the additional time for loss recovery besides the first $T_{unit}$. 
% that ideally (i.e., the resent data will not be re-lost) consumes only one \textsf{T}$_\textsf{unit}$. 
% Thus, \textsf{IRT} can also be expressed by formula \ref{t_waiting_irt}. 
% \xu{I feel that this formula is highly misleading because we have never used it to calculate recovery latency. We have always relied on its definition for calculations. Should we consider removing it?}
% \xu{whether to remove this equation}
% \vspace{-3mm}
% \begin{equation}\label{t_waiting_irt}
%     \textsf{recovery latency} = \textsf{T}_\textsf{waiting} - \textsf{T}_\textsf{unit}
%     \vspace{0mm}
% \end{equation}
In Fig. \ref{retrans_fig_1}, recovery latency is 0. 
Fig. \ref{retrans_fig_2} shows the resent data is detected lost again and another recovery packet (\emph{that will be successfully received} at $t_{7}$) is sent at $t_{6}$, where recovery latency = $t_{6}$ - $t_{3}$. 
% Besides, \textsf{IRT} should also contain the delayed time (\textsf{T}$_\textsf{drst}$) of resending a lost packet, which will become larger once facing sender-side limited \textsf{cwnd} or sending rate. 
Besides, the \emph{maximum recovery latency} is employed to evaluate the largest recovery latency for loss recoveries. 

% \item \textbf{Recovery deterioration rate (\textsf{RDR})} is defined as the ratio of lost data amount (\textsf{D}$_\textsf{k}$), which takes two or more \textsf{T}$_\textsf{unit}$ to be recovered, to the amount of all lost data. 
% For example, there are 2 lost packets - one is in Fig. \ref{retrans_fig_1} and the other is in Fig. \ref{retrans_fig_2}, which actually require 1 and 2 \textsf{T}$_\textsf{unit}$ for their successful recoveries (at $t_{4}$ and $t_{7}$), respectively. 
% Then, we can learn \textsf{D}$_\textsf{k}$ \textsf{= 1} and \textsf{RDR = 50\%}. 
% \end{itemize}

We make large-scale measurements and collect the transmission logs of 50 million live streams. 
% that are delivered from our CDN servers to clients. 
% We find one or more packet losses are actually detected in 54.2\% of these streams while others have not suffered from any loss recovery. 
% \footnote{In this measurement, a packet that is identified as lost is based on whether it will be acknowledged by its receiver instead of the existing detection scheme \cite{iyengar2020quic}, which can filter the false-positive loss detection.}. 
% We then classify the average values of our measured \textsf{recovery latency} and \textsf{maximum recovery latency} based on the ranges of loss rate (\textsf{loss\_rate}) and smooth RTT (\textsf{SRTT}), as Table \ref{lrq_table} shows. 
We then classify the average values of measured recovery latency and maximum recovery latency based on the ranges of loss rate and Smooth Round-Trip Time (SRTT), as Table \ref{lrq_table} shows. 
% i.e., 0\% $\sim$ 3\%, 3\% $\sim$ 10\%, 10\% $\sim$ 30\% and 30\% $\sim$ 50\%, and gain 68, 32, 20 and 14 streams, respectively. 
% The following observations can be obtained.  

\vspace{1ex} \noindent \textbf{Observation \#1: 
The current recovery latency of live streaming is far from satisfactory.}
For the lost data that requires more than one retransmission, traffic senders waste 123.2 ms (i.e., recovery latency = 123.2 ms), on average, before sending out the recovery packets that their receivers will indeed acknowledge. 
Furthermore, the average maximum recovery latency is up to 279.3 ms, which can easily cause video freezing.

\vspace{1ex} \noindent \textbf{Observation \#2: Higher loss rate and larger SRTT can easily introduce more deteriorated recovery latency.} 
As Table \ref{lrq_table} shows, with the increase in loss rate, the recovery latency increases accordingly.
This is because the recovery data may experience another packet loss in the networks with a higher loss rate. 
By contrast, SRTT has serious impacts on recovery latency. For example, maximum recovery latency will be improved to $\sim$1 s, on average, under the SRTT of 500 $\sim$ 2000 ms, which is unacceptable for the timeliness of loss recovery. 
% By contrast, \textsf{SRTT} has serious impacts on \textsf{recovery latency}, whose larger values caused by network congestion might bring more deteriorated \textsf{recovery latency}. 
% By contrast, \textsf{SRTT} has serious impacts on \textsf{recovery latency}. 
% For example, \textsf{maximum recovery latency} will be improved to $\sim$1s, on average, under the \textsf{SRTT} of 500 $\sim$ 2000ms, which is unbearable for the timeliness of loss recovery. 

Therefore, an enhanced recovery scheme is highly required to further optimize the current unsatisfied recovery latency and promote the recovery timeliness of live streams. 
% , especially with higher loss rates and larger transmission delays. 
\vspace{-1.5mm}

% \section{Large-Scale Measurements}
% \label{measurement_sec}
% \input{chapters/3-measurement.tex}

\section{The \name{} Overview}
\label{framework_sec}
% \subsection{Key Idea}
% \label{key_idea_sec}
\begin{figure}[!t]
\includegraphics[width=0.48\textwidth]{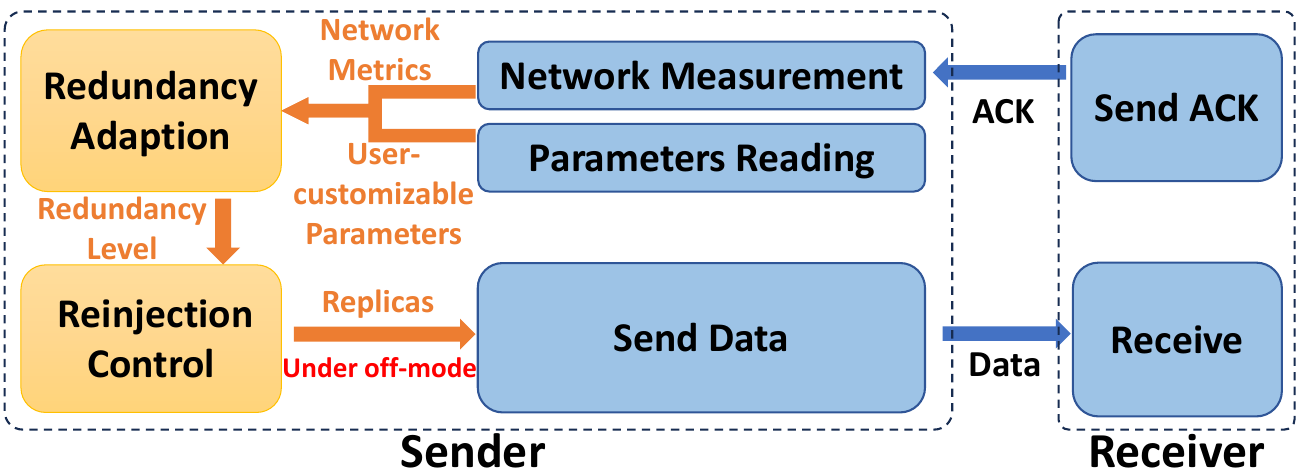}
\vspace{-0.2cm}
\centering
\caption{The architecture of \name{}.}\label{architecture_fig}
  \vspace{-0.4cm}
\end{figure}

In this section, we first discuss the design principles of \name{}, whose architecture overview will then be given.
\vspace{-2mm}
%, which is affected by the current unsatisfied \textsf{LRQ}. 
\subsection{Design Principles}
\label{insight_sec}
\name{} regards the control-unfriendly on-off mode as an essential opportunity for enhancing the recovery latency of live streaming and mitigating the negative effects on player freezing caused by potential HOL blocking. Our goal is to accelerate loss recovery to meet user demands whenever
possible while keeping the overhead controllable.
% send ``few but enough" replicas for each loss. 
\name{} should follow two design principles for optimizing the timeliness of loss recovery in live streams.

\vspace{1ex} \noindent \textbf{Principle \#1: The overhead introduced by \name{} must be controllable.}
If the overhead is not carefully managed, \name{} could result in negative returns. From the perspective of a CDN vendor, \name{} primarily introduces two types of overhead. The first type is the traffic cost incurred by replicas under traffic billing conditions, which increases costs. The second type is the potential goodput reduction caused by redundant replicas, which can lead to a decrease in QoE and thus reduce revenue. Therefore, \name{} should have clear and simple control logic to manage the overhead introduced by injecting replicas.

\vspace{1ex} \noindent \textbf{Principle \#2: \name{} should ensure that recovery latency is reduced to meet user demands whenever possible.}
Different application scenarios have different requirements for recovery latency. \name{} should allow for the setting of recovery latency requirement and, while strictly adhering to overhead constraints, inject few but enough packets to reduce the recovery latency to meet the recovery latency requirement whenever possible.

\subsection{The Architecture of \name{}}
\label{overview_sec}
\name{} is a sender-side modification to the protocol stack whose key modules are illustrated in Fig.~\ref{architecture_fig}. Particularly, \name{} adopts redundancy adaption to compute the number of replicas of a lost packet that should be retransmitted next periodically. Given the number of replicas, \name{} then adopts reinjection control to determine the specific order and time for sending out each replica from the sender. 

\vspace{1ex} \noindent \textbf{Redundancy adaptation.} This module answers the question of how many replicas should be sent to accelerate loss recovery. The \emph{redundancy level} (denoted by $K$, $K \in\mathbb{N}$) is defined as the number of replicas that should be resent for a specific lost packet. 
% We define the \emph{reinjection overhead} as the total number of replicas that are sent during transmission. 
To adapt to the dynamics of loss, we incorporate redundancy adaptation to allow the redundancy level to vary dynamically. To achieve this, we introduce the \emph{Redundancy Adapter}, which carefully selects the most appropriate redundancy level for the lost packets. 
% This ensures that \name{} can adapt to the dynamics of packet loss while minimizing the redundancy cost.

\vspace{1ex} \noindent \textbf{Reinjection control.} This module answers the question of how to send the given number of replicas determined by the Redundancy Adapter. For each lost packet, more than one replica might be injected into the network. To avoid bandwidth contention for unlost data transmission, we introduce the \emph{Reinjection Controller} to enable sending replicas of the lost packet if the stream is in off-mode. To further reduce the loss recovery latency, the Reinjection Controller also opportunistically captures the ideal chance to reinject replicas even when the stream is not in the off-mode. 
% The Reinjection Controller can also keep the bandwidth contention for unlost data transmission within a safe limit. 
This ensures that \name{} can accelerate packet loss recovery even when off-mode is absent or unevenly distributed.

\vspace{-3mm}

\iffalse
\begin{figure}[!t]
\includegraphics[width=0.4\textwidth]{figures/autorec_overview_fig_v5.pdf}
\vspace{-0.2cm}
\centering
\caption{The workflow of \name{}.}\label{overview_fig}
  \vspace{-0.4cm}
\end{figure}
\fi

\section{Detailed Design}
\label{details_sec}
In this section, we will depict the detailed designs of \name{} for optimizing loss recovery.

\vspace{-3mm}

\subsection{Redundancy Adapter}
\label{redundancy_adaption_sec}

\begin{figure}[t]
  \centering
  \includegraphics[width=0.48\textwidth]{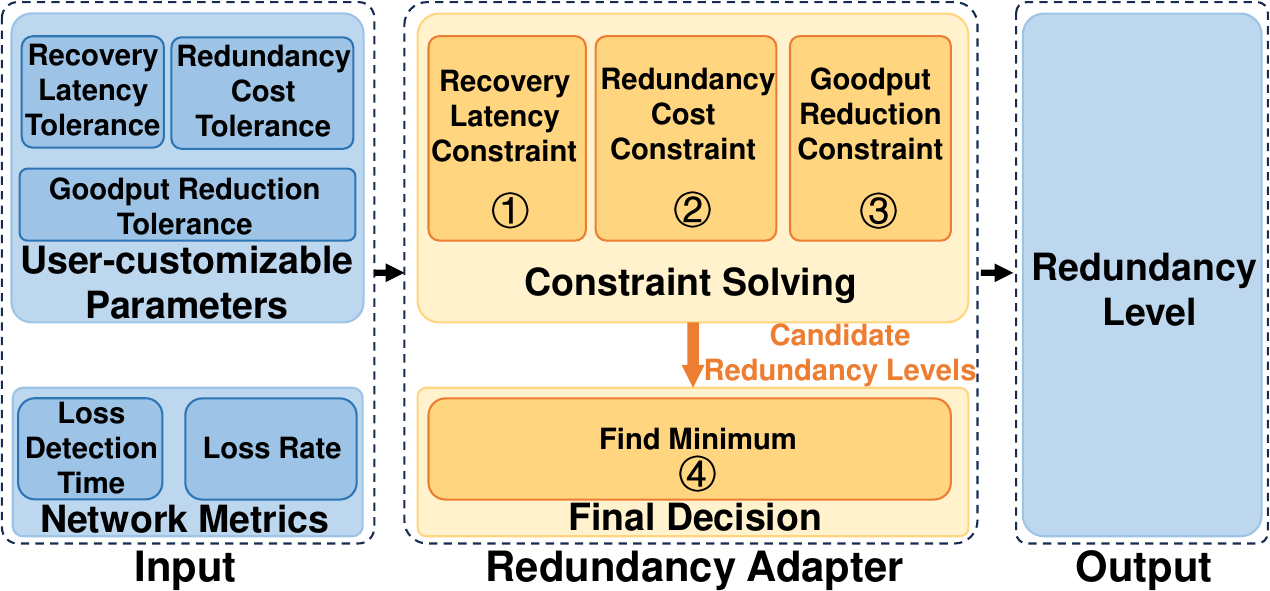}\\
  \caption{The redundancy adapter of \name{}.}\label{redundancy_adapter_fig}
\vspace{-0.4cm}
\end{figure}

%We introduce the Redundancy Adapter to apply a model-based scheme to adjust the redundancy level (denoted by $\textsf{A}_\textsf{thres}$) for each loss recovery dynamically. The goal is to reinject ``few but enough" replicas for optimizing data timeliness affected by packet losses. 

We first define \emph{redundancy cost} as the ratio of the total number of injected replicas to the total number of sent packets (excluding retransmitted packets and injected replicas). \emph{Goodput reduction} is defined as the reduction percentage in goodput when injected replicas compared to the goodput without injected replicas during transmission. The Redundancy Adapter provides three user-customizable parameters: the recovery latency tolerance (denoted by $\alpha$), the redundancy cost tolerance (denoted by $\beta$), and the goodput reduction tolerance (denoted by $\gamma$).

The redundancy level is determined periodically\footnote{The decision interval (denoted by $D$) is usually set several Round-Trip Times (RTTs), e.g., $D = 5$ RTTs.}.
\name{} will measure the network metrics within each decision interval. The network metrics include the average loss detection time (\S \ref{loss_recovery_quality_sec}) and the average loss rate. When there is no packet loss (i.e., loss rate $= 0$) in the last decision interval, the Redundancy Adapter directly sets the redundancy level to 0. Otherwise (i.e., 0 $<$ loss rate $<$ 1), taking the user-customizable parameters and network metrics as input, the Redundancy Adapter will execute the following four steps to calculate the output, which is the redundancy level, as Fig. \ref{redundancy_adapter_fig} shows. \textbf{Step \#1:} Calculate the minimum redundancy level satisfying the recovery latency constraint (denoted by $K_{\alpha}$). \textbf{Step \#2:} Calculate the maximum redundancy level satisfying the redundancy cost constraint (denoted by $K_{\beta}$). \textbf{Step \#3:} Calculate the maximum redundancy level satisfying the goodput reduction constraint (denoted by $K_{\gamma}$). \textbf{Step \#4:} Determine the final redundancy level (denoted by $K_{\theta}$) according to the above $K_{\alpha}$, $K_{\beta}$, $K_{\gamma}$. Next, we will elaborate step by step.

% Specifically, for each decision interval(denoted by \textsf{DI}),  \name{} performs the following two steps. \textbf{Step \#1:} \name{} utilizes the packet loss rate and $T_{unit}$ measured in the previous decision interval to calculate the redundancy level that ensures the actual redundancy cost does not exceed the preset redundancy cost (denoted to as RL1), the redundancy level that ensures the actual goodput reduction does not exceed the preset goodput reduction (denoted to as RL2), and the redundancy level that ensures the actual recovery latency does not exceed the preset recovery latency (denoted to as RL3). \textbf{Step \#2:} \name{} takes the minimum value among RL1, RL2, and RL3 as the redundancy level for the current decision interval. It is worth noting that when the packet loss rate is equal to 0 or 1, \name{} skips the above steps and directly sets the redundancy level to 0.

\vspace{1ex} \noindent \textbf{Step \#1: Calculate the minimum redundancy level satisfying the recovery latency constraint.} Let $K$ denote the redundancy level, $F(K)$ denote the average recovery latency for lost packets\footnote{In this paper, we only focus on the lost packets that are retransmitted over once.} under network conditions with the maximum average recovery latency (i.e., the entire transmission process remains in on-mode), and $\alpha$ denote the recovery latency tolerance. The formal description of the problem to calculate the minimum redundancy level that satisfies the recovery latency constraint (denoted by $K_{\alpha}$) is as follows:
\begin{equation}\label{RL1_equation}
\begin{split}
     K_{\alpha}=\arg\min_{K} K \quad \text{s.t.} ~ F(K)\leq \alpha
     ~ \text{and} ~ K\in\mathbb{N} \nonumber
    % \vspace{-2mm}
\end{split}
\end{equation}

Given the loss rate (denoted by $R$), the loss detection time $T_{unit}$ and the redundancy level $K$, when $0<R<1$, $T_{unit} \geq 0$ and $K\in\mathbb{N}$, $F(K)$ can be computed as follows (please refer to Appendix B for the detailed derivation):
\begin{equation}
\begin{split}
\label{recovery_latency_constraint_equation}
    F(K)=&\frac{(1+KR^K)}{(1-R)(1+K)}T_{unit}\\
    \text{where } K\in\mathbb{N} &\text{ , } 0<R<1 \text{ and } T_{unit}\geq 0
\vspace{-1mm}
\end{split}
\end{equation}

We further find that $F(K)$ is \emph{monotonically decreasing} for $K \in\mathbb{N}$  (please refer to Appendix C for the detailed proof). In this case, we can determine $K_{\alpha}$ using the binary search method.

\vspace{1ex} \noindent \textbf{Step \#2: Calculate the maximum redundancy level satisfying the redundancy cost constraint.} 
Let $K$ denote the redundancy level, $G(K)$ denote the average redundancy cost, and $\beta$ denote the redundancy cost tolerance. The formal description of the problem to calculate the maximum redundancy level that satisfies the redundancy cost constraint (denoted by $K_{\beta}$) is as follows:
\begin{equation}\label{RL2_equation}
    K_{\beta}=\arg\max_{K}{K} \quad \text{s.t.} ~ G(K)\leq \beta   ~ \text{and} ~ K\in\mathbb{N} \nonumber
\end{equation}

Given the loss rate (denoted by $R$) and the redundancy level $K$, when $0<R<1$ and $K\in\mathbb{N}$, $G(K)$ can be computed as follows (please refer to Appendix D for the detailed derivation):

\vspace{-5mm}

\begin{equation}\label{redundancy_ratio_constraint_equation}
\begin{split}
    &G(K)=KR \\
    \text{where } K&\in\mathbb{N} \text{ and } 0<R<1
    \vspace{-2mm}
\end{split}
\end{equation}

We further find that $G(K)$ is \emph{monotonically increasing} for $K \in\mathbb{N}$  (please refer to Appendix E for the detailed proof). In this case, we can determine $K_{\beta}$ using the binary search method.

\vspace{1ex} \noindent \textbf{Step \#3: Calculate the maximum redundancy level satisfying the goodput reduction constraint.}
Let $K$ denote the redundancy level, $H(K)$ denote the average goodput reduction under network conditions with the maximum average goodput reduction (i.e., the entire transmission process remains in on-mode), and $\gamma$ denote the goodput reduction tolerance. The formal description of the problem to calculate the maximum redundancy level that satisfies the goodput reduction constraint is (denoted by $K_{\gamma}$) as follows:
\begin{equation}\label{RL3_equation}
    K_{\gamma}=\arg\max_{K}{K} \quad \text{s.t.} ~ H(K)\leq \gamma   ~ \text{and} ~ K\in\mathbb{N} \nonumber
    % \vspace{-2mm}
\end{equation}

\vspace{-1mm}

Given the loss rate (denoted by $R$) and the redundancy level $K$, when $0<R<1$ and $K\in\mathbb{N}$, $H(K)$ can be computed as follows (please refer to Appendix F for the detailed derivation):
\begin{equation}
\label{goodput_reduction_ratio_constraint_equation}
\begin{split}
    % H(K)=\frac{KR-(1+K)R^{2}+R^{K+2}}{1+KR-(1+K)R^{2}+R^{K+2}}\\
    H(K) = &1 - \frac{1}{1+KR-(1+K)R^{2}+R^{K+2}}\\
    &\text{where } K\in\mathbb{N} \text{ and } 0<R<1
    \vspace{-1mm}
\end{split}
\end{equation}
We further find that $H(K)$ is \emph{monotonically increasing} for $K \in\mathbb{N}$  (please refer to Appendix G for the detailed proof). In this case, we can determine $K_{\gamma}$ using the binary search method.

\vspace{1ex} \noindent \textbf{Step \#4: Determine the redundancy level for this decision interval.} 
Since AutoRec aims to satisfy the constraint of recovery latency whenever possible while strictly satisfying the constraints of the redundancy cost and the goodput reduction, the Redundancy Adapter determines the $K_{\theta}$ as follows.
\begin{equation}\label{RL3_equation}
    K_{\theta} = min \{K_{\alpha}, K_{\beta}, K_{\gamma}\}
    \vspace{-2mm}
\end{equation}

\subsection{Reinjection Controller}
\label{reinjection_controller_sec}

We use $D_{ul}$ to denote the lost data that has been resent but unacknowledged by its receiver. 
The data $D_{ul}$ is added and stored in a reinjection queue that is managed at the sender. 
To decide when to retransmit the replicas, the Reinjection Controller is introduced, which reinjects replicas when the current stream enters off-mode. 
The Reinjection Controller also adopts opportunistic reinjection to further reduce loss recovery latency while keeping the bandwidth contention for unlost data transmission within acceptable bounds.

%Specifically, all $D_{ul}$ is ranked according to the timestamps of their latest retransmission, which will be fetched from the head of the reinjection queue and then resent out.

%once the sender enters the off-mode or the configured reinjection timer (\textsf{T}$_\textsf{timer}$) expires.

%The \name{} sender maintains a reinjection queue for each live stream, which records the lost data $D_{ul}$ that has been resent but unacknowledged by its receiver. 

%In the reinjection queue, all $D_{ul}$ is sorted according to the timestamps of their latest retransmission/reinjection, 
% where the newly resent data will be inserted to the end of reinjection queue. 
% Then, we will detail the establishment and maintenance of sender-side reinjection queue. 

 \begin{figure}[!t]
  \centering
  \includegraphics[width=8.3cm]{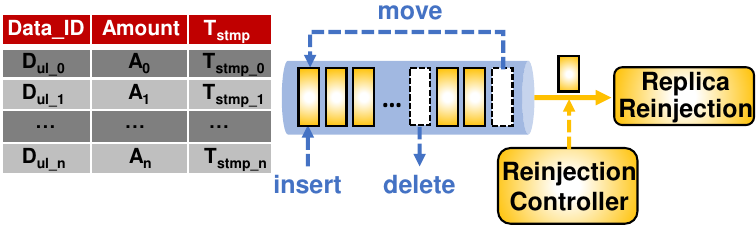}\\
  \caption{Sender-side reinjection queue.}\label{queue_fig}
\vspace{-0.4cm}
\end{figure}

 \begin{figure}[t]
  \centering
  \includegraphics[width=0.4\textwidth]{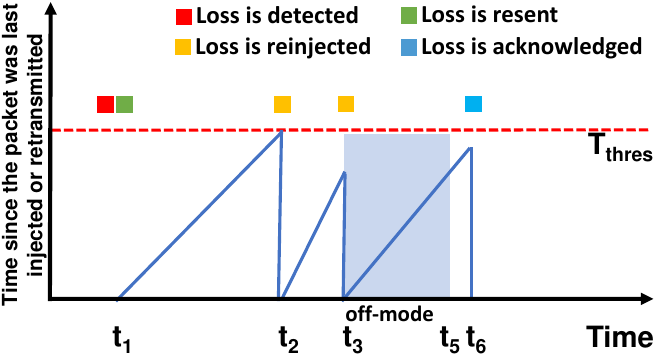}\\
  \caption{An example of how the Reinjection Controller handles lost injections.}\label{timer_rein_fig}
\vspace{-5mm}
\end{figure}

\vspace{1ex} \noindent \textbf{Queue management.}
In \name{}, a reinjection queue will be created by a traffic sender when a new connection of live streaming is established, which will be removed once this connection is closed. \name{} enables senders to update the reinjection queue (using the following operations) for each live stream whenever lost data is retransmitted or acknowledged, as Fig. \ref{queue_fig} shows. 
% as Fig. \ref{queue_fig} shows, which can be achieved by the following operations. 
% Reinjection queue reinjection queue should be updated if any lost data is resent or acknowledged on on-/off-streaming state, which can be achieved by the following operations. 
The detected lost data will be inserted into the end of the reinjection queue after it has been retransmitted. 
The $D_{ul}$ will be deleted from the reinjection queue when (i) the resent $D_{ul}$ is acknowledged by its receiver, or (ii) the performed reinjection times exceeds the decided $K_{\theta}$. 
The $D_{ul}$ will be moved from the head to the end of the reinjection queue if it has been resent again, which is sorted by the timestamp of its last reinjection or retransmission.  
% one of the following conditions can be met: 
% (i) it has been retransmitted again during on- or off-streaming state; 
% (ii) its reinjected retransmission times (\textsf{T}$_\textsf{rein}$) has reached its threshold ($T_{thres}$) that will be described shortly. 
% Note that the \textsf{move} operation can be achieved by \textsf{insert} and \textsf{delete} operations. 
In addition, the \name{} sender maintains a status table for each reinjection queue, which records the reinjection times ($A_{i}$) that has been performed, timestamp ($T_{stmp}$) of $D_{ul}$'s last reinjection or retransmission and $D_{ul}$ identification ($Data\_ID$)\footnote{In this paper, $Data\_ID$ can be recognized as the packet number ($pkt\_num$) in TCP or the stream offset ($stream\_offset$) in QUIC.}, as Fig. \ref{queue_fig} shows. 

% remove by yanxu
% For multiplexed transport, e.g., QUIC\cite{iyengar2020quic} and HTTP/2\cite{belshe2015hypertext}, an \name{} sender enables a reinjection queue for each connection. 
% Besides, reinjection queue can also choose to be established and released based on whether $D_{ul}$ actually exists in a live stream. 
% In this case, reinjection queue establishment occurs only if some lost has been resent but unacknowledged, which will be released once all lost data has successfully arrived at its receiver. 

% As a basic solution, the proposed off-mode reinjection still faces the challenge that the detected loss occurs often with limited (no) off-streaming mode due to the uneven distribution of on-off switchings (\S \ref{onoff_sec}), which is addressed in \S \ref{loss_rein_timer_sec}.  
% In this case, the reinjection timer \textsf{T}$_\textsf{timer}$ is introduced for ensuring the timely reinjection of any loss duplicates (\S \ref{loss_rein_timer_sec}). 
\iffalse
\begin{figure}[!t]
\includegraphics[width=0.4\textwidth]{figures/rein_queue_fig_v4.pdf}
%\vspace{-0.6cm}
\centering
\caption{Sender-side reinjection queue reinjection queue.}\label{queue_fig}
  \vspace{-0.3cm}
\end{figure}

\vspace{-1mm}
\fi

% \subsection{Loss Reinjection Timer}
\label{loss_rein_timer_sec}

\vspace{1ex} \noindent \textbf{Opportunistic reinjection.} To address the issue of unevenly distributed on-off mode (\S \ref{onoff_sec}) and conduct replica reinjection more timely, \name{} enables traffic senders to inject packets even during on-mode.
As illustrated in Fig.~\ref{queue_fig}, in the status table of the reinjection queue, $T_{stmp}$ can be used to compute $D_{ul}$'s ``silence" duration since its last retransmission or reinjection. 
Once it exceeds the threshold $T_{thres}$, $D_{ul}$ will be fetched from the reinjection queue and then be resent regardless of whether the off-mode is entered or not. 
% Note that for achieving well-distributed loss reinjections,
% % (the challenge in \S \ref{onoff_sec})
% the threshold $T_{thres}$ is updated based on the latest-determined $K_{\theta}$ as follows: 
To achieve well-distributed loss reinjections, at each decision interval, the Reinjection Controller will use the average loss detection time (denoted by $T_{unit}$) measured by AutoRec and the $K_{\theta}$ calculated by the Redundancy Adapter (\S \ref{redundancy_adaption_sec}) as input, and calculate $T_{thres}$ through the following formula:
\begin{equation}\label{rein_timer_func}
    T_{thres} = \frac{T_{unit}}{K_{\theta} + 1}
\end{equation}
% where $T_{unit}$ is defined in \S \ref{loss_recovery_quality_sec}. 

% It is also worth noting that the smaller the $T_{thres}$, the larger the reinjection overhead. 
% The goodput might decrease due to the bandwidth contention from the reinjected packets. 
% However, the designed Redundancy Adapter in \name{} has already considered the impact of reinjection on goodput (see Eq. \ref{RL3_equation}), which helps \name{} maintain the incurred overhead within an acceptable range.

Fig. \ref{timer_rein_fig} illustrates an example of how the Reinjection Controller performs lost injections with several critical events. Assume the redundancy level determined by the Redundancy Adapter is 2.
% when stepping into off-mode or \textsf{T}$_\textsf{timer}$ expires. 
In this example, a lost packet is detected and resent at $t_{1}$. At time $t_{2}$, the time elapsed since the last transmission of this packet has exceeded threshold $T_{thres}$. Therefore, the Reinjection Controller will reinject a replica of this packet even if the sender is currently not in off-mode.
From $t_{3}$ to $t_{5}$, the \name{} sender enters off-mode, which enables another loss replica to be reinjected at $t_{3}$. After this point, no further replicas of this packet are injected, because for a single packet, we cannot inject more replicas than the redundancy level determined by the Redundancy Adapter.
At $t_{6}$, one of the reinjected replicas is acknowledged, which means that the loss recovery process of the packet is completed. Note that only the last injected replica is checked for loss. Even if all replicas are lost, the existing ARQ mechanism will retransmit the last injected replica when it detects the loss of the last injected replica. 
% Therefore, there is no need to worry about the packet not being retransmitted by ARQ or the packet being retransmitted excessively by ARQ.
Thus, the packet will not fail to be retransmitted by ARQ, nor will it be retransmitted excessively.

\vspace{-1mm}

\vspace{-3mm}

\section{Implementation}
\label{implement_sec}
% \vspace{-2mm}
We integrate all the components described in Section \ref{details_sec} into only sender-side upgrades and implement \name{} prototype based on the user-space QUIC protocol (with the version of LSQUIC Q043) \cite{lsquic_2023} 
% \footnote{For more convenient deployment and evaluation, we select QUIC protocol that is implemented in user-space as the transport protocol.} 
and NGINX architecture (with the version of \textsf{ias-nginx} 1.17.3) \cite{iasnginx_2023}. 
Our implementation consists of 900+ lines of code without any client-side modification. Please refer to Appendix H for details on the \name{} implementation.

\begin{figure}[t]
  \centering
  \includegraphics[width=0.4\textwidth]{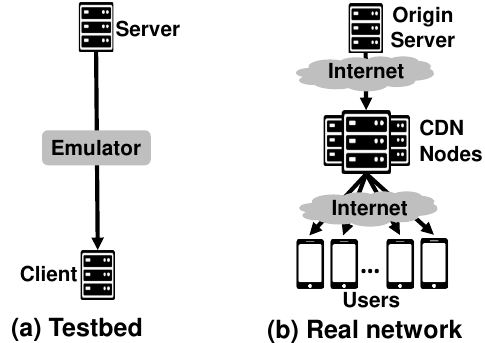}\\
  \caption{Testbed and real-world deployments.}\label{deploy_fig}
\vspace{-0.4cm}
\end{figure}
\vspace{-3mm}

\begin{figure*}[tbp]
    \centering
    \subfigure[BBR]{
        \centering
        \includegraphics[width=0.31\linewidth]{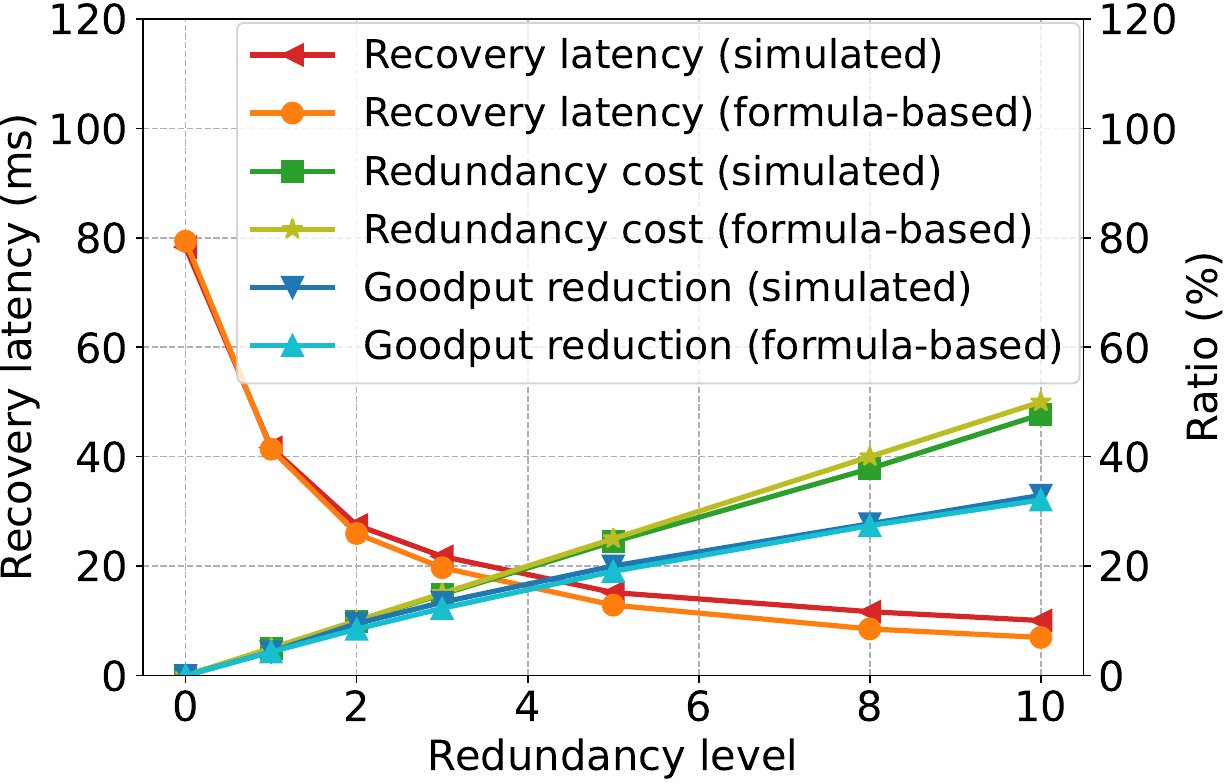}
        \label{model_testing_bbr_fig}
    }
    \hfill
    \subfigure[Copa]{
        \centering
        \includegraphics[width=0.31\linewidth]{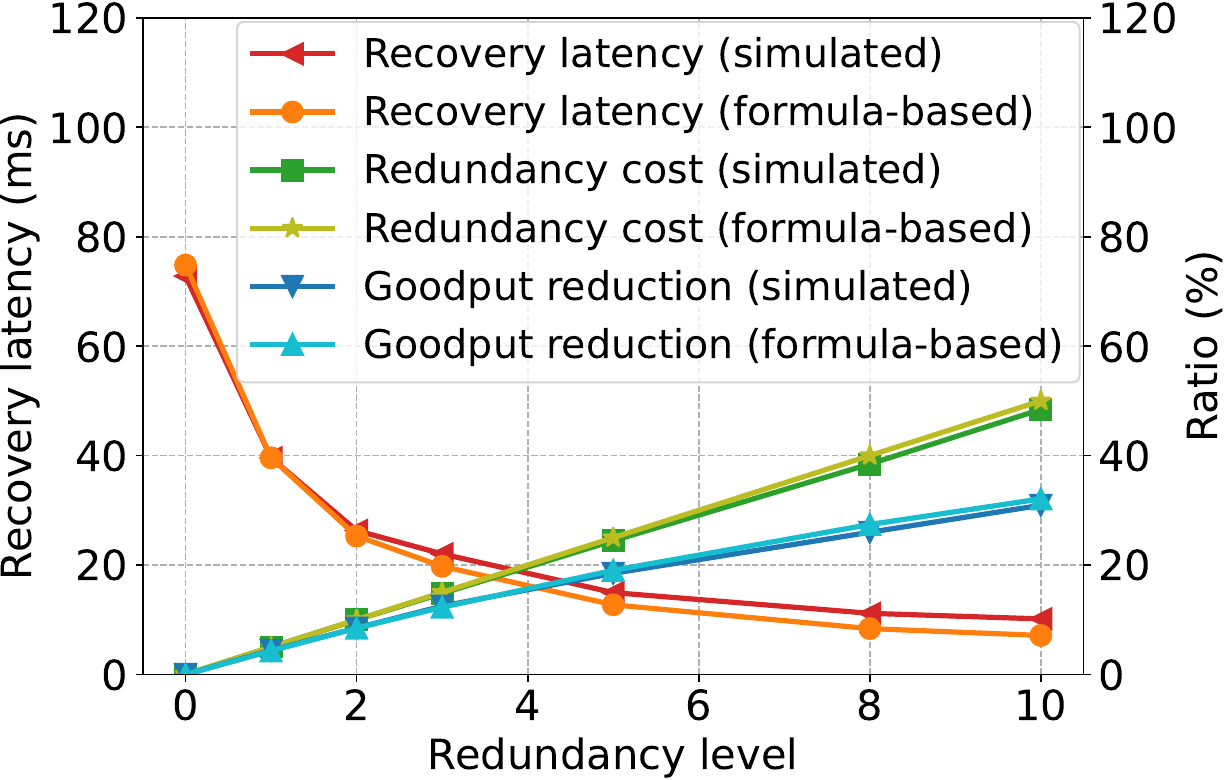}
        \label{model_testing_copa_fig}
    }
    \hfill
    \subfigure[CUBIC]{
        \centering
        \includegraphics[width=0.31\linewidth]{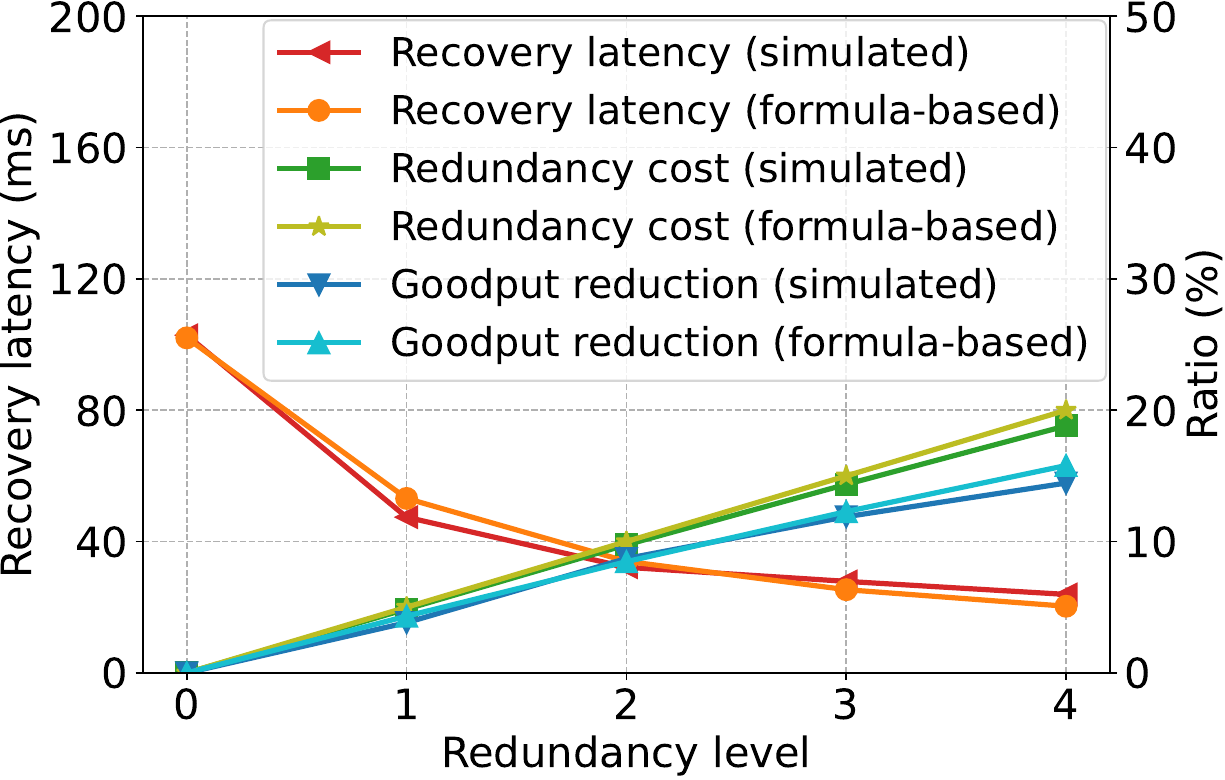}
        \label{model_testing_cubic_fig}
    }
    \vspace{-3mm}
    \caption{Formula accuracy validation.}
    \label{model_correctness_validation_fig}
    \vspace{-5mm}
\end{figure*}
% \vspace{-2mm}

\section{Experimental Evaluation}
\label{experimental_sec}

We perform the experimental evaluations on our established testbed and the real networks, respectively. 
In the testbed experiments, servers deploying \name{} generate live-streaming data and transmit it to clients through a network environment simulated by the simulator, as Fig. \ref{deploy_fig}(a) shows. In real network experiments, CDN nodes deploying \name{} deliver live-streaming data obtained from the origin servers to real user clients, as Fig. \ref{deploy_fig}(b) shows. 
%whose traffic senders (deploying \name{}) can pull and transmit the requested live streaming from our live CDN to its real-network users.

% We integrate all the components described in Section \ref{details_sec} into only sender-side upgrades and 
% The \name{} prototype is implemented based on QUIC protocol \cite{iyengar2020quic} (with LSQUIC Q043) \cite{iasnginx_2023}
% % % \footnote{For more convenient deployment and evaluation, we select QUIC protocol that is implemented in user-space as the transport protocol.} 
% and NGINX architecture (with \textsf{nginx} 1.17.3) \cite{lsquic_2023}, which consists of 900+ lines of code without any client-side modification. 
% Our testbed is composed of (i) two commodity servers that act as the server and the client of live streaming, and (ii) one network emulator that simulates various network environments. 
The testbed server and client are running on CentOS Linux release 7.9 with Intel(R) Xeon(R) CPU E5-2670 v3 @ 2.30GHz (E5-2620 v3 @ 2.40GHz), 48 (24) processors, 62 GB memory, and 1000 Mbps NIC. 
The employed network emulator is HoloWAN ultimate 2600u which supports 0$\sim$1000 Mbps bandwidth, 0$\sim$10 s delay, 0$\sim$100\% loss rate and 0$\sim$1000 GB buffer length. 
In this section, unless otherwise declared, the BBR (with version 1) \cite{cardwell2017bbr} scheme is leveraged for congestion controls. Additionally, to enhance the robustness of \name{}, we limit the value range of the redundancy level to [0, 10].
% compiled and built based on NGINX architecture, where QUIC protocol and BBR (with version 1)\cite{cardwell2017bbr} scheme  are leveraged to achieve traffic transmission and congestion control, respectively. 
% In this experiment, we will compare the \textsf{LRQ} benefits and reinjection overhead of baseline and \name{} variants, and then analyze the tradeoff of these two metrics. 
The baseline scheme is the typical ARQ paradigm that will retransmit one recovery packet once a packet is detected lost. 
% a series of \name{} versions with the configured \textsf{T}$_\textsf{thres}$ = 0 (baseline), 1, 2, and the online learning based adjustment for \textsf{T}$_\textsf{thres} \in$ [0, 4], respectively. 
In this section, each stream's recovery latency is displayed only for the lost data, whose recovery requires at least two retransmissions. 
%remove by xu
% \begin{figure}[t]
%   \centering
%   \includegraphics[width=7cm,height=4cm,angle=0]{figures/utility_overhead_testbed_fig_v2.pdf}\\
%   \caption{Utility and overhead in testbed experiments.}\label{utility_testbed_fig}
% \vspace{-3mm}
% \end{figure}
% \vspace{2mm}

To better evaluate \name{}, we define a new metric called \emph{recovery deterioration rate}, which refers to the ratio between the amount of lost data ($D_{k}$) that takes two or more $T_{unit}$ (i.e., recovery latency $\geq$ $T_{unit}$) to be recovered, to the amount of all lost data. 
For example, there are 2 packets (one is in Fig. \ref{retrans_fig_1} and the other is in Fig. \ref{retrans_fig_2}) detected lost, which require one $T_{unit}$ and two $T_{unit}$ for their successful recoveries (at $t_{4}$ and $t_{7}$), respectively. 
Then, we can learn $D_{k} = 1$ and recovery deterioration rate $=$ $50\%$.

\vspace{-2mm}
\subsection{Testbed Evaluation}
\label{testbed_sec}
The testbed experiments are performed to validate the accuracy of the formulas used by \name{}, the parameter sensitivity of \name{}, the effectiveness of \name{}'s opportunistic reinjection, the advantages compared to the ART~\cite{li2023art}, the performance of \name{} in Real-Time Communication (RTC) scenarios, the robustness of \name{} across diverse packet loss models and loss rate magnitudes, and the performance comparison of \name{} against multiple FEC strategies~\cite{holmer2013handling,2020webrtc88,meng2024hairpin,michael2023tambur}.
% by evaluating its performance under different network environments, specially-designed functions and the sensitivity to some pre-set parameters in \S \ref{details_sec}. 
% Besides, the specially-designed functions (in \S \ref{retran_queue_sec}, \S \ref{loss_rein_timer_sec}, \S \ref{activation_sec} and \S \ref{smart_scheduler_sec}) and the sensitivity of pre-set parameters are also further explored. 
The basic environment is configured as follows: 5\% loss rate, 60 ms Round-Trip Time (RTT), 12 Mbps bandwidth. Unless otherwise declared, we use recovery latency tolerance $\alpha = 30$ ms, redundancy cost tolerance $\beta = 30\%$, and goodput reduction tolerance $\gamma = 20\%$, without loss of generality. The QUIC's loss recovery mechanism employs Tail Loss Probe (TLP)~\cite{10.17487/RFC8985}, Retransmission Timeout (RTO)~\cite{10.17487/RFC6298}, and Forward Acknowledgement (FACK)~\cite{mathis1996forward} for packet loss detection. Upon detecting lost packets, it immediately initiates retransmissions. Crucially, QUIC imposes no maximum retransmission limit per packet to fulfill its reliable data delivery guarantee.
%and 500KB network buffer. 
Each obtained metric is the average value of over 100 sets of experiments, in which each live stream will last for 60 seconds.

\begin{figure}[tp]
\centering
\subfigure[Benefits with different recovery latency tolerance.]{
\includegraphics[width=4.0cm]{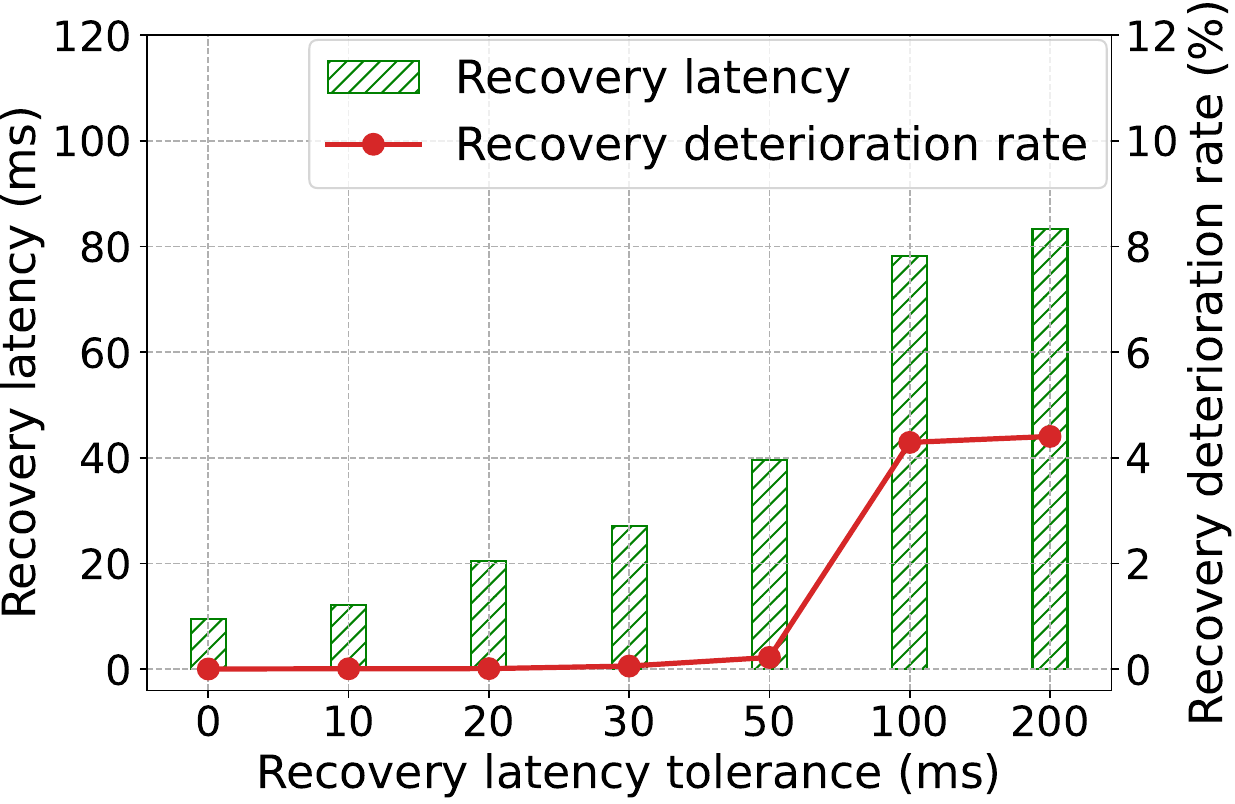}
\label{parameter_sensitivity_recovery_latency_benefit_cutted_fig}
}
\hspace{1mm}
\subfigure[Overhead with different recovery latency tolerance.]{
\includegraphics[width=4.0cm]{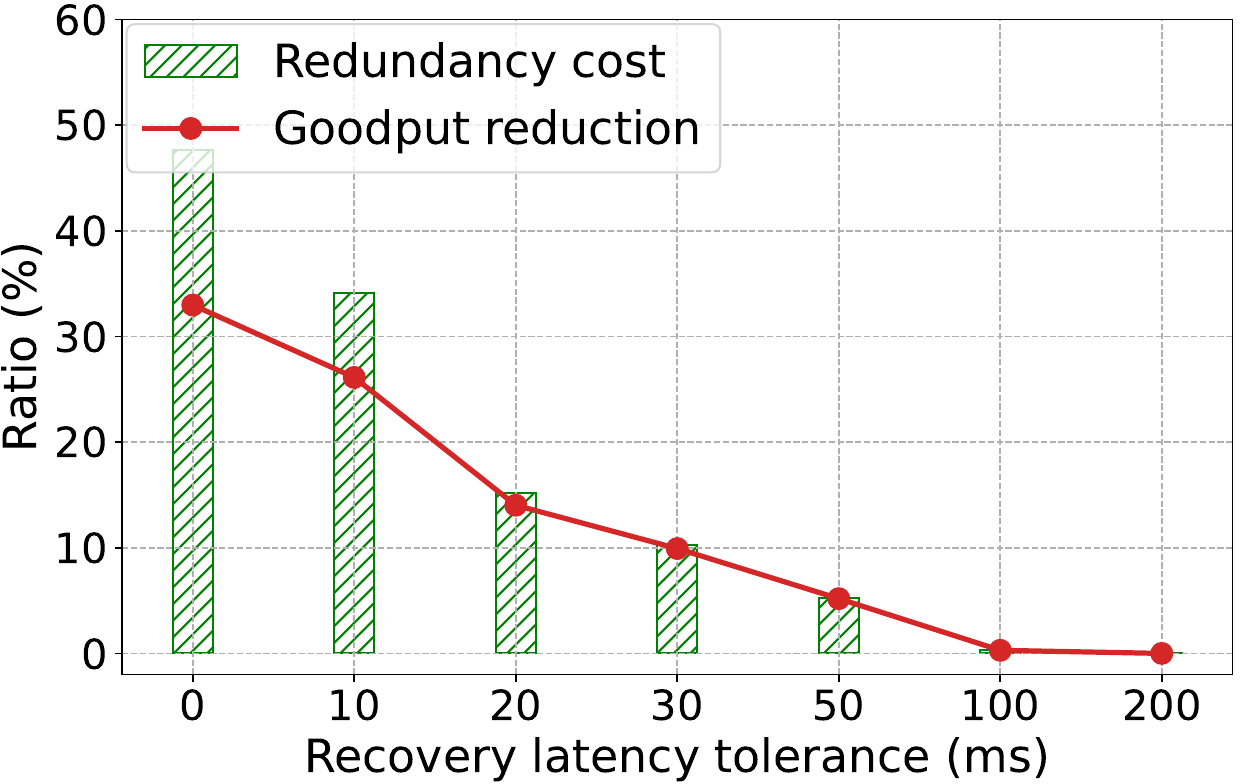}
\label{parameter_sensitivity_recovery_latency_cost_cutted_fig}
}
\vspace{-4mm}
\caption{Recovery latency tolerance ($\alpha$) sensitivity validation.}\label{pre-set_recovery_latency_sensitivity_validation_fig}
\vspace{-4mm}
\end{figure}

\begin{figure}[t]
\centering
\subfigure[Benefits with different redundancy cost tolerance.]{
\includegraphics[width=4.0cm]{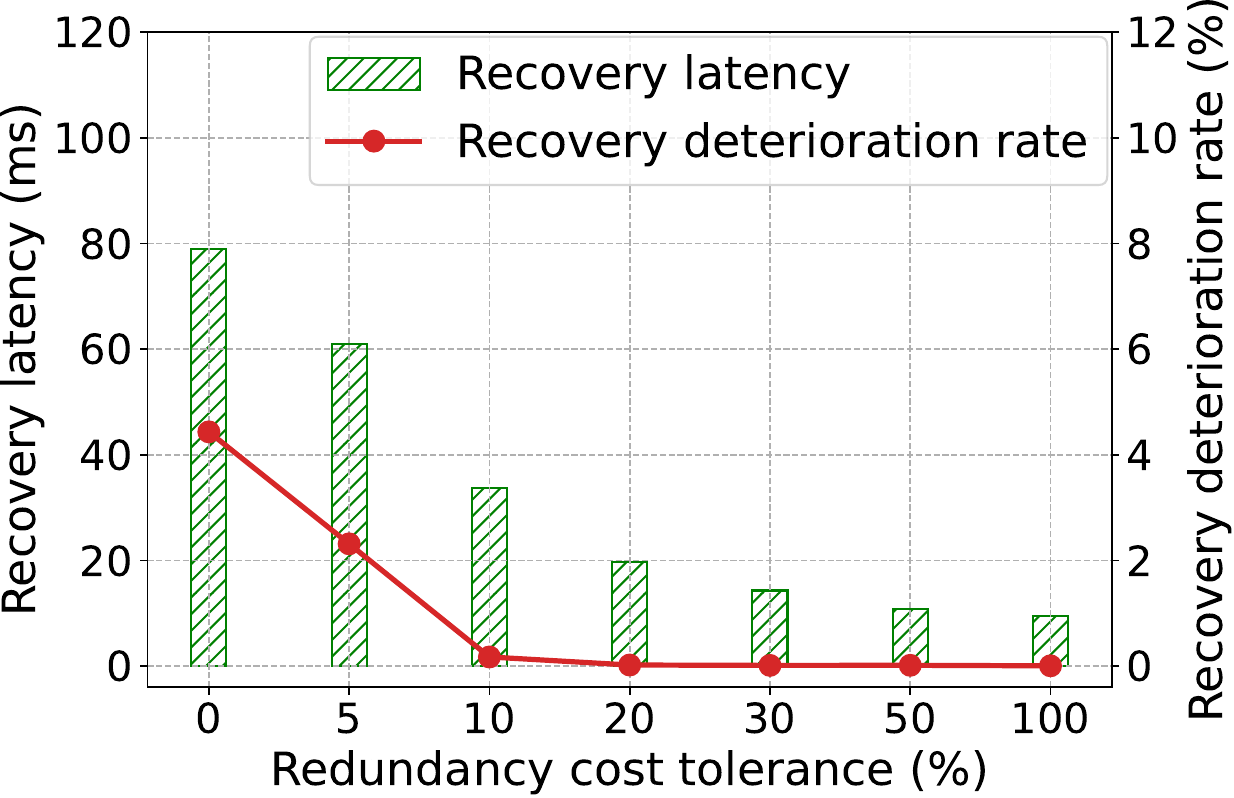}
\label{parameter_sensitivity_cost_ratio_benefit_cutted_fig}
}
\hspace{1mm}
\subfigure[Overhead with different redundancy cost tolerance.]{
\includegraphics[width=4.0cm]{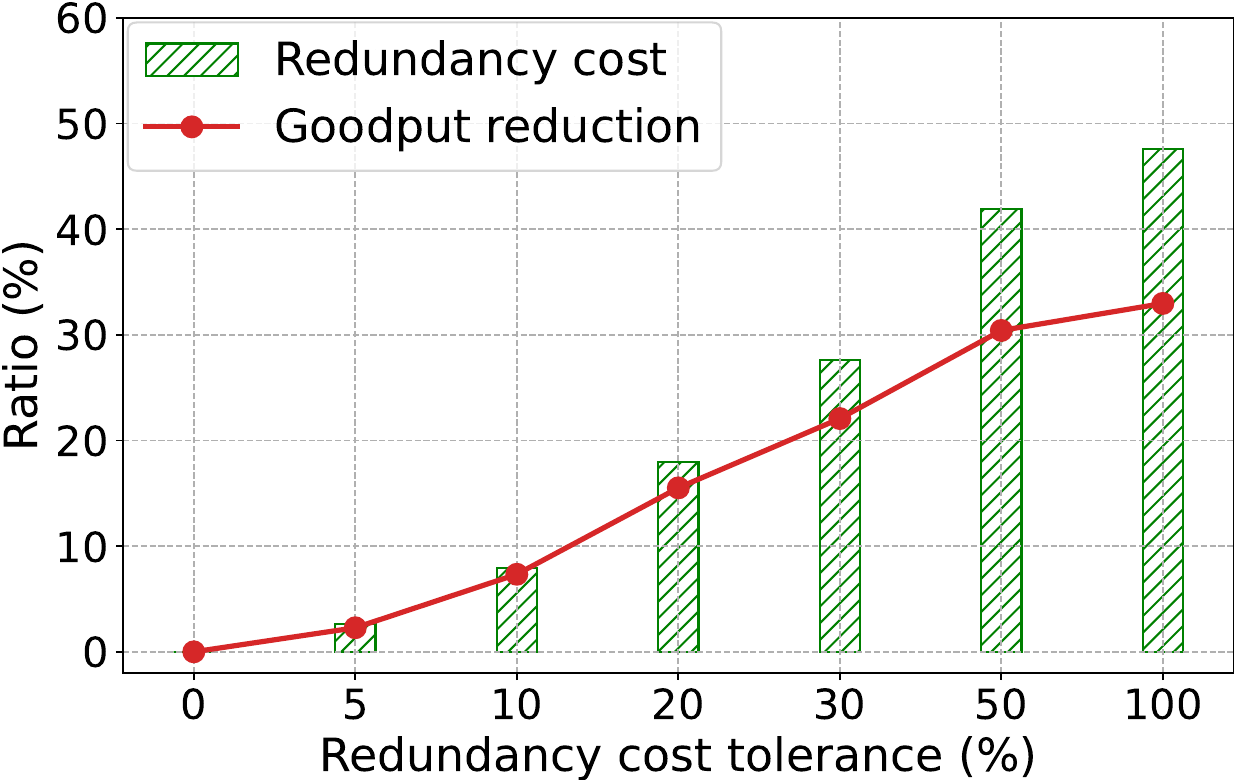}
\label{parameter_sensitivity_cost_ratio_cost_cutted_fig}
}
\vspace{-4mm}
\caption{Redundancy cost tolerance ($\beta$) sensitivity validation.}\label{pre-set_cost_ratio_sensitivity_validation}
\vspace{-5mm}
\end{figure}

\begin{figure}[tp]
\centering
\subfigure[Benefits with different goodput reduction tolerance.]{
\includegraphics[width=4.0cm]{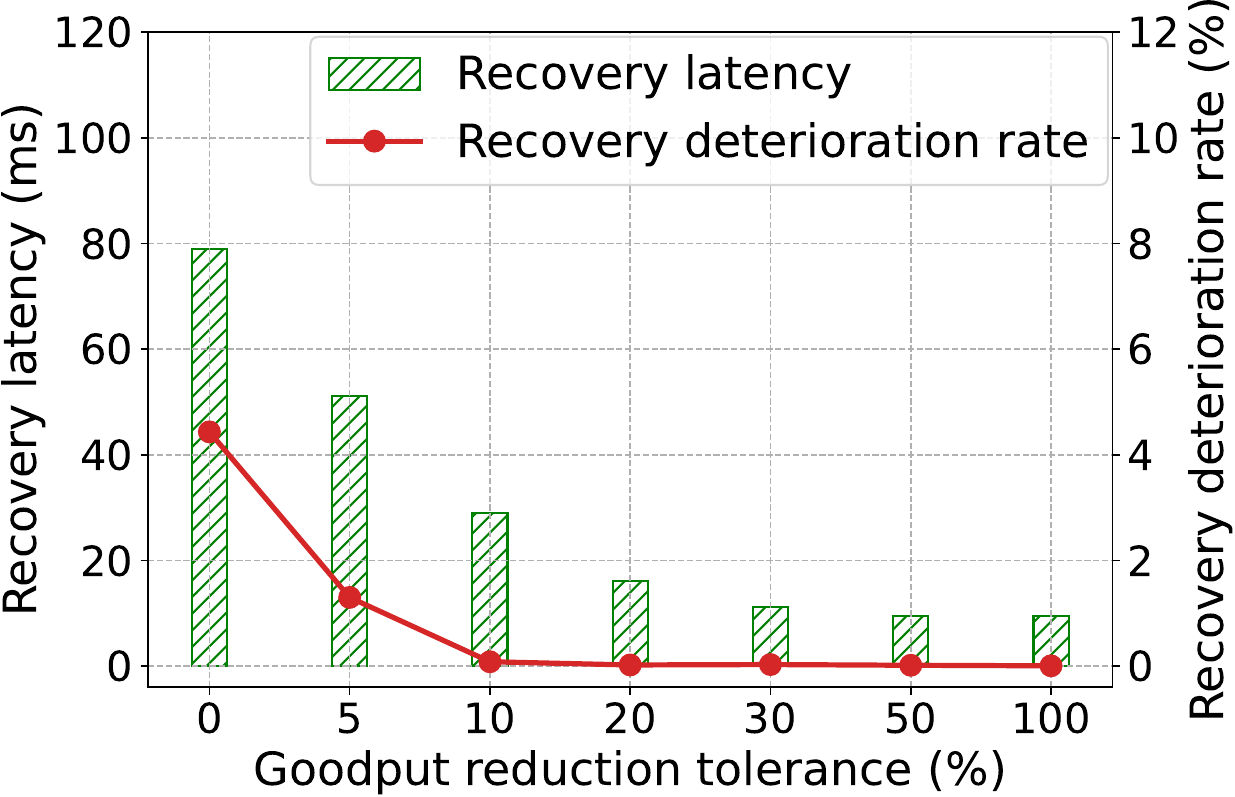}
\label{parameter_sensitivity_goodput_decline_ratio_benefit_cutted_fig}
}
\hspace{1mm}
\subfigure[Overhead with different goodput reduction tolerance.]{
\includegraphics[width=4.0cm]{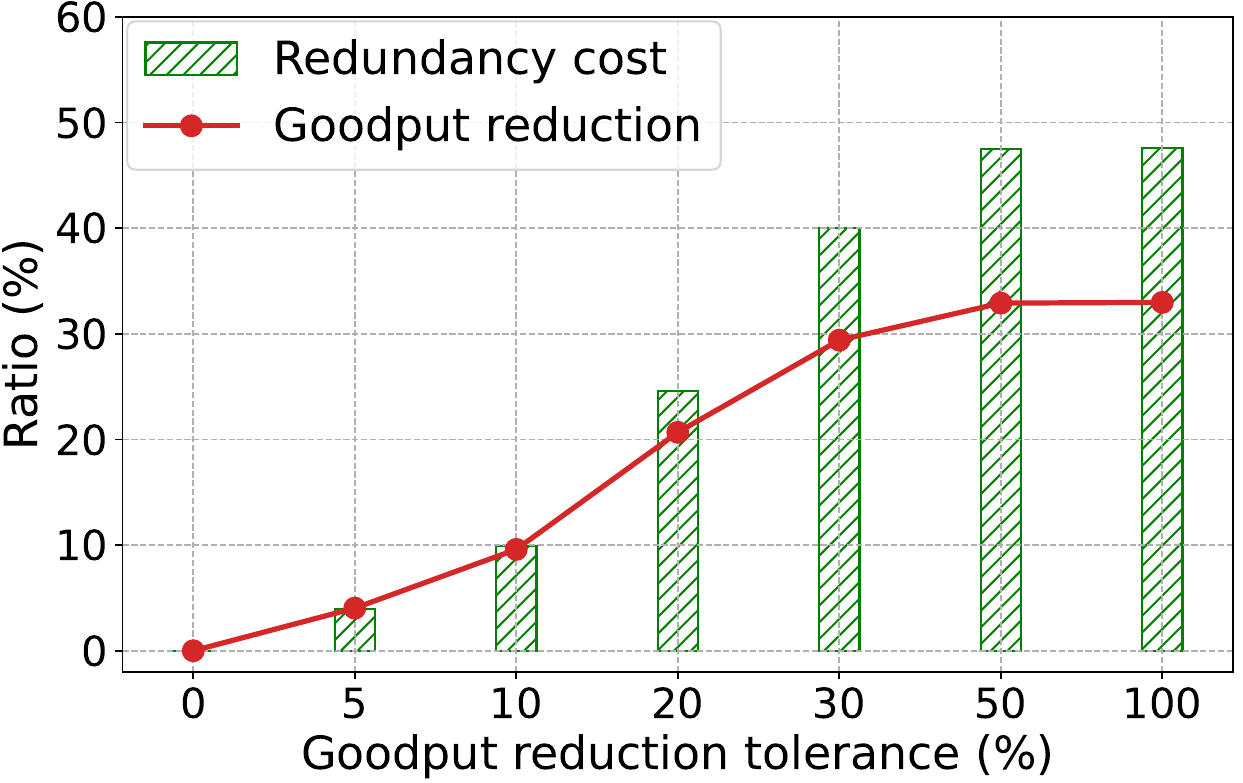}
\label{parameter_sensitivity_goodput_decline_ratio_cost_cutted_fig}
}
\vspace{-2mm}
\caption{Goodput reduction tolerance ($\gamma$) sensitivity validation.}\label{pre-set_goodput_decline_ratio_sensitivity_validation_fig}
\vspace{-6mm}
\end{figure}

\vspace{-1mm}
\vspace{1ex} \noindent \textbf{Formulas accuracy validation.}
\label{benefit_condition_sec}
% \begin{table}[tbp]
%   \small
%   \centering
%   \caption{The introduced overhead and obtained utility.}
%   \begin{threeparttable}
%   \begin{tabular}{|c|c|c|c|}
%    \hline
%    \hline
%    &\cellcolor{lightgray!20}Goodput&Retran\_ratio&\cellcolor{lightgray!20}Utility\\
%    \hline
%    Bitrate&\cellcolor{lightgray!20}-5.1\%$\sim$-9.5\%&4.3\%$\sim$4.4\%&\cellcolor{lightgray!20}29.8\%$\sim$39.9\%\\
%    \hline
%    Loss rate&\cellcolor{red!20}-4.8\%$\sim$-48.5\%&2.7\%$\sim$8.8\%&\cellcolor{red!20}-39.9\%$\sim$35.9\%\\
%    \hline
%    RTT&\cellcolor{lightgray!20}-6.5\%$\sim$-11.2\%&3.5\%$\sim$5.1\%&\cellcolor{lightgray!20}21.3\%$\sim$34.4\%\\
%     \hline
%    BW&\cellcolor{lightgray!20}-3.8\%$\sim$-9.2\%&4.3\%$\sim$4.4\%&\cellcolor{lightgray!20}32.0\%$\sim$37.7\%\\
%    \hline
%    Buffer&\cellcolor{lightgray!20}-6.0\%$\sim$-9.2\%&4.0\%$\sim$4.3\%&\cellcolor{lightgray!20}29.4\%$\sim$37.3\%\\
%    \hline
%    \hline
%   \end{tabular}\label{cost_utility_testbed_table}
%   \end{threeparttable}
%  \end{table}
% The model used by \name{} is accurate as Fig. \ref{model_correctness_validation_fig} shows. 
In this experiment, we validate the accuracy of three formulas used by \name{} for calculating the benefits and overhead varying with the redundancy level (i.e., Formula \ref{recovery_latency_constraint_equation}, Formula \ref{redundancy_ratio_constraint_equation}, and Formula \ref{goodput_reduction_ratio_constraint_equation} in \S \ref{redundancy_adaption_sec}). We use a live stream with a bitrate of 12 Mbps to ensure that the entire transmission process of the stream remains in on-mode, thereby maximizing the negative impact of redundancy injection on goodput. Additionally, we slightly modify the Redundancy Adapter so that the redundancy level remains a fixed preset value throughout the transmission process, allowing us to measure the benefits and overhead of \name{} at the specified redundancy level. We use the formulas employed by \name{}, which describe the benefits and overhead varying with the redundancy level, to calculate the benefits and overhead of \name{} at each fixed redundancy level. Then, we use the testbed to simulate and measure the benefits and overhead of \name{} at each fixed redundancy level. By comparing the values calculated using the formulas with those measured through simulation, we can validate the accuracy of the formulas used by \name{} for calculating the benefits and overhead varying with the redundancy level. Due to the significant impact of congestion control algorithms on transmission control, we conduct formulas accuracy validation based on different congestion control algorithms separately.

As shown in Fig. \ref{model_correctness_validation_fig}, whether using BBR, Copa, or Cubic, the calculated values from the formulas and the testbed simulation values for recovery latency, redundancy cost, and goodput reduction for a given redundancy level are in close agreement. When using BBR and Copa, the absolute error of recovery latency does not exceed 3.2 ms, and the absolute errors of redundancy cost and goodput reduction do not exceed 2.4\%. When using CUBIC, due to the larger values of recovery latency, the absolute error of recovery latency is slightly higher but does not exceed 5.8 ms, and the absolute errors of redundancy cost and goodput reduction do not exceed 1.4\%. As the redundancy level increases, differences between the calculated values from the formulas and the testbed simulation values increase. This is because the actual amount of injected replicas is limited by the sending rate determined by the congestion control algorithm, meaning that the amount of replicas injected within a $T_{unit}$ is capped. The larger the redundancy level, the greater the difference between the actual average amount of replicas injected per packet and the redundancy level, resulting in a larger difference between the simulated measurement values and the calculated values from the formulas for \name{}'s benefits and overhead. 

In summary, the formulas used by \name{} can accurately describe the benefits and overhead of \name{} at each fixed redundancy level.

\vspace{-2mm}

\vspace{1ex} \vspace{1ex} \noindent \textbf{The sensitivity of user-customizable parameters.}
In this experiment, we test the performance of \name{} under different user-customizable parameters (i.e., recovery latency tolerance, redundancy cost tolerance, and goodput reduction tolerance in \S \ref{redundancy_adaption_sec}). We use a live stream with a bitrate of 12 Mbps to ensure that the entire transmission process remains in on-mode, thereby maximizing the impact of injection on goodput. We first set both the redundancy cost tolerance and the goodput reduction tolerance to 100\% to observe how \name{}'s benefits and overhead change with the recovery latency tolerance. Next, we set recovery latency tolerance to 0 ms and observe how \name{}'s benefits and overhead change with the redundancy cost tolerance or goodput reduction tolerance by setting the goodput reduction tolerance to 100\% or the redundancy cost tolerance to 100\%, respectively. 

As shown in Fig. \ref{pre-set_recovery_latency_sensitivity_validation_fig}, the benefits and overhead of \name{} gradually decrease as the recovery latency tolerance increases. As shown in Fig. \ref{pre-set_cost_ratio_sensitivity_validation} and Fig. \ref{pre-set_goodput_decline_ratio_sensitivity_validation_fig}, as the redundancy cost tolerance or goodput reduction tolerance increases, \name{}'s benefits and overhead also increase. When the recovery latency tolerance is set to 0 ms and both the redundancy cost tolerance and goodput reduction tolerance are set to 100\%, \name{}'s benefits and overhead are at their maximum, with the values of recovery latency, recovery deterioration rate, redundancy cost, and goodput reduction being 9.5 ms, $\sim$0\%, 47.6\%, and 33.0\%, respectively. 

As shown in Fig. \ref{pre-set_recovery_latency_sensitivity_validation_fig}, due to factors such as the limit on the maximum redundancy level, \name{} cannot achieve the theoretical minimum recovery latency. When the recovery latency tolerance is set between 10 ms and 200 ms, \name{} can inject appropriate replicas to control the recovery latency within no more than 3 ms of the recovery latency tolerance. When the recovery latency tolerance exceeds 100 ms, the recovery latency can already meet the recovery latency tolerance requirements without injecting redundant replicas by \name{}. At this point, \name{} injects negligible replicas, and the redundancy cost and goodput reduction approach 0\%.

As shown in Fig. \ref{pre-set_cost_ratio_sensitivity_validation} and Fig. \ref{pre-set_goodput_decline_ratio_sensitivity_validation_fig}, regardless of the values set for the redundancy cost tolerance or goodput reduction tolerance, \name{} can maintain the measured redundancy cost or goodput reduction no more than 1\% of the redundancy cost tolerance or goodput reduction tolerance. When the redundancy cost tolerance or goodput reduction tolerance gradually increases from 0\% to 30\%, \name{} can quickly reduce the recovery latency to below 20 ms. 
% When the redundancy cost tolerance or goodput reduction tolerance exceeds 30\%, the overhead of \name{} is relatively high, and the optimization effect on recovery latency is not significantly improved compared to when the redundancy cost tolerance or goodput reduction tolerance is set to 30\%.
Further increasing the redundancy cost or goodput reduction tolerance beyond 30\% provides diminishing returns in terms of recovery latency optimization, while incurring higher overhead.

In summary, \name{} can accelerate recovery latency to meet user-customizable recovery latency tolerance whenever possible while satisfying the constraints of the user-customizable redundancy cost tolerance and goodput reduction tolerance.

\begin{figure}[tbp]
\centering
\subfigure[Benefits with different bandwidth.]{
\includegraphics[width=4.0cm]{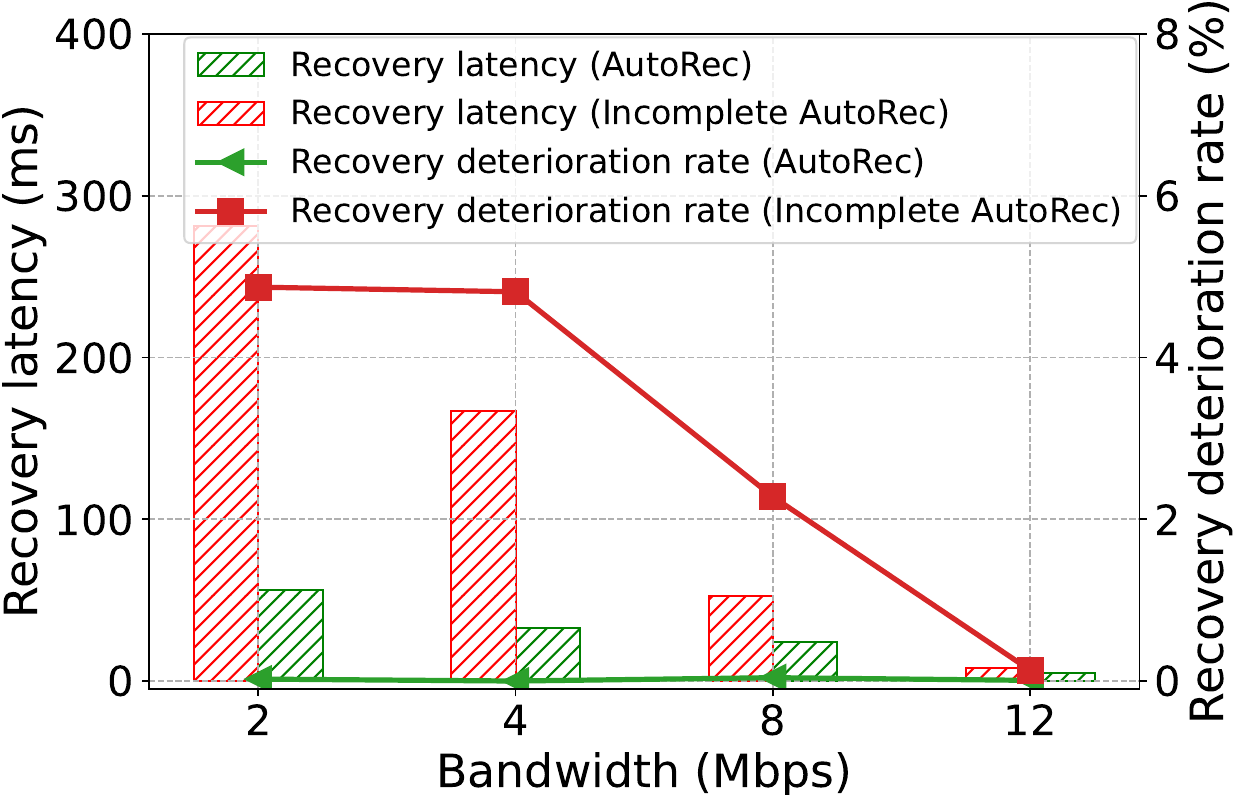}
\label{opportunistic_reinjection_effectiveness_validation_benefit_fig}
}
\hspace{1mm}
\subfigure[Overhead with different bandwidth.]{
\includegraphics[width=4.0cm]{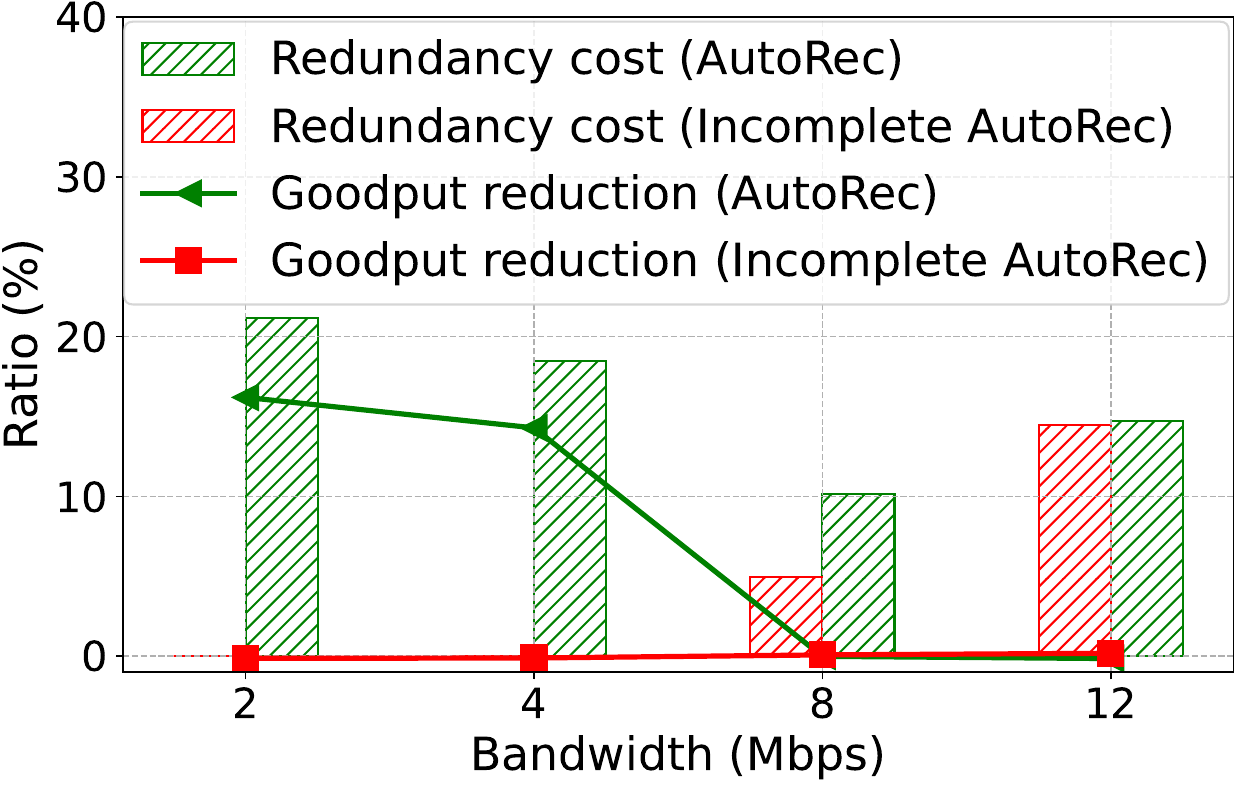}
\label{opportunistic_reinjection_effectiveness_validation_overhead_fig}
}
\vspace{-4mm}
\caption{Opportunistic reinjection effectiveness validation.}\label{opportunistic_reinjection_effectiveness_validation_fig}
\vspace{-6mm}
\end{figure}

\vspace{-2mm}

\vspace{1ex} \vspace{1ex} \noindent \textbf{Opportunistic reinjection effectiveness validation.} In this experiment, we verify the effectiveness of using opportunistic reinjection in the Reinjection Controller (\S \ref{reinjection_controller_sec}). We implement two versions of \name{}: one with the opportunistic reinjection mechanism and one without. We set the bitrate of the live stream to 4 Mbps and observe how the benefits and overhead of the two versions of \name{} change as the bandwidth gradually increases from 2 Mbps to 12 Mbps.

When the bandwidth is between 2 Mbps and 4 Mbps, the entire transmission process is in on-mode. As shown in Fig. \ref{opportunistic_reinjection_effectiveness_validation_fig}, in this situation, the version of \name{} without the opportunistic reinjection mechanism injects negligible replicas, resulting in higher recovery latency and recovery deterioration rate, which are 281.4 ms $\sim$ 166.7 ms and 4.9\% $\sim$ 4.8\%, respectively, while its redundancy cost and goodput reduction are almost 0. In contrast, \name{} with the opportunistic reinjection mechanism injects an appropriate amount of redundant replicas, optimizing the recovery latency to 56.1 ms $\sim$ 32.9 ms and the recovery deterioration rate to nearly 0\%. This optimization effect is significantly better compared to \name{} without the opportunistic reinjection mechanism. Furthermore, its redundancy cost and goodput reduction reach 21.2\% $\sim$ 18.5\% and 16.2\% $\sim$ 14.3\%, respectively, which are higher than those of \name{} without the opportunistic reinjection mechanism. Nevertheless, it still meets the constraints of the redundancy cost tolerance and goodput reduction tolerance.

When the bandwidth is between 8 Mbps and 12 Mbps, part of the transmission process will be in off-mode. As shown in Fig. \ref{opportunistic_reinjection_effectiveness_validation_fig}, the version of \name{} without the opportunistic reinjection mechanism can inject more redundant replicas as the bandwidth increases, optimizing the recovery latency and recovery deterioration rate to 52.3 ms $\sim$ 7.9 ms and 2.3\% $\sim$ 0.1\%, respectively, with its redundancy cost and goodput reduction being 5.0\% $\sim$ 14.5\% and nearly 0\%. Similarly, the version of \name{} with the opportunistic reinjection mechanism can also inject an appropriate amount of redundant replicas, optimizing the recovery latency and recovery deterioration rate to 24.3 ms $\sim$ 5.1 ms and nearly 0\%, respectively, which is better than that of the version without the opportunistic reinjection mechanism. Furthermore, its redundancy cost and goodput reduction are 10.2\% $\sim$ 14.7\% and nearly 0\%, respectively. Although the overhead is higher compared to the version without the opportunistic reinjection mechanism, they still meet the constraints of redundancy cost tolerance and goodput reduction tolerance.

In summary, compared to the version of \name{} without the opportunistic reinjection mechanism, the version with the opportunistic reinjection mechanism can inject an appropriate amount of replicas to accelerate loss recovery when there is no off-mode or when the off-mode distribution is uneven.

\begin{figure}[tbp]
\centering
\subfigure[Recovery latency benefits with different RTT.]{
\includegraphics[width=4.0cm]{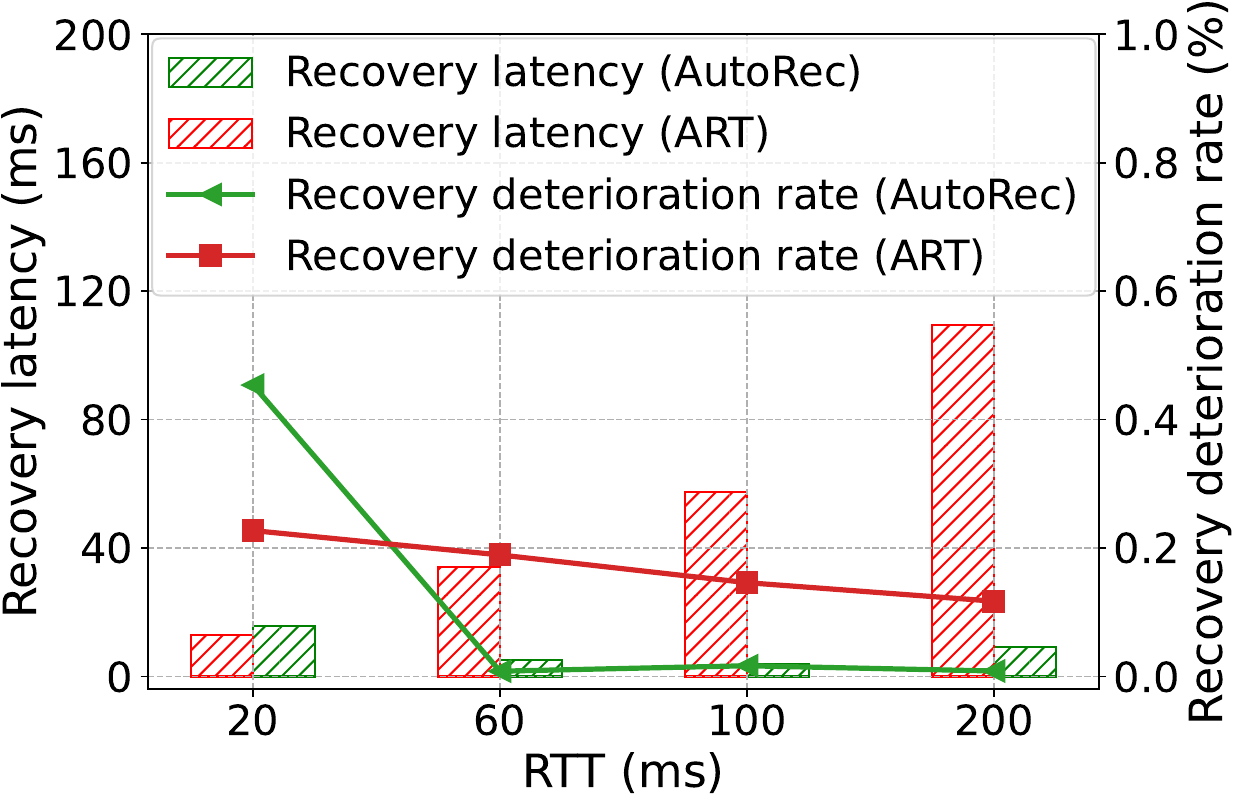}
\label{comparison_with_art_benefit_cutted_fig}
}
\hspace{1mm}
\subfigure[Overhead with different RTT.]{
\includegraphics[width=4.0cm]{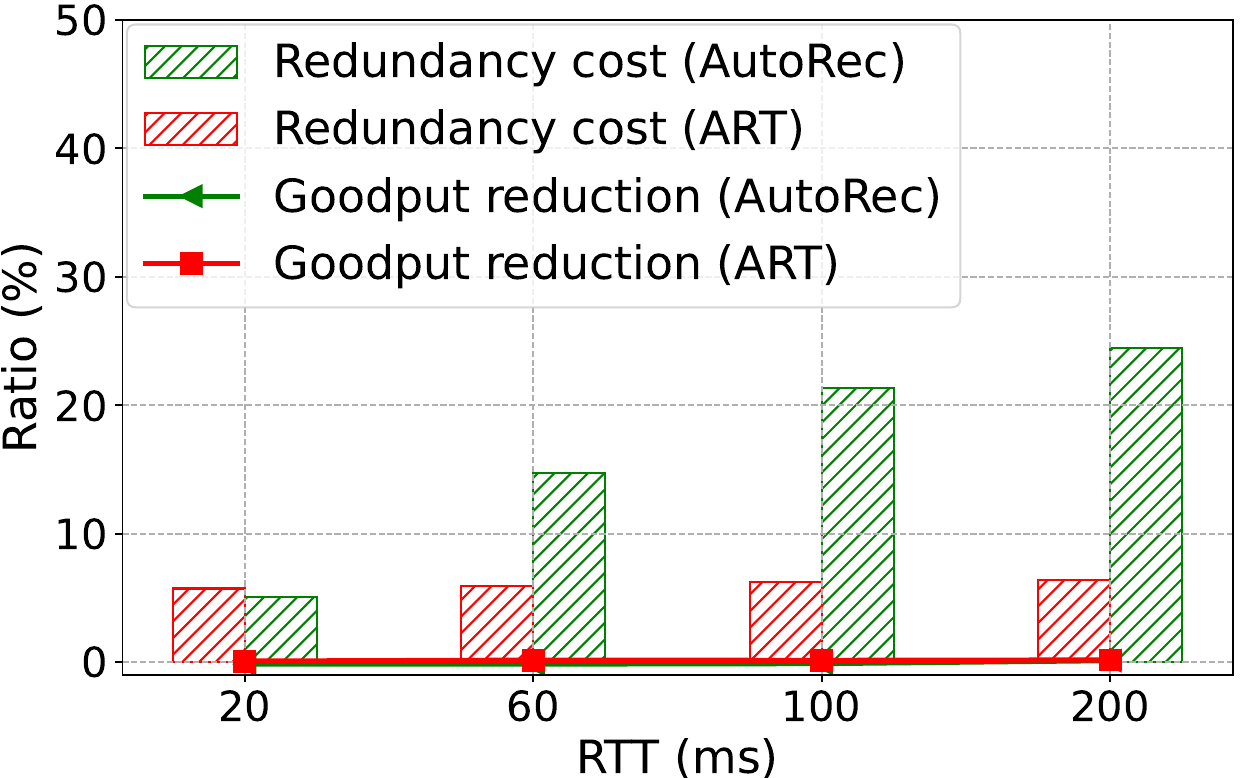}
\label{comparison_with_art_cost_cutted_fig}
}
\vspace{-4mm}
\caption{\name{} vs. ART.}\label{\name{}_vs._ART_fig}
\vspace{-5mm}
\end{figure}

\vspace{-2mm}
\vspace{1ex} \vspace{1ex} \noindent \textbf{\name{} vs. ART.}
This experiment compares two loss recovery acceleration schemes: \name{} and ART~\cite{li2023art}. We set the bitrate of the live stream to 4 Mbps and observe how the benefits and overhead of \name{} and ART change as the RTT gradually increases from 20 ms to 200 ms.

As Fig. \ref{\name{}_vs._ART_fig} shows, when the RTT is 20 ms, ART injects more replicas than \name{}, optimizing the recovery latency and recovery deterioration rate to 13.02 ms and 0.23\%, respectively. Its redundancy cost and goodput reduction are 5.73\% and nearly 0\%, respectively. In this case, \name{} optimizes the recovery latency and recovery deterioration rate to 15.76 ms and 0.45\%, respectively. Although the optimization effect is slightly worse than ART, it meets the recovery latency tolerance. Its redundancy cost and goodput reduction are 5.10\% and nearly 0\%, respectively, which is slightly lower compared to ART.

As Fig. \ref{\name{}_vs._ART_fig} shows, when the RTT is between 60 ms and 200 ms, ART's recovery latency increases from 34.03 ms to 109.44 ms as the RTT increases, with the recovery deterioration rate between 0.12\% and 0.19\%. Its redundancy cost and goodput reduction are 5.93\% $\sim$ 6.37\% and nearly 0\%, respectively.
% with benefits and overhead not significantly different from when the RTT is 20 ms. 
On the other hand, \name{} injects more redundant replicas than ART, maintaining the recovery latency below 30 ms and the recovery deterioration rate nearly 0\%. Its redundancy cost increases with the RTT, ranging from 14.72\% to 24.44\%, while its goodput reduction remains nearly 0\%.

In summary, compared to ART, \name{} adjusts the amount of redundant replica injection based on different RTTs to optimize the loss recovery latency.

\vspace{-2mm}

\vspace{1ex} \vspace{1ex} \noindent
\textbf{\name{} vs. FEC.}
% In this experiment, we implemented \name{} on two Real Time Communication(RTC) simulation platforms and compared its effectiveness against multiple FEC strategies across different packet loss models and loss rates.
% 并于多种FEC策略对比，以此评估AutoRec在RTC场景中不同丢包模型和丢包率下的表现。
In this experiment, we implement \name{} on two RTC simulation platforms to: (1) validate its performance in RTC scenarios, (2) assess its robustness across diverse packet loss models and loss rate magnitudes, and (3) benchmark its effectiveness against multiple FEC strategies.

This experimental evaluation consists of the following mechanisms:

\vspace{-1mm}

\begin{itemize}
\item $\mathbf{\name{}}$: Our proposed packet loss recovery mechanism. In RTC scenarios, we use parameter settings recovery latency tolerance $\alpha=0$ ms (minimizing recovery latency), redundancy cost tolerance $\beta=20\%$ , and goodput reduction tolerance $\gamma=20\%$ as the standard configuration.
\item $\mathbf{\name{}_{LC}}$: The \name{} mechanism with $\alpha=0$ ms, $\beta=5\%$, and $\gamma=5\%$. Compared to standard \name{} ($\beta=20\%$, $\gamma=20\%$), \name{} under this configuration aims to reduce bandwidth cost by accepting higher susceptibility to video freezing.
\item $\mathbf{\name{}_{HC}}$: The \name{} mechanism with $\alpha=0$ ms, $\beta=50\%$, and $\gamma=50\%$. Compared to standard \name{} ($\beta=20\%$, $\gamma=20\%$), 
\name{} under this configuration aims to reduce video freezing by tolerating higher bandwidth costs.

\item $\mathbf{RTX}$: A fundamental packet loss recovery mechanism that detects lost packets via ACK sequencing and retransmits them.

\item $\mathbf{WebRTC_{'14}}$: The original Web Real Time Communication (WebRTC)~\cite{webrtcorg} FEC mechanism proposed in Google's 2014 paper ~\cite{holmer2013handling}, which dynamically adjusts redundancy based on RTT measurements.

% \item \textbf{WebRTC$\mathbf{_{'14}}$}: The original WebRTC FEC mechanism proposed in Google's 2014 paper [46], which dynamically adjusts redundancy based on RTT measurements.

\item $\mathbf{WebRTC_{'NOW}}$: The current industry-standard WebRTC FEC implementation (used in Stadia~\cite{di2021analysis}, Meet~\cite{brianhu2021meet}, etc.) that employs more aggressive redundancy than WebRTC$_{'14}$.
% , implemented based on Chromium m88~\cite{2020webrtc88}.

\item $\mathbf{Hairpin}$: A novel packet loss recovery mechanism for edge-based interactive streaming that adaptively applies FEC encoding to lost packets~\cite{meng2024hairpin}. The standard coefficient $\lambda$ of Hairpin is set to $10^{-4}$.

\item $\mathbf{Hairpin_{LC}}$: The Hairpin mechanism configured with $\lambda = 10^{-1}$. Compared to standard Hairpin ($\lambda = 10^{-4}$), Hairpin under this configuration aims to reduce bandwidth cost by accepting higher susceptibility to video freezing.

\item $\mathbf{Hairpin_{HC}}$: The Hairpin mechanism configured with $\lambda = 10^{-7}$. Compared to standard Hairpin ($\lambda = 10^{-4}$), Hairpin under this configuration aims to reduce video freezing by tolerating higher bandwidth costs.
\xu{\item $\mathbf{Tambur}$: An efficient loss recovery mechanism for videoconferencing that leverages streaming codes—a specialized FEC variant optimized for burst losses~\cite{michael2023tambur}.
We set the maximum FEC redundancy ratio to $25\%$ by default.

\item $\mathbf{Tambur_{LC}}$: The Tambur mechanism was configured with its maximum redundancy parameter set to the minimum value, resulting in the generation of exactly one FEC redundancy packet per video frame. Compared to standard Tambur (maximum redundancy ratio $ = 25\%$), Tambur under this configuration aims to reduce bandwidth cost by accepting higher susceptibility to video freezing.

\item $\mathbf{Tambur_{HC}}$: The Tambur mechanism configured with maximum redundancy ratio $ = 50\%$. Compared to standard Tambur (maximum redundancy ratio $ = 25\%$), Tambur under this configuration aims to reduce video freezing by tolerating higher bandwidth costs.
}

\end{itemize}

We evaluate the following performance metrics: 
\begin{itemize}
\item $\mathbf{Deadline}$ $ \mathbf{miss}$ $\mathbf{rate}$ $\mathbf{(DMR)}$: Proportion of video frames failing to render before their deadline due to packet loss.

\item $\mathbf{Bandwidth}$ $ \mathbf{cost}$ $\mathbf{(BWC)}$: The ratio of non-video-frame bytes (e.g., reinjection redundancy bytes, FEC redundancy bytes) to total transmitted bytes.

\item $\mathbf{Percent}$ $ \mathbf{of}$ $\mathbf{video}$ $\mathbf{spent}$ $\mathbf{frozen}$: 
% The cumulative duration of video freezing events divided by the total video playback time, expressed as a percentage.
The ratio of the cumulative duration of video freezing events to the total video playback time, expressed as a percentage.

\end{itemize}

% First, we evaluate the deadline miss rate and bandwidth cost of \name{}, \name{}$\mathsf{_{LC}}$, \name{}$\mathsf{_{HC}}$, WebRTC$\mathsf{_{'14}}$, WebRTC$\mathsf{_{'NOW}}$, Hairpin, Hairpin$\mathsf{_{LC}}$, and Hairpin$\mathsf{_{HC}}$ under various fixed packet loss rates using an ns3-based WebRTC simulator~\cite{Soonyangzhang2020webrtcns3}

We configure the environment with 12 Mbps bandwidth, 4 Mbps bitrate, 60 fps frame rate, 30 ms RTT, and 100 ms playback deadline.

\begin{figure}[tbp]
\centering
\subfigure[Loss rate = 0.5\%.]{
\includegraphics[width=4.1cm]{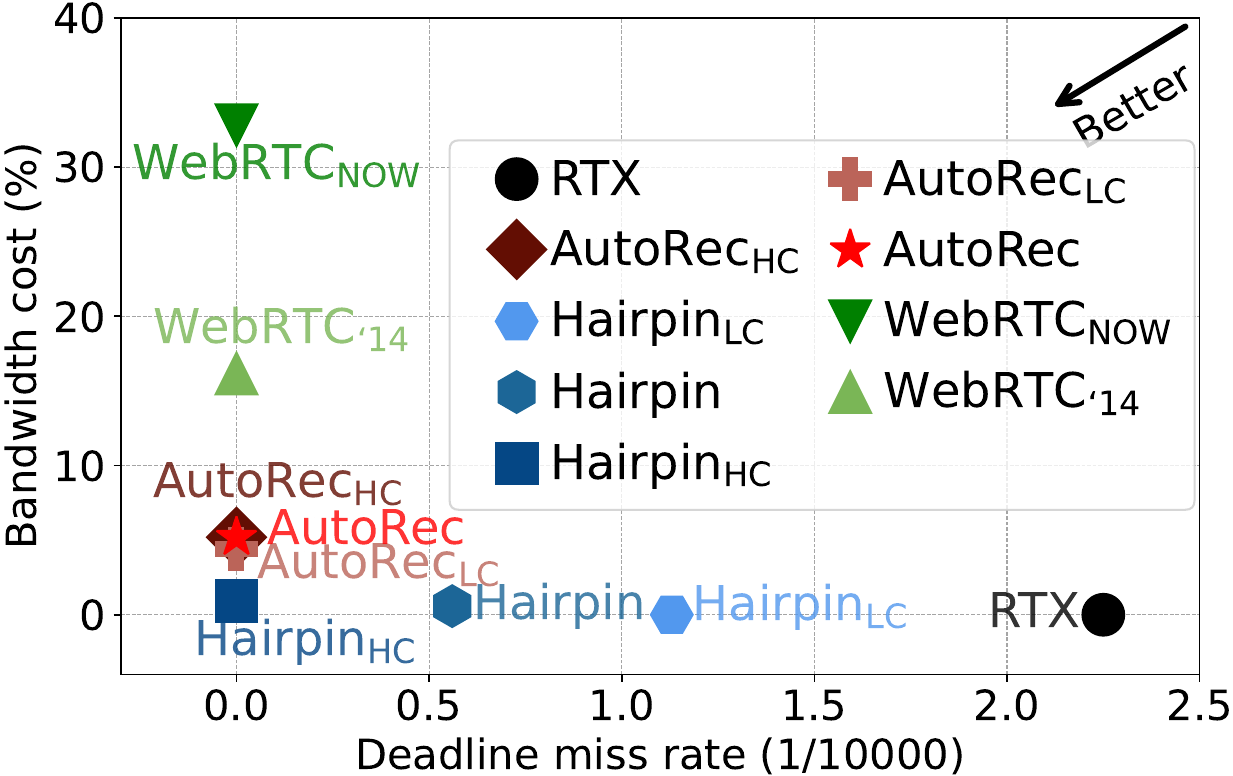}
\label{autorec_vs_fec_0.5}
}
% \hspace{1mm}
\subfigure[Loss rate = 1\%.]{
\includegraphics[width=4.1cm]{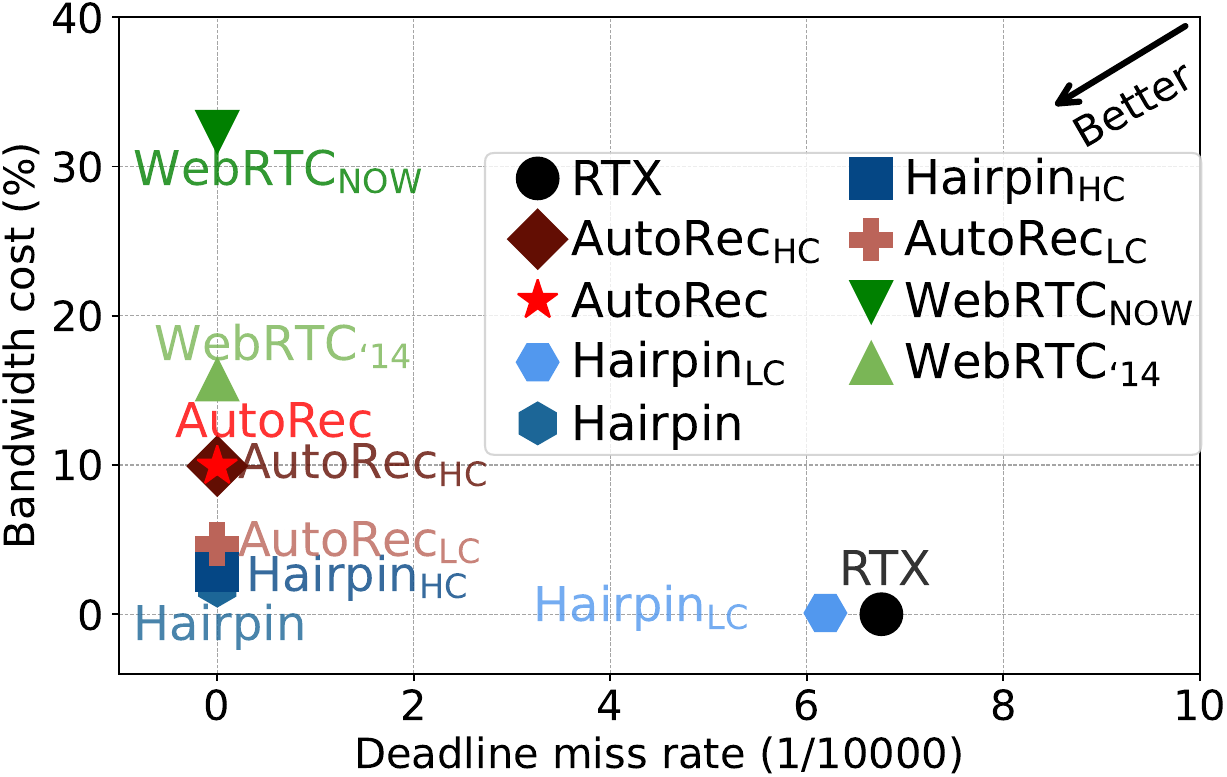}
\label{autorec_vs_fec_1}
}
\subfigure[Loss rate = 5\%.]{
\includegraphics[width=4.1cm]{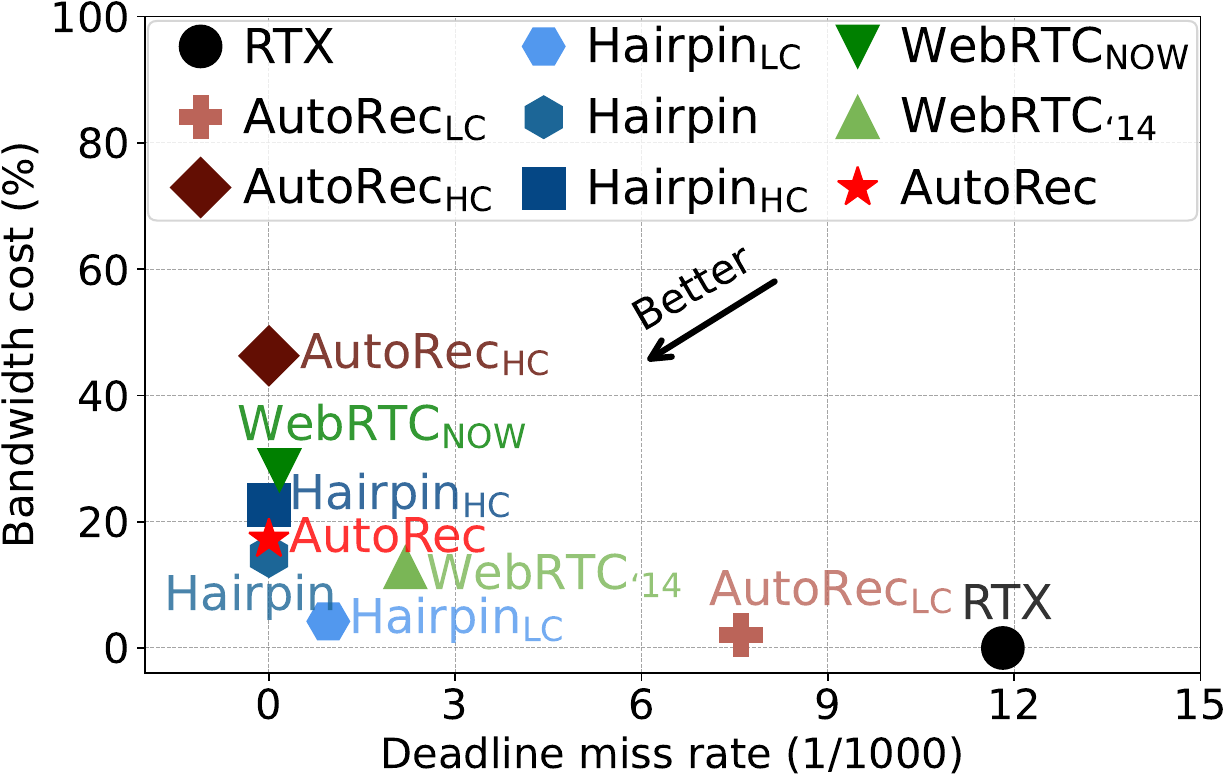}
\label{autorec_vs_fec_5}
}
% \hspace{1mm}
\subfigure[Loss rate = 10\%.]{
\includegraphics[width=4.1cm]{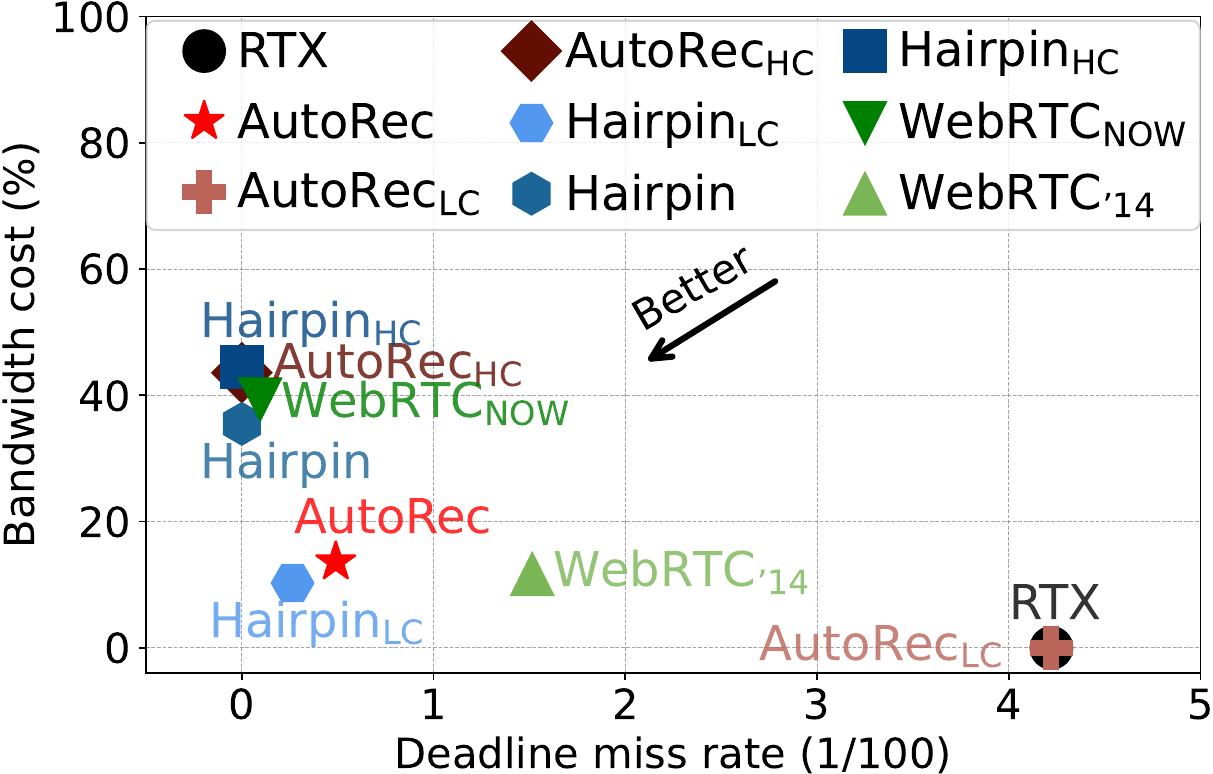}
\label{autorec_vs_fec_10}
}

\subfigure[Trace \#1.]{
\includegraphics[width=4.1cm]{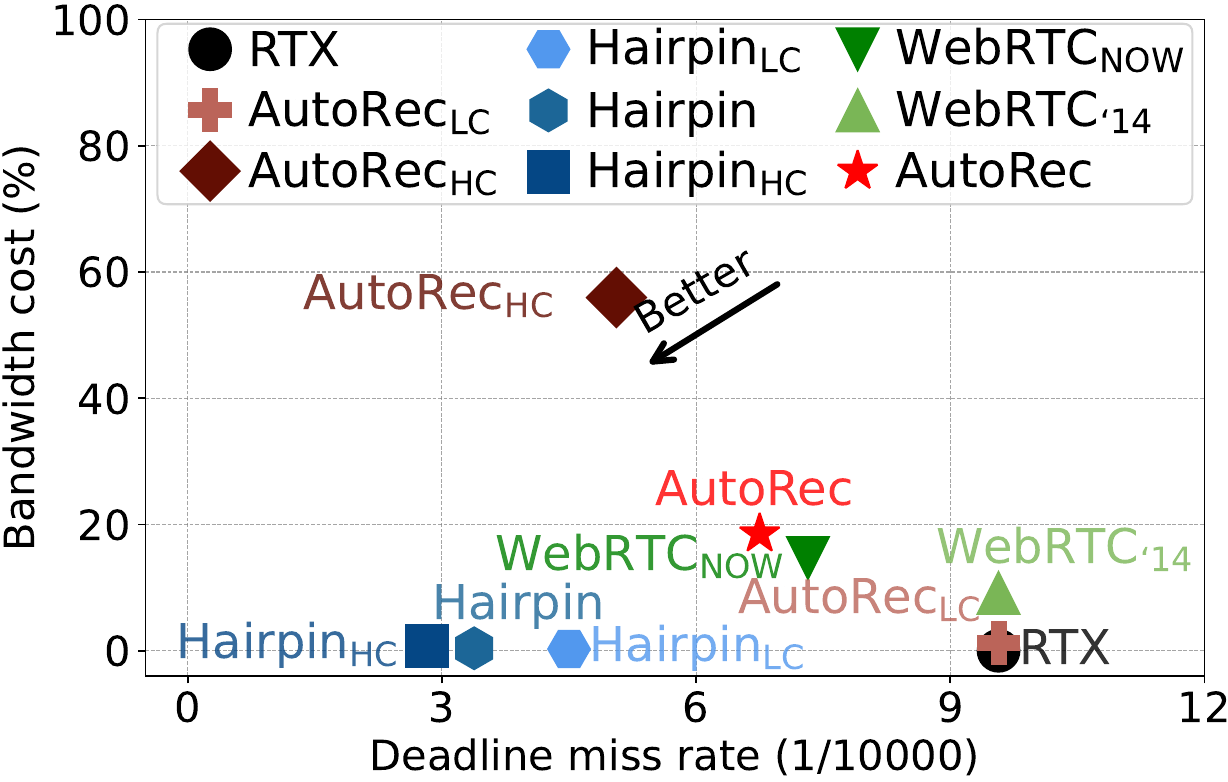}
\label{autorec_vs_fec_sample_tr}
}
\subfigure[Trace \#2.]{
\includegraphics[width=4.1cm]{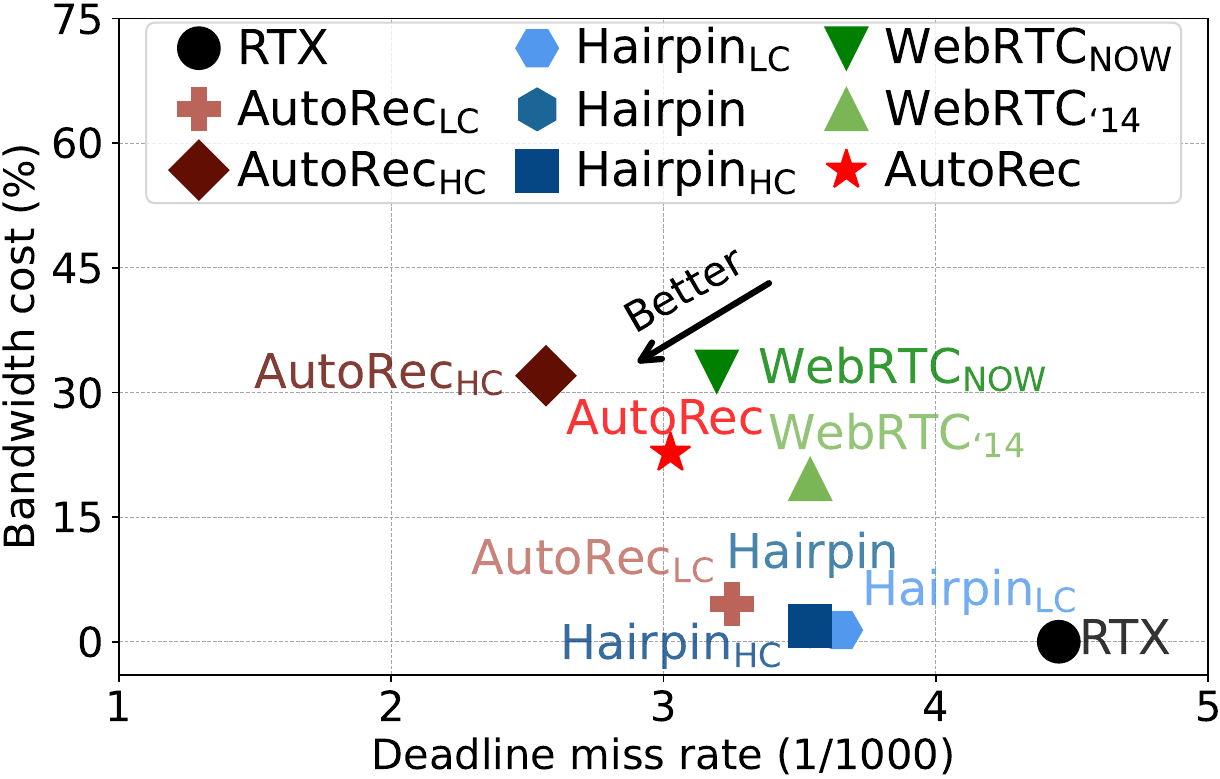}
\label{autorec_vs_fec_trace4cell}
}
% \hspace{1mm}
\vspace{-4mm}
\caption{The performance of the \name{} and FEC mechanisms under both controlled fixed packet loss rates and real-world network traces.}\label{autorec_vs_fec}
\vspace{-4mm}
\end{figure}

\xu{
% First, we evaluate the deadline miss rate and bandwidth cost of \name{}, \name{}$\mathsf{_{LC}}$, \name{}$\mathsf{_{HC}}$, WebRTC$\mathsf{_{'14}}$, WebRTC$\mathsf{_{'NOW}}$, Hairpin, Hairpin$\mathsf{_{LC}}$, and Hairpin$\mathsf{_{HC}}$ under both four fixed packet loss rates (0.5\%, 1\%, 5\%, 10\%) and two real-world network traces using an ns3-based WebRTC simulator~\cite{Soonyangzhang2020webrtcns3}.
First, using an ns3-based WebRTC simulator~\cite{Soonyangzhang2020webrtcns3}, we evaluate the deadline miss rate and bandwidth cost of \name{}, \name{}$\mathsf{_{LC}}$, \name{}$\mathsf{_{HC}}$, WebRTC$\mathsf{_{'14}}$, WebRTC$\mathsf{_{'NOW}}$, Hairpin, Hairpin$\mathsf{_{LC}}$, and Hairpin$\mathsf{_{HC}}$ under a variety of conditions, including four fixed packet loss rates (0.5\%, 1\%, 5\%, 10\%) and two real-world network traces.
Specifically, Trace \#1 represents a high-loss, stable-latency environment, characterized by an average bandwidth of 11.3 Mbps, an average RTT of 13.7 ms, an RTT variance of 13, an average packet loss rate of 10.1\%, and a 95th percentile packet loss rate of 75.0\%. In contrast, Trace \#2 exemplifies a low-loss, high-jitter RTT scenario, with an average bandwidth of 4.6 Mbps, a longer average RTT of 38.1 ms, a substantially larger RTT variance of 178, an average packet loss rate of 2.8\%, and a 95th percentile packet loss rate of 25.0\%.
}

\xu{Fig. \ref{autorec_vs_fec} illustrates the experimental results. As shown in Fig. 
\ref{autorec_vs_fec_0.5} and 
Fig. \ref{autorec_vs_fec_1}, at low packet loss rates (0.5\% and 1\%), \name{}$\mathsf{_{HC}}$, \name{}, and \name{}$\mathsf{_{LC}}$ all achieve 0\% deadline miss rate with modest bandwidth cost, where the highest cost are 9.92\%, 9.92\%, and 4.39\% respectively, satisfying redundancy cost constraint.} While WebRTC$\mathsf{_{NOW}}$ and WebRTC$\mathsf{_{'14}}$ also achieve 0\% deadline miss rate, they incur substantially higher bandwidth costs reaching 32.81\% and 16.16\%. Hairpin$\mathsf{_{HC}}$ achieves 0\% deadline miss rate at both loss rates, whereas Hairpin fails at 0.5\% loss and Hairpin$\mathsf{_{LC}}$ fails at both rates. Notably, Hairpin in all three configurations maintains minimal bandwidth costs under 3\%.

% At high loss rates (5\% and 10\%), \name{}$\mathsf{_{HC}}$ maintains 0\% deadline miss rate with bandwidth cost up to 46.28\%. However, \name{} fails to achieve 0\% deadline miss rate at 10\% loss rate (17.23\% bandwidth cost), while \name{}$\mathsf{_{LC}}$ fails to achieve 0\% deadline miss rate completely, performing comparably to RTX at 10\% loss rate - all within redundancy cost constraint. WebRTC$\mathsf{_{'NOW}}$ achieves 0\% deadline miss rate (bandwidth cost $\leq$ 39.38\%) but WebRTC$\mathsf{_{'14}}$ fails at both loss rates (bandwidth cost $\leq$  12.77\%). Hairpin$\mathsf{_{HC}}$ and standard Hairpin achieve 0\% deadline miss rate with bandwidth cost up to 44.48\% and 35.48\% respectively, while Hairpin$\mathsf{_{LC}}$ fails. Importantly, at 10\% loss, Hairpin$\mathsf{_{LC}}$ achieves greater deadline miss rate reduction than \name{} using less bandwidth cost.

\xu{As shown in Fig. 
\ref{autorec_vs_fec_5} and 
Fig. \ref{autorec_vs_fec_10}, at high loss rates (5\% and 10\%), \name{}$\mathsf{_{HC}}$ maintains a 0\% deadline miss rate with bandwidth cost up to 46.28\%. }However, \name{} fails to achieve 0\% deadline miss rate at 10\% loss rate while incurring up to 17.23\% bandwidth cost. \name{}$\mathsf{_{LC}}$ completely fails to achieve 0\% deadline miss rate and performs comparably to RTX at 10\% loss rate. \name{} in all three configurations satisfies its respective redundancy cost constraints. Neither WebRTC$\mathsf{_{NOW}}$ nor WebRTC$\mathsf{_{'14}}$ achieves a 0\% deadline miss rate under either loss rate, with bandwidth costs below 39.38\% and 12.77\% respectively. Hairpin$\mathsf{_{HC}}$ and standard Hairpin achieve 0\% deadline miss rate with bandwidth cost up to 44.48\% and 35.48\% respectively, while Hairpin$\mathsf{_{LC}}$ fails with bandwidth cost up to 10.26\%. Notably, at 10\% loss rate, Hairpin$\mathsf{_{LC}}$ reduces deadline miss rate more significantly than \name{} while consuming less bandwidth.

\xu{
% Second, we evaluate the deadline miss rate and bandwidth cost of \name{}, \name{}$\mathsf{_{LC}}$, \name{}$\mathsf{_{HC}}$, WebRTC$\mathsf{_{'14}}$, WebRTC$\mathsf{_{'NOW}}$, Hairpin, Hairpin$\mathsf{_{LC}}$, and Hairpin$\mathsf{_{HC}}$ under real-world traces using an ns3-based WebRTC simulator~\cite{Soonyangzhang2020webrtcns3}.

As shown in Fig. \ref{autorec_vs_fec_sample_tr}, under Trace \#1, \name{} performs considerably worse than Hairpin but outperforms WebRTC. Specifically, in this high-loss environment, the \name{}$\mathsf{_{LC}}$ instance injects negligible redundant packets to avoid exceeding the overhead limit, resulting in minimal improvement to the deadline miss rate. In contrast, the \name{} and \name{}$\mathsf{_{HC}}$ instances reduce the deadline miss rate to 0.067\% and 0.051\%, with associated bandwidth costs of 18.5\% and 55.9\%, respectively. However, owing to the superior efficiency of FEC over simple retransmission, even the best-performing \name{} instance (\name{}$\mathsf{_{HC}}$) is surpassed by Hairpin$\mathsf{_{LC}}$. The latter achieves a lower deadline miss rate of 0.045\% at a significantly lower bandwidth cost of only 0.27\%. Among the WebRTC versions, WebRTC$\mathsf{_{'NOW}}$ achieves the best result, yet its deadline miss rate (0.073\%) remains higher than that achieved by the \name{} instance, with a bandwidth cost of 14.7\%.

As shown in Fig. \ref{autorec_vs_fec_trace4cell}, under Trace \#2, \name{} demonstrates stronger performance than both Hairpin and WebRTC. Owing to the low loss rate and high RTT variability, Hairpin overestimates the number of remaining retransmission opportunities, leading to an insufficient number of FEC packets. Consequently, the best-performing instance of Hairpin (Hairpin$\mathsf{_{HC}}$) only reduces the deadline miss rate to 0.35\% with a bandwidth cost of 1.85\%. In comparison, the least effective instance of \name{} (\name{}$\mathsf{_{LC}}$) achieves a lower deadline miss rate of 0.33\% with a bandwidth cost of 4.56\%. Among the WebRTC versions, WebRTC$\mathsf{_{'NOW}}$ achieves the best result, yet its deadline miss rate and bandwidth cost are both worse than those of \name{}.
}

These results demonstrate \name{}'s effectiveness in RTC scenarios. Under low packet loss conditions, \name{} outperforms these FEC mechanisms, achieving comparable stall reduction with similar or lower bandwidth cost. However, as packet loss rates increase, \name{}'s effectiveness diminishes relative to these FEC mechanisms. To match their stall reduction at higher loss rates, \name{} incurs greater bandwidth costs that increasingly exceed those of the FEC mechanisms.

\begin{figure}[tbp]
\centering
\subfigure[Percent of video spent frozen vs. packet loss rates in the bad state of Gilbert-Elliott model.]{
\includegraphics[width=4.1cm]{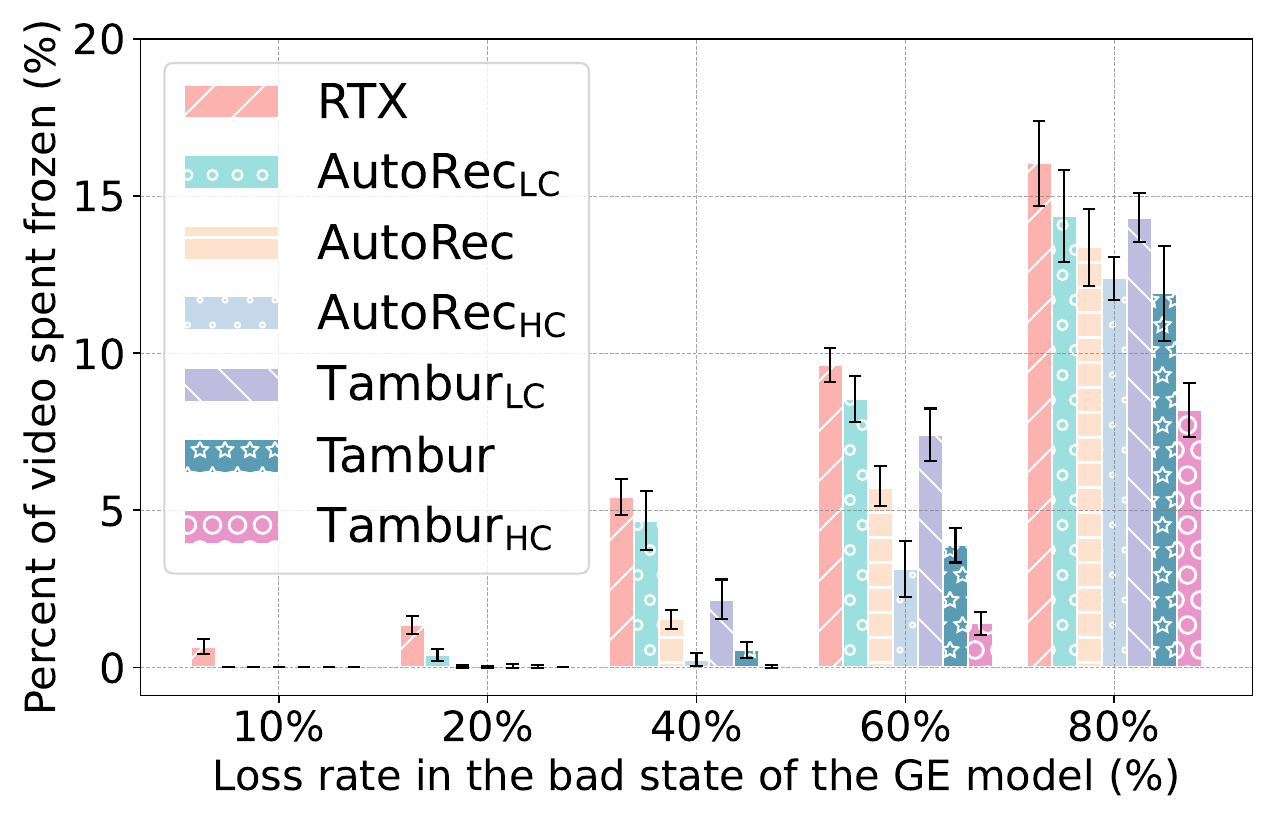}
\label{freezing_duration_autorec_vs_tambur}
}
% \hspace{1mm}
\subfigure[Bandwidth cost vs. packet loss rates in the bad state of Gilbert-Elliott model.]{
\includegraphics[width=4.1cm]{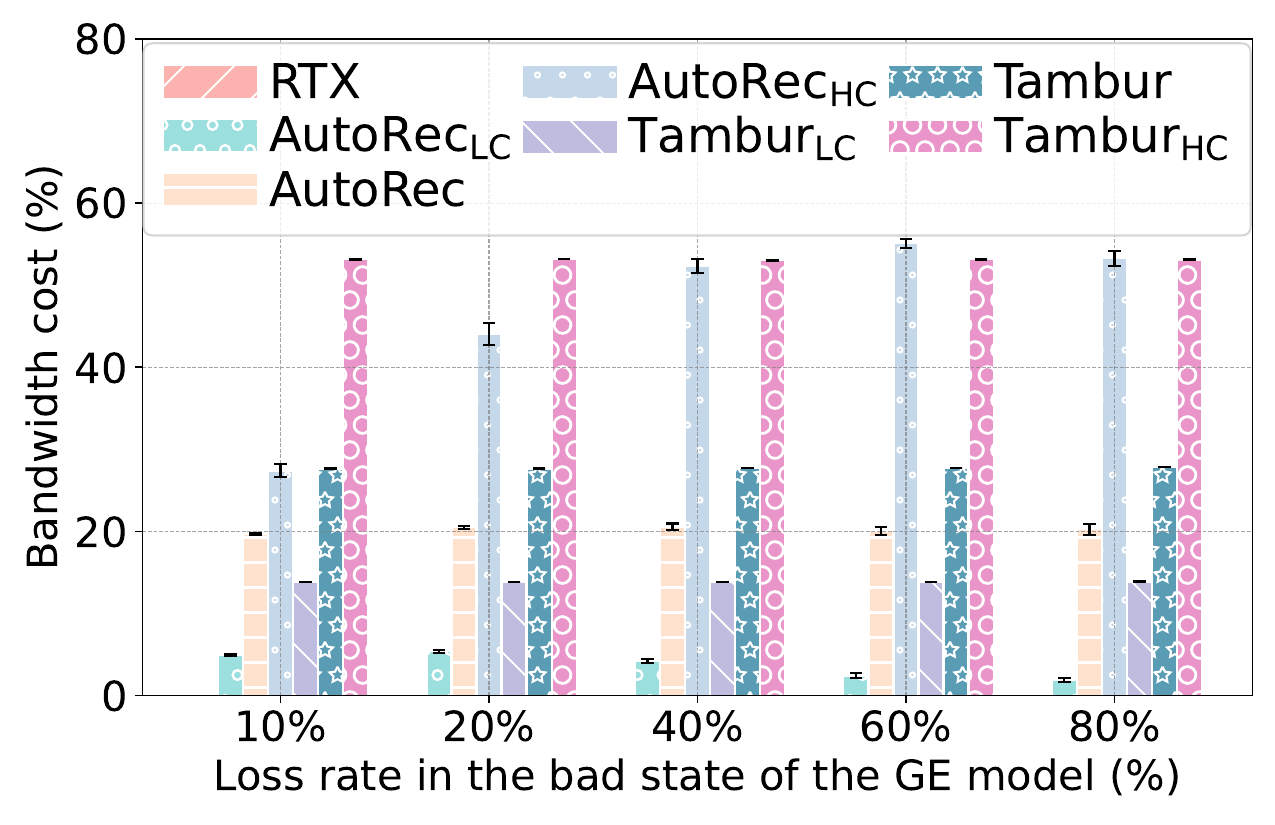}
\label{bandwidth_cost_autorec_vs_tambur}
}
\vspace{-3mm}
\caption{The performance of the \name{} and Tambur under varying packet loss rates in the bad state of Gilbert-Elliott model.}\label{autorec_vs_tambur}
\vspace{-5mm}
\end{figure}

Second, we evaluate the percent of video spent frozen and bandwidth cost of \name{}, \name{}$\mathsf{_{LC}}$, \name{}$\mathsf{_{HC}}$, Tambur, Tambur$\mathsf{_{LC}}$, and Tambur$\mathsf{_{HC}}$ in Gilbert-Elliott loss model~\cite{Elliott1963Estimates} using Ringmaster (an extensible videoconferencing research platform for FEC benchmarking, replacing WebRTC for controlled experiments)~\cite{Microsoft2020ringmaster}.

\xu{
The Gilbert-Elliott (GE) model is a two-state Markov chain widely used to simulate burst packet losses in networks, where the channel alternates between good (low-loss) and bad (high-loss) states. 
We implement a Gilbert-Elliott loss model with state transitions every 30 ms, fixed transition probabilities (good to bad: 20\%; bad to good: 80\%), fixed good state loss rate (1\%), and varying bad state loss rates (10\%, 20\%, 40\%, 60\%, 80\%) to evaluate \name{}'s performance under varying burst loss intensities.
Ringmaster utilizes a constant bitrate transmission scheme without congestion control. It does not employ a timeout-based termination mechanism. Instead, if the first unacknowledged datagram remains outstanding for more than one second, the sender forces the next frame to be a key frame and continues transmission.

Fig. \ref{autorec_vs_tambur} illustrates the experimental results. At low bad state loss rate (10\%), \name{}, \name{}$\mathsf{_{LC}}$, and \name{}$\mathsf{_{HC}}$ can reduce the percent of video spent frozen to 0\%, with \name{}$\mathsf{_{LC}}$ operating under the strictest redundancy cost constraint and incurring only 4.93\% bandwidth cost. Meanwhile, Tambur, Tambur$\mathsf{_{LC}}$, and Tambur$\mathsf{_{HC}}$ achieve 0\% of video spent frozen. Tambur$\mathsf{_{LC}}$ (the one with the smallest bandwidth cost among Tambur, Tambur$\mathsf{_{LC}}$, and Tambur$\mathsf{_{HC}}$) requires over 13.81\% bandwidth cost, significantly exceeding \name{}'s bandwidth cost. At moderate bad state loss rates (20\% to 40\%), the performance in reducing the percent of video spent frozen of both \name{} and Tambur across configurations consistently correlates with their bandwidth cost rankings. Under high bad state loss rates (60\% and 80\%), Tambur achieves 3.88\% and 11.91\% of video spent frozen at a bandwidth cost of $\sim$28\%; by contrast, even \name{}$\mathsf{_{HC}}$ (the best-performing one for freeze reduction among \name{}, \name{}$\mathsf{_{LC}}$, and \name{}$\mathsf{_{HC}}$) achieves only 3.13\% and 12.38\% of video spent frozen at a bandwidth cost of $\sim$53\%.

These results demonstrate that \name{} delivers strong performance even under burst packet loss conditions. While \name{} outperforms Tambur at low burst loss rates, its effectiveness in reducing video freezes diminishes as the burst loss rate increases. Under high burst loss rates, \name{} falls short of Tambur's performance. 

In summary, \name{} effectively reduces video freezes within redundancy cost constraints for RTC applications. It consistently mitigates freezes under all tested packet loss conditions, including varying loss magnitudes (low to high) and types (random and burst), though its effectiveness decreases as loss rates increase. Crucially, \name{} outperforms multiple FEC schemes at low loss rates, matches their performance at moderate loss rates, but is surpassed by multiple FEC mechanisms in high-loss scenarios.
}

\vspace{-1mm}

\begin{figure}[tbp]
\centering
\subfigure[The average and quantile of the benefits.]{
\includegraphics[width=4.0cm]{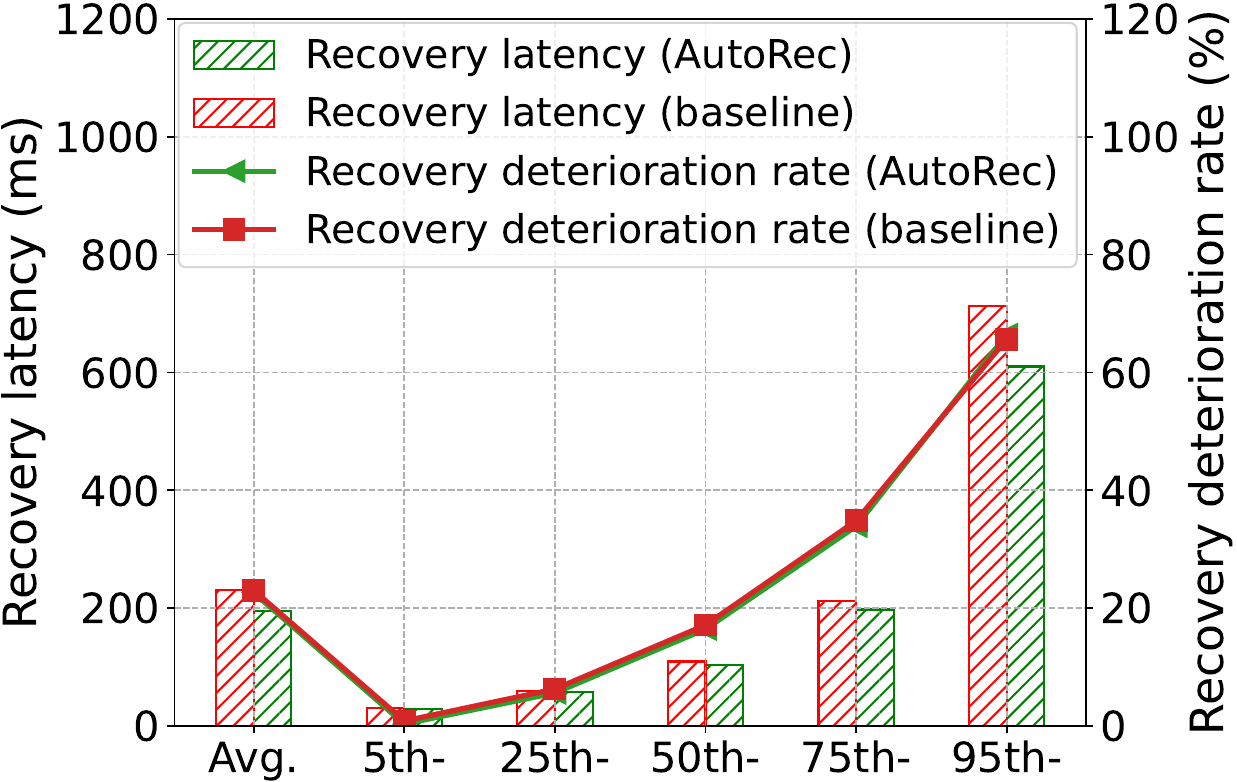}
\label{benefit_real_fig}
}
\hspace{1mm}
\subfigure[The average overhead and the quantile of the goodput.]{
\includegraphics[width=4.0cm]{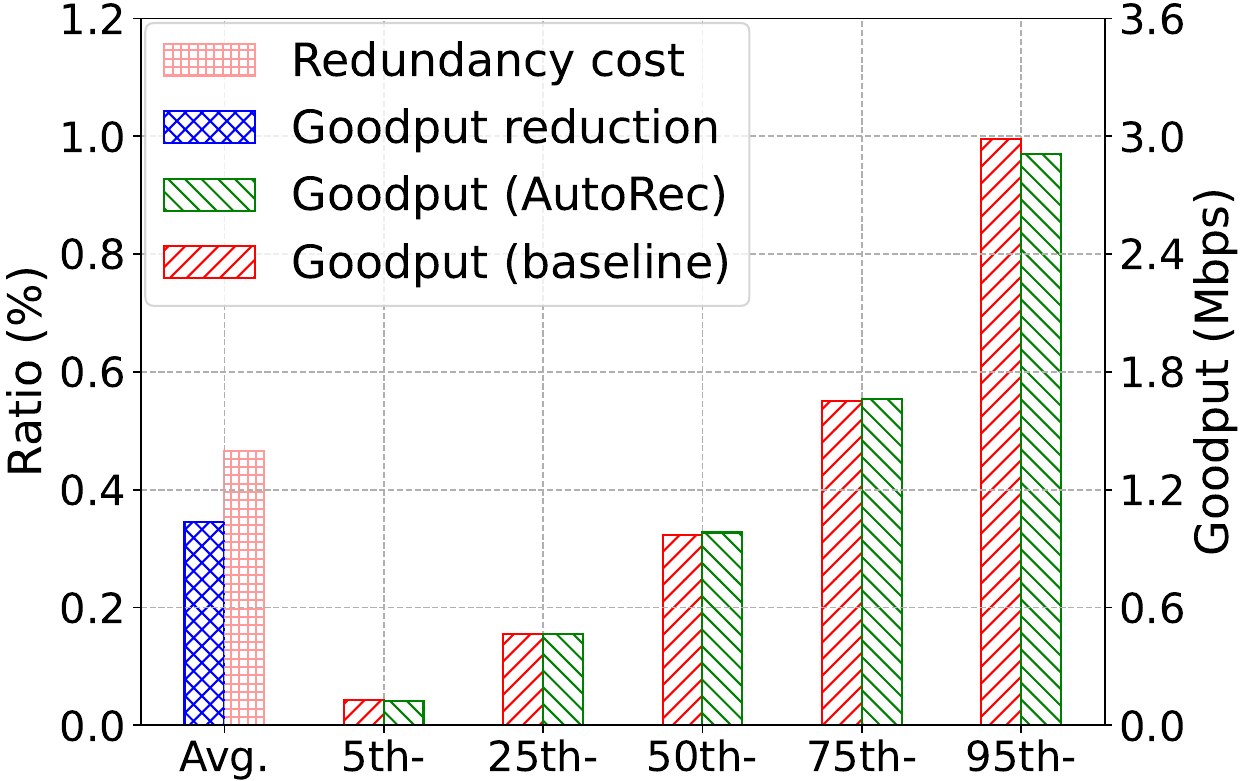}
\label{overhead_real_fig}
}
\vspace{-4mm}
\caption{\name{}'s benefits and overhead in the real network.}\label{benefits_and_overhead_real_fig}
\vspace{-4mm}
\end{figure}

% \vspace{-3mm}
\subsection{Real-Network Evaluation}
\label{real_network_sec}
To further explore \name{}'s performance, 
% the real-network experiments are carried out, where the L7-player freezing is also evaluated in addition to \textsf{LRQ} and reinjection cost. 
% To evaluate the data timeliness changes of live streaming, 
we deploy \name{} prototype in our CDN proxy and evaluate video freezing metrics, loss recovery benefits, and reinjection overhead.

\vspace{1ex} \noindent \textbf{Parameter configuration.}
\label{deloy_experience_sec}
To facilitate the deployment of \name{} in real-world networks,
% the following rules are obeyed, which are based on our experiences for optimizing player video freezing. 
% To address the issue that BBRv1 is hard and slow to exit its startup phase \cite{cardwell2019bbrv2}, some conditions are configured, which contain (i) the loss rate threshold of 5\%, and (ii) the ratio threshold (i.e., 15) of monitored \textsf{SRTT} to the minimum RTT. 
the user-customizable parameters in the Redundancy Adapter (\S \ref{redundancy_adaption_sec}), namely recovery latency tolerance ($\alpha$), redundancy cost tolerance ($\beta$), and goodput reduction tolerance ($\gamma$), are set to 200 ms, 50\%, and 50\%, respectively.
% $\theta_\textsf{thres}$ and $\alpha$ of \name{} are configured as 30ms and 0.1, respectively. 
% To avoid the issues caused by higher \textsf{loss rate}, the deployed \name{} is recommended to be turned off or kept inactivated if the newly monitored \textsf{loss rate} $>$ 10\%. 

% \vspace{-1mm}

\vspace{1ex} \noindent \textbf{\name{}'s loss recovery benefits and overhead.}
\label{lrq_benefit_real_network_sec}
% \vspace{-1mm}
% \name{} keeps continuous optimizations for recovery latency and recovery deterioration rate in the real network, as Fig. \ref{benefit_real_fig} shows.
Fig. \ref{benefit_real_fig} demonstrates the performance of \name{} in the real network, showing consistent improvements in recovery latency and recovery deterioration rate.
% We can learn the average recovery latency and recovery deterioration rate can be lowered by the ratio of 15.8\% and 1.7\%, whose values are reduced from 231.2 ms and 23.0\% to 194.6 ms and 22.6\%, respectively. 
The average recovery latency is reduced by 15.8\%, from 231.2 ms to 194.6 ms, while the average recovery deterioration rate decreases by 1.7\%, from 23.0\% to 22.6\%.
% In particular, high-percentile (i.e., 95th-) recovery latency can be decreased by 102.2 ms, which means 5\% receivers take 102.2 ms less to wait to recover some loss if the first-time resent data is detected lost again. 
Notably, the 95th percentile recovery latency is reduced by 102.2 ms. This means that for the worst-affected 5\% of receivers, the waiting time for recovering from the loss of a first retransmission is cut by 102.2 ms.
These results indicate that \name{} effectively reduces the recovery latency below the recovery latency tolerance, which satisfies the recovery latency constraints.

The overhead of \name{} under the evaluated parameter configuration is shown in Fig. \ref{overhead_real_fig}, which demonstrates that the introduced overhead is negligible. Specifically, the average redundancy cost of \name{} is only 0.47\%, indicating that the total traffic increases by at most 0.47\%. Meanwhile, the average goodput reduction of \name{} is just 0.34\%. Compared to the baseline, the reduction in goodput is more pronounced at higher percentiles, with a decrease of 2.58\% at the 95th percentile, from 2.99 Mbps to 2.91 Mbps. From the above results, it is evident that \name{}'s redundancy cost and goodput reduction are both strictly within the redundancy cost tolerance and goodput reduction tolerance, meeting the constraints on overhead.

In summary, in the real network, the deployed \name{} accelerates loss recovery to meet user demands whenever possible while keeping the overhead controllable.

\begin{figure}[tbp]
\centering
\subfigure[Video freezing times.]{
\includegraphics[width=4.0cm]{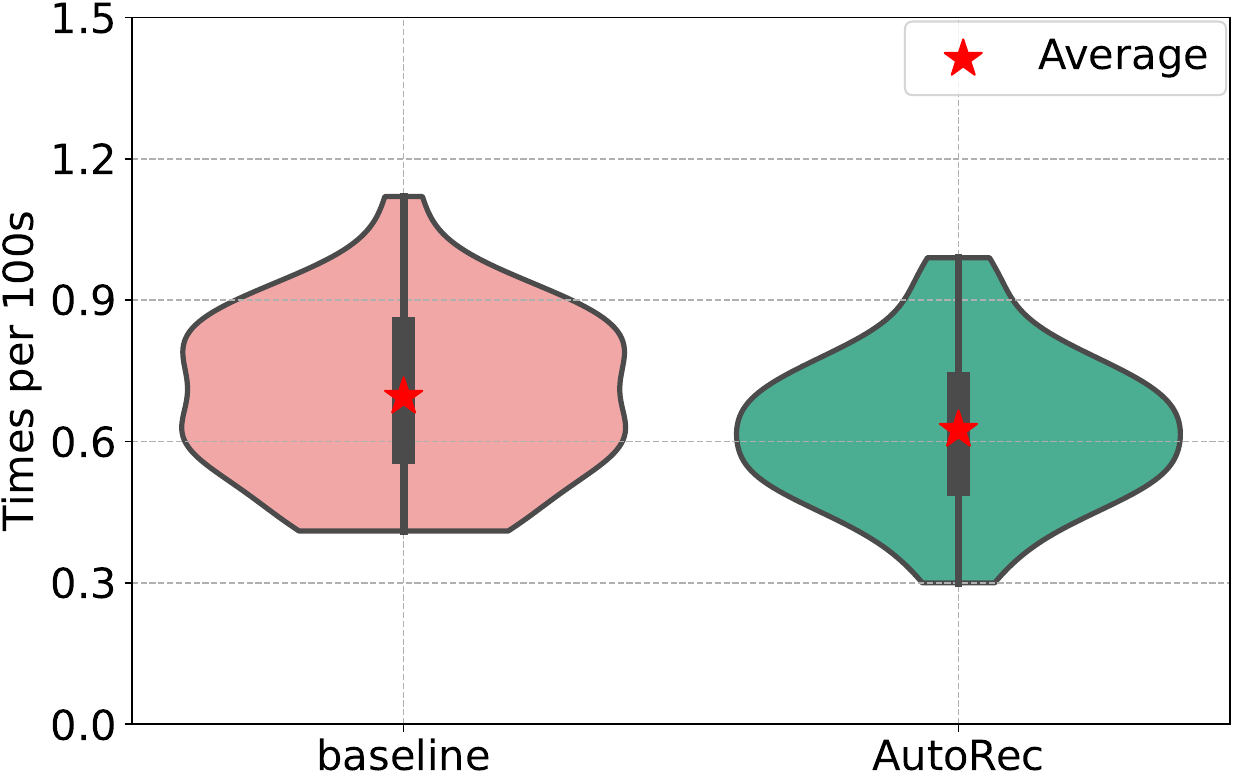}
\label{l7_freezing_times_real_fig}
}%\hspace{1mm}
\subfigure[Video freezing duration.]{
\includegraphics[width=4.0cm]{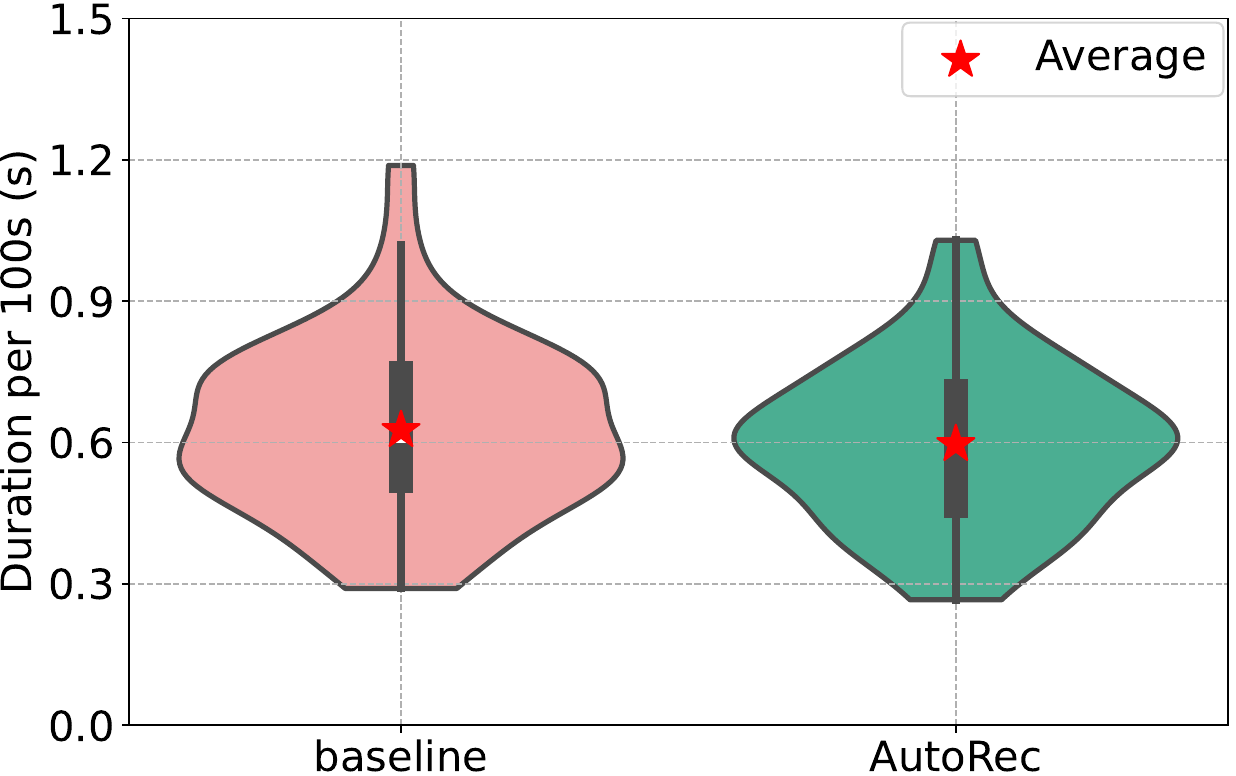}
\label{l7_freezing_duration_real_fig}
}
\vspace{-1mm}
\caption{The video freezing changes in the real network.}\label{l7_player_freezing_evaluation_fig}
\vspace{-2mm}
\end{figure}

\vspace{1ex} \noindent \textbf{Video freezing.}
\label{video_freezing_evaluate_sec}
The \name{} performance can be further evaluated by the observed client-side video freezing, including its frequency and duration. 
As Fig. \ref{l7_freezing_times_real_fig} shows, \name{} reduces the average freezing times (per 100s) from 0.69 to 0.62, representing an improvement of 10.10\%. The minimum and maximum values of video freezing times (per 100s) are reduced by 26.83\% and 11.61\%, respectively.
% As Fig. \ref{l7_freezing_times_real_fig} shows, the average freezing times (per 100s) can be optimized by 10.1\% from 0.69 to 0.62, whose min and 95th-percentile can be lowered by 24.4\% and 34.1\%, respectively. 
Furthermore, \name{} also reduces freezing duration (per 100s) by 4.74\%, in which the maximum value is reduced by 159.03 ms (with the ratio of 13.39\%), as Fig. \ref{l7_freezing_duration_real_fig} shows. 

% In summary, \name{} is worthwhile for optimizing the timeliness of loss recovery and client-side video freezing. 
In summary, the results demonstrate that \name{} effectively enhances loss recovery timeliness, leading to a measurable reduction in both the frequency and duration of client-side video freezing.

% \vspace{-3mm}

\section{Discussion}
\label{discussion_sec}

% \vspace{-2mm}

\subsection{What about the overhead of \name{}?}
\label{computational_and_storage_overhead_sec}

The computational and storage overhead of the \name{} remains
minimal even under extreme conditions.

% \vspace{-1mm}
\vspace{1ex} \noindent 

\textbf{Computational overhead.}
The computational overhead of \name{} consists of the following two aspects: 
%(i) The calculation of the average values of network metrics in the Redundancy Adapter;
(i) the calculation of the redundancy level in the Redundancy Adapter; (ii) the computational operations of insertion, deletion, and movement of the reinjection queue in the Reinjection Controller. 

%(iv) The calculation of $T_{thres}$ in the Reinjection Controller. 
% For aspect (i), regarding network metrics, it only requires recording information and updating the corresponding cumulative variables each time a packet is sent, lost, retransmitted, or received. Then, at the end of a decision cycle, the average can be calculated using the cumulative variables, without the need for excessive calculations. 
For aspect (i), the redundancy level is periodically calculated at each decision interval. The time complexity of this calculation is $O(log(n))$, where $n$ is the maximum allowable value of $K_\theta$ (\S \ref{redundancy_adaption_sec}), typically set to 10. Given the small value of $n$ and the logarithmic time complexity, the computational overhead is negligible.

For aspect (ii), since the insertion, deletion, and movement operations do not affect the order that the entire reinjection queue should maintain, i.e., an order where the time of their last transmission is from the earliest at the front to the latest at the back, there is no need for 
% computationally expensive
sorting operations. Insertion and movement operations do not require traversing the reinjection queue. Instead, they only access the head and tail of the reinjection queue, which means no excessive computation is required. However, deletion requires traversing the reinjection queue to find the acknowledged packets and then delete them. The time complexity here is $O(n)$, where $n$ is the number of packets in the reinjection queue. In the worst-case scenario, namely that all sent packets are lost, the reinjection queue will contain all the packets sent within one RTT. According to our measurement results, the average bitrate of live streams is 984.5 kbps, the average SRTT is 95.5 ms, and the amount of data carried by each packet is approximately 1300 bytes. 
% Therefore, in the worst-case scenario, the average number of packets in the reinjection queue of each connection is calculated as follows:
% \begin{equation}
%  984.5 \times 10^3 \times 95.5 \times 10 ^{-3} \div 8 \div 1300  \approx 9
% \end{equation}
Therefore, in the worst-case scenario, the average number of packets in the reinjection queue of each connection is approximately 9.
So, the overhead required to traverse the reinjection queue once is extremely small.

% \xu{
% In conclusion, the computational overhead of the \name{} remains minimal even under extreme conditions.
% }
\vspace{-1mm}
\vspace{1ex} \noindent

 \textbf{Storage overhead.} The storage overhead of \name{} mainly consists of the following two aspects: (i) the storage overhead required for measuring the loss detection time in the Redundancy Adapter; (ii) the storage overhead needed to implement the reinjection queue in the Reinjection Controller.

For aspect (i), we need to store the first transmission time of each currently unacknowledged packet so that the loss detection time of the packet can be obtained when it is first retransmitted. 
As analyzed in \S \ref{computational_and_storage_overhead_sec}, the average number of packets in the reinjection queue of each connection is 9, therefore, the storage overhead of storing the sending time of each currently unacknowledged packet will not be large either.

For aspect (ii), we need to store each packet that is currently being retransmitted but has not been acknowledged and whose number of injection times is less than $K_{\theta}$ (\S\ref{redundancy_adaption_sec}) in the reinjection queue. Additionally, we have to record some information for these packets. Evidently, compared with the overhead of storing packets, the overhead of recording information for these packets can be neglected. As analyzed in \S \ref{computational_and_storage_overhead_sec}, the average number of packets in reinjection queue at any given moment is 9. Therefore, in the worst-case scenario, the overhead of storing packets in reinjection queue will not be significant either.

% \vspace{-2mm}
% \xu{
% In summary, the storage overhead of the \name{} remains minimal even under extreme conditions.
% }

\subsection{Will \name{} exacerbate congestion?}
In fact, the number of packets injected by \name{} will not be so large as to exacerbate network congestion. Firstly, \name{} can adaptively adjust the number of injections for each packet. When there is congestion, the packet loss rate will increase. At this time, \name{} will adaptively reduce the number of injections for each lost packet in the next decision interval according to the goodput reduction tolerance and redundancy cost tolerance, so that the total number of injected packets will not exceed a certain threshold. In addition, as analyzed in \S \ref{computational_and_storage_overhead_sec}, the average number of packets that are sent but not acknowledged is very small. As shown in Fig. \ref{loss_feature_wild}, the average packet loss rate in the real network is about 4\%, that is, the lost packets account for a very small part of the total number of sent packets. And we only inject packets that are retransmitted but not acknowledged, so the number of injected packets will not be large to exacerbate congestion.

% \vspace{-2mm}

\subsection{How \name{} integrates with QUIC's existing loss-recovery mechanisms?}
QUIC's native loss-recovery mechanism exclusively tracks the last transmission attempt (whether initial transmission, retransmission, or reinjection) for each packet. It utilizes this final attempt's information to determine packet loss and initiate retransmissions, thereby ensuring reliable data delivery. \name{}, conversely, performs reinjection of previously retransmitted or injected packets only under specific conditions
% (during application-limited states or the time elapsed since the last transmission of the packet has exceeded threshold)
. Notably, both reinjection and retransmission operations require first renumbering the original data packet and then sending the renumbered packet. Crucially, \name{} fundamentally advances retransmission timing by executing these packet retransmissions before loss occurs. This process operates without interfering with QUIC's native loss-recovery mechanism or compromising reliability.

Illustrative example (AutoRec's redundancy level = 2): When QUIC's native loss-recovery mechanism detects that packet $P_1$ (carrying application data $D$) is lost, it retransmits $P_1$ by renumbering it as $P_2$ and sending $P_2$. QUIC then stops monitoring $P_1$ for loss and begins monitoring $P_2$ for loss. If AutoRec's conditions are met, it reinjects $P_2$ by renumbering $P_2$ as $P_3$ and sending $P_3$. QUIC consequently stops monitoring $P_2$ for loss and begins monitoring $P_3$ for loss. When AutoRec’s conditions are met again, it reinjects $P_3$ by renumbering $P_3$ as $P_4$ and sending $P_4$. QUIC then stops monitoring $P_3$ for loss and begins monitoring $P_4$ for loss. Upon a third trigger, AutoRec skips reinjection of $P_4$ because the packet carrying data $D$ has been reinjected twice, reaching the maximum allowed by the redundancy level. Should $P_4$ be lost, QUIC's native loss-recovery mechanism detects this loss, retransmits $P_4$ by renumbering $P_4$ as $P_5$ and sending $P_5$, then stops monitoring $P_4$ for loss and begins monitoring $P_5$ for loss. Thereafter, QUIC's existing loss-recovery mechanism ensures data $D$ is successfully delivered to the receiver through this persistent cycle of loss detection and retransmission. Notably, if any packet from $P_1$ to $P_5$ is ACKed, the loss detection for $P_5$ is terminated to prevent infinite retransmission loops.

\vspace{-1mm}

\section{Related Work}
\label{relatedwork_sec}
%In this section, we discuss the related research on loss tolerance control. 

%\textbf{FEC-based coding.}

%\textbf{Semi-reliability control.}

%\textbf{Multi-path retransmission.} 

%\textbf{ARQ variants.} 
% \vspace{-2mm}
\vspace{1ex} \noindent \textbf{Adaption to on-off traffic pattern.} To the best of our knowledge, almost all prior works~\cite{wierman2003unified, wierman2003modeling, esteban2012interactions,kupka2012performance,  de2013elastic, huang2016tuning, zhao2017off, arun2018copa,li2020tack, yanev2022does,wu2022autoplex,yan2023poster,wu2024reducing} fall into the category of adapting congestion control to on-off traffic patterns. For example, Zhang et al. \cite{huang2016tuning} proposed a TCP variant to overcome the challenges, in which the on-off traffic pattern disturbs the increase of the TCP congestion window and triggers packet loss at the beginning of the on-mode. 
% This paper does not focus on the congestion control issues of live streaming. Instead, as TADOC~\cite{zhang2021tadoc} optimizes text analytics by combining repetitive textual features with compression, we leverage the on-off pattern of live streaming to accelerate packet loss recovery.
This paper does not focus on the congestion control issues of live streaming. Instead, in a manner analogous to how TADOC~\cite{zhang2021tadoc} leverages repetitive textual features to optimize text analytics, we leverage the on-off pattern to accelerate packet loss recovery.

% we take a first step toward taming the loss tolerance control under the on-off pattern for live streaming. 

% \vspace{1ex} \noindent \textbf{Loss tolerance control for live streaming.} 
% Many studies have been proposed to enhance the data timeliness of live streaming to achieve efficient loss recovery and optimize client-side video freezes. The key ideas of these works include injecting supplement data (e.g., FEC~\cite{chan2006video,michel2019quic,michael2023tambur,meng2024hairpin} and multi-path retransmissions~\cite{zheng2021xlink,chen2018fuso,li2018measurement}), ignoring some non-critical losses (e.g., application-level controls \cite{vamanan2012deadline,zhou2021deadline,zhang2015more} and semi-reliable transmissions \cite{palmer2021voxel,pauly2018unreliable}). However, in commercial large-scale live-streaming product networks, the CDN vendor is mainly responsible for optimizing the end-to-end transmission performance and has control rights only on the sender side rather than the client side. Therefore, the above arts that apply client-side modification cannot meet the requirements of real-world deployment. In this paper, we propose a sender-side approach from the perspective of CDN vendors.

\vspace{1ex} \noindent
\textbf{Loss tolerance control for live streaming.} 
Many studies have been proposed to enhance the data timeliness of live streaming to achieve efficient loss recovery and optimize client-side video freezes. The key ideas of these works include injecting supplement data (e.g., FEC~\cite{chan2006video,michel2019quic,michael2023tambur,meng2024hairpin} and multi-path retransmissions~\cite{zheng2021xlink,chen2018fuso,li2018measurement}), ignoring some non-critical losses (e.g., application-level controls \cite{vamanan2012deadline,zhou2021deadline,zhang2015more}, semi-reliable transmissions \cite{palmer2021voxel,pauly2018unreliable, cheng2024grace}). For instance, Tambur~\cite{michael2023tambur} employs streaming codes, a specialized FEC variant optimized for burst losses. In contrast, \name{} optimizes loss recovery for diverse loss patterns by selectively reinjecting only lost packets, significantly reducing bandwidth overhead under low-loss conditions. Whereas GRACE~\cite{cheng2024grace} preserves QoE across varying packet losses via a novel neural video codec implemented through specialized application-layer frame encoding and decoding, \name{} operates at the transport layer by reinjecting lost packets. The mechanism most close to \name{} is Hairpin~\cite{meng2024hairpin}, which adaptively applies FEC encoding to lost packets specifically for edge-based interactive streaming. In fact, Hairpin can be viewed as an advanced variant of \name{} designed for scenarios where modifying both client and server endpoints is feasible, since under these conditions, FEC is more efficient than simple packet reinjection.
% Conversely, when both endpoints cannot be modified simultaneously, we can only resort to simple packet reinjection as implemented by \name{}. 
However, in commercial large-scale live-streaming product networks, the CDN vendor is mainly responsible for optimizing the end-to-end transmission performance and has control rights only on the sender side. Therefore, the above arts that apply client-side modification cannot meet the requirements of real-world deployment. In this paper, we propose a sender-side approach from the perspective of CDN vendors.

\vspace{1ex} \noindent \textbf{Advancements upon ARQ.} 
Modern CDN vendors simply apply the ARQ for loss tolerance control in commercial live-streaming services. However, the legacy ARQ-based loss recovery is far from satisfactory. There also exist many studies~\cite{bakshi1997improving,barman2004tcp,chen2009towards,paxson2011computing,xie2022revisiting,liu2013improving,liu2019reducing,li2023art} that improve the performance of ARQ by introducing redundancy to loss recovery. These works, however, suffer from obviously deteriorated goodput since the inserted extra packets occupy the sender-side and in-network resources. This paper overcomes the above challenges by taking full advantage of the off-mode of live streaming. 

\vspace{1ex} \noindent 
\textbf{Upstream path.} 
The data transmission paths for live streaming can generally be simply divided into the upstream path from the live streamer to the server and the downstream path from the server to the user. Currently, many research studies ~\cite{kim2020neural,li2022onuploading,ray2019Vantage,siekkinen2017optimized} are dedicated to improving the transmission quality of the upstream path. In contrast to these efforts, \name{} primarily focuses on the optimization of the downstream path performance, with particular emphasis on reducing packet loss recovery latency. Since the downstream path is directly relevant to the users, optimizing the packet loss recovery latency using \name{} on the downstream path rather than the upstream path is more likely to result in an improvement of users' Quality of Experience.

\vspace{-1mm}

% \vspace{-5mm}
% width=3.8cm,height=2.22cm,angle=0

\section{Conclusion}
\label{conclusion_sec}
\xu{
Slow loss recovery is not primarily caused by the retransmission process itself, but rather by the loss of retransmitted packets, which constitutes the fundamental challenge. To address this issue, we propose \name{}, a mechanism that accelerates loss recovery by enabling senders to reinject lost packet duplicates in a strategic and controlled manner. \name{} leverages the on-off mode switching prevalent in live streaming to improve recovery timeliness without disrupting ongoing data transmission. The effectiveness of \name{} is validated through extensive testbed experiments and real-world deployment. It has been integrated into Tencent's global CDN \cite{tencentcdn_2024} and its EdgeOne platform \cite{tencenteo_2024}, serving billions of live-streaming users worldwide.

}

\bibliographystyle{unsrt}
\balance
\bibliography{reference}

\vspace{-10mm}

\begin{IEEEbiographynophoto}{Tong Li}(Member, IEEE) received his Ph.D. degree from the Department of Computer Science and Technology of Tsinghua University, China in 2017. He worked as a Chief Engineer at Huawei before 2022, and currently, he serves as an associate professor at Renmin University of China. His research interests include network protocols, distributed systems, and network security.
\end{IEEEbiographynophoto}

\vspace{-10mm}

\begin{IEEEbiographynophoto}{Xu Yan}
received his B. S. degree from the Department of Computer Science and Technology of Shandong University, Qingdao, China in 2023.
He is currently pursuing his M.S. degree at Renmin University of China, Beijing, China. His research interests include network transport protocols and congestion control.
\end{IEEEbiographynophoto}

\vspace{-10mm}

\begin{IEEEbiographynophoto}{Bo Wu} received his Ph.D. degree from the Department of Computer Science and Technology of Tsinghua University, Beijing, China in 2019. Currently, he serves as a research fellow in the Department of Cloud Architecture and Platform in Tencent Technologies. 
His research areas include next-generation Internet, network security, network transport, and congestion control.
\end{IEEEbiographynophoto}

\vspace{-10mm}

\begin{IEEEbiographynophoto}{Cheng Luo}
received his M.S. degree from Zhejiang University, Hangzhou, China, in 2015.
Currently, he leads the Application Framework Group at the Department of Cloud Architecture and Platform in Tencent Technologies. 
His research areas include Internet architecture, transmission optimization. 
\end{IEEEbiographynophoto}

\vspace{-10mm}

\begin{IEEEbiographynophoto}{Fuyu Wang}
received his M.S. degree in  Electromagnetic Field and Microwave Technology from UESTC, China, in 2013. Currently, he serves as a chief engineer in the Department of Cloud Architecture and Platform in Tencent Technologies. His research areas include Internet architecture, network transport protocols, and congestion control.  
\end{IEEEbiographynophoto}

\vspace{-10mm}

\begin{IEEEbiographynophoto}{Jiuxiang Zhu} received his B.S. degree from the Department of Computer Science and Engineering of Central South University, Changsha, China in 2024. He is working toward his M.S. degree at Renmin University of China, Beijing, China. His research interests include high-performance web proxy and QUIC optimization.
\end{IEEEbiographynophoto}

\vspace{-10mm}

% \begin{IEEEbiographynophoto}{Haoyi Fang} is working toward his bachelor's degree at Renmin University of China, Beijing, China. His research areas include networking and big data.

\begin{IEEEbiographynophoto}{Haoyi Fang} received his B.S. degree from Renmin University of China, Beijing, China. He is working toward his M.S. degree at Renmin University of China, Beijing, China. His research areas include networking and big data.
\end{IEEEbiographynophoto}

\vspace{-10mm}

\begin{IEEEbiographynophoto}{Xinle Du} received a B.E. degree from the Department of Computer Science and Technology of Xidian University, Xi'an, China in 2018, and a Ph.D. degree from the Department of Computer Science and Technology, Tsinghua University, Beijing, China, in 2023. Currently, he has worked as Chief Engineer at Computer Network and Protocol Lab, Huawei Technologies since 2023. His research interests include networking and LLM systems.
\end{IEEEbiographynophoto}

\vspace{-10mm}

\begin{IEEEbiographynophoto}{Ke Xu}(Fellow, IEEE) received the Ph.D. degree from Tsinghua University, Beijing, China. He is currently a Full Professor with the Department of Computer Science and Technology, Tsinghua University. He has published more than 200 technical articles in the research areas of next-generation internet, blockchain systems, and network security. He won the IWQoS 24 Best Paper Award and the Distinguished Paper Award at USENIX Security 23/24, CCS 25, and NDSS 25.
\end{IEEEbiographynophoto}

% \vfill

\end{document}